\documentclass[fleqn,usenatbib]{mnras}

\usepackage[T1]{fontenc}

\DeclareRobustCommand{\VAN}[3]{#2}
\let\VANthebibliography\thebibliography
\def\thebibliography{\DeclareRobustCommand{\VAN}[3]{##3}\VANthebibliography}

\usepackage{graphicx}	
\usepackage{amsmath}	
\usepackage{amssymb}	
\usepackage{xcolor}
\usepackage{lipsum}
\usepackage[english]{babel}
\usepackage{blindtext}
\usepackage{scalefnt}
\usepackage{soul}
\usepackage{multirow}
\usepackage{makecell}
\usepackage{arydshln}
\usepackage{float}
\usepackage{siunitx}

\newcommand{\Lbol}{$L_{\rm{bol}}$}
\newcommand{\fedd}{$\lambda_{\rm{Edd}}$}
\newcommand{\fgas}{$f_{H_2}$}

\newcommand{\fGD}{$f_{\rm{GD}}$}

\newcommand{\LbolEq}[2]{\ensuremath{L_{\rm{bol}}{#1} 10^{#2}\rm{\ erg\ s^{-1}}}}
\newcommand{\feddEq}[2]{\ensuremath{\lambda_{\rm{Edd}}{#1} 10^{#2}}}
\newcommand{\fgasEq}[2]{\ensuremath{f_{H_2}{#1} 10^{#2}}}
\newcommand{\MstarEq}[2]{\ensuremath{M_{\star}{#1} 10^{#2}\rm{\ M_{\odot}}}}
\newcommand{\ts}{\textsuperscript}

\usepackage{newtxtext,newtxmath}

\title[AGN feedback in cosmological simulations]{Cosmological simulations predict that AGN preferentially live in gas-rich, star-forming galaxies despite effective feedback}

\author[S.R. Ward et al.]{
S. R. Ward,$^{1,2,3}$\thanks{E-mail: sward@eso.org}
C. M. Harrison,$^{4}$\thanks{E-mail: christopher.harrison@newcastle.ac.uk} 
T. Costa$^{5}$ and
V. Mainieri$^{1}$
\\
$^{1}$European Southern Observatory, Karl-Schwarzschild-Straße 2, 85748 Garching bei München, Germany\\
$^{2}$Excellence Cluster ORIGINS, Boltzmannstraße 2, 85748 Garching bei München, Germany\\
$^{3}$Ludwig Maximilian Universität, Professor-Huber-Platz 2, 80539 München, Germany \\
$^{4}$School of Mathematics, Statistics and Physics, Newcastle University, Newcastle upon Tyne, NE1 7RU, UK \\
$^{5}$Max-Planck Institut für Astrophysik, Karl-Schwarzschild-Straße 1, 85748 Garching bei München, Germany
\\
}

\date{Accepted XXX. Received YYY; in original form ZZZ}

\pubyear{2022}

\begin{document}
\label{firstpage}
\pagerange{\pageref{firstpage}--\pageref{lastpage}}
\maketitle

\begin{abstract}

Negative feedback from active galactic nuclei (AGN) is the leading mechanism for the quenching of massive galaxies in the vast majority of modern galaxy evolution models. However, direct observational evidence that AGN feedback causes quenching on a population scale is lacking. Studies have shown that luminous AGN are preferentially located in gas-rich and star-forming galaxies, an observation that has sometimes been suggested to be in tension with a negative AGN feedback picture. We investigate three of the current cosmological simulations ({\sc IllustrisTNG}, EAGLE and SIMBA) along with post-processed models for molecular hydrogen gas masses and perform similar tests to those used by observers. We find that the simulations predict: (i) no strong negative trends between {\Lbol} and {\fgas} or sSFR; (ii) both high-luminosity ({\LbolEq{\geq}{44}}) and high-Eddington ratio ({\fedd}$\geq 1\%$) AGN are preferentially located in galaxies with high molecular gas fractions and sSFR; and (iii) that the gas-depleted and quenched fractions of AGN host galaxies are lower than a control sample of non-active galaxies. These three findings are in qualitative agreement with observational samples at $z=0$ and $z=2$ and show that such results are not in tension with the presence of strong AGN feedback, which all simulations we employ require to produce realistic massive galaxies. However, we also find quantifiable differences between predictions from the simulations, which could allow us to observationally test the different subgrid feedback models.
\end{abstract}

\begin{keywords}
galaxies: evolution -- galaxies: active-- quasars: supermassive black holes -- methods: numerical
\end{keywords}



\section{Introduction}
\label{sec:intro}

The population of galaxies in our universe exhibits a bimodal distribution, split into star-forming (or `blue cloud') galaxies, which have high levels of star formation and blue colours due to young massive stars; and quiescent (or `red sequence') galaxies with low star formation, and red colours from their older stellar population \citep[e.g.][]{Baldry2004,Schawinski2014,Bluck2020}. To create the population of quiescent galaxies, some process is required to `quench' star formation. For star formation to occur, gas needs to cool, forming molecular hydrogen ($H_2$) clouds, and collapse gravitationally. Thus, any mechanism that acts to quench a galaxy must remove the gas, prevent it from cooling, stabilise it against collapse, or destroy the densest gas phase. A strong candidate for this mechanism is the energy released by the supermassive black holes (SMBHs) that are located at the centres of all massive galaxies \citep{Kormendy2013}. If these SMBHs have a readily available fuel supply of cold gas, they can grow rapidly through gas accretion and light up to become `Active Galactic Nuclei' (AGN). Such AGN are able to release vast amounts of energy ({\LbolEq{\simeq}{42-48}}) and even if only a small fraction of this power can couple to the gas in a galaxy, it has the potential to unbind gas, prevent cooling and have a significant influence on the evolution of their host galaxies through a process known as `AGN feedback' \citep{Fabian2012}.

Indeed, AGN feedback is essential to our current theoretical understanding of galaxy evolution and is deeply embedded in cosmological models and simulations \citep[e.g.][]{Somerville2008,Schaye2015,Khandai2015,Dubois2016,Weinberger2018,Dave2019}. This process is necessary in the models to reproduce the observed galaxy bimodality; without it, purely environmental effects, or self-regulation from star formation, are not enough to sufficiently suppress star formation in the most massive galaxies \citep{Bower2006, McCarthy2011,Beckmann2017}. Furthermore, AGN feedback is required in cosmological models and simulations to explain a variety of other observational results, such as a solution to the `cooling problem' in galaxy clusters, reproducing observed galaxy sizes, and it could be critical in determining galaxy structures and dynamics \citep{Sijacki2007,Cattaneo2009,Choi2018,vanderVlugt2019,Irodotou2021}. 

Despite the success of AGN feedback, its numerical implementation varies considerably across cosmological simulations and there is no clear consensus as to the most realistic model \citep[][for more details see Section~\ref{Simulations}]{Crain2015,Weinberger2018,Dave2019, Costa2020}. Observational tests are crucial to refine or rule out different models, as well as establish whether AGN feedback is effective at quenching galaxies, as theoretical models have long suggested \citep{Silk1998,King2003}.

From the observational perspective, many studies have shown that luminous AGN  (\LbolEq{\gtrsim}{44}) can drive large-scale multi-phase outflows and inject turbulence into the host galaxy's interstellar medium \citep[ISM;][]{Sturm2011,Cicone2018,Baron2018,Baron2019,Veilleux2020,Girdhar2022}. However, this evidence does not directly establish that luminous AGN have a significant, and lasting, impact on the global star formation or molecular gas content of the host galaxies \citep{Harrison2017}. 

There has been a large array of observational studies over the last decade measuring the star formation and molecular gas properties of luminous AGN populations to search for evidence of an impact of AGN feedback. A variety of approaches have been taken; for example: (1) investigating the trends of star formation rates or $H_2$ content, with the luminosity of the AGN \citep[e.g.][]{Page2012,Harrison2012,Rosario2013aa,Stanley2015,Kakkad2017,Shangguan2019,Jarvis2020,Zhuang2021,Kim2022}; and (2) comparing the star formation rates and molecular gas properties of AGN host galaxies to similar galaxies without an AGN, either by looking at averages properties, or by comparing distributions \citep[e.g.][]{Bernhard2016,Rosario2018,Scholtz2018,Schulze2019,Kirkpatrick2019,Florez2020,Circosta2021,Bischetti2021,Valentino2021,Scholtz2021}. 

However, the observational studies taking these types of approaches have not provided definite evidence for strong AGN feedback. Across most studies, trends of star formation rates and molecular gas content with {\Lbol} are either found to be flat or positive \citep{Lutz2010,Mullaney2012,Stanley2015,Azadi2015,Gurkan2015,Shimizu2017,Scholtz2018,Kirkpatrick2019,Ramasawmy2019,Zhuang2021} with some studies finding a positive correlation only in the most luminous ({\LbolEq{\gtrsim}{45}}), although this may be due to underlying mass trends \citep[][]{Rosario2012,Stanley2017}. Additionally, once matched for galaxy morphology and stellar mass, AGN and non-AGN hosts are typically found to have similar levels of star formation and molecular gas \citep{Mainieri2011,Husemann2017,Stanley2017,Rosario2018,Shangguan2018,Smirnova-Pinchukova2021,Valentino2021,RamosAlmeida2022}. 

On the other hand, there are a few studies of high redshift AGN, $z\gtrsim1$, that suggest there is reduced molecular gas in AGN hosts \citep[e.g.][]{Kakkad2017,Perna2018,Circosta2021,Bischetti2021}, and a possible negative connection between AGN-driven outflows and star formation rates at lower redshifts \citep[][]{Wylezalek2016,Chen2022}. The situation is further complicated by the use of different AGN selection methods (e.g. X-ray, IR or radio) which can bias the sample of galaxies taken, affecting the final results for the distributions of host galaxy properties \citep[e.g.][and references within.]{Azadi2017,Harrison2017,Ji2022} Nonetheless, the broad consensus is that luminous AGN are preferentially observed in gas-rich and highly star-forming galaxies \citep[e.g.][]{Rosario2013aa,Bernhard2016,Jarvis2020,Florez2020,Xie2021,Koss2021,Vietri2021}. This apparent lack of unanimous evidence for reduced star formation or molecular gas in luminous AGN has led some recent studies to speculate that the evidence for feedback by luminous AGN is weak \citep[e.g.][]{Trump2015,Shangguan2018,Ramasawmy2019,Schulze2019,Shangguan2020b,Koss2021,Valentino2021,Ji2022}. 

Despite the aforementioned conclusion of some studies, observationally connecting the observed AGN luminosity with the long-term impact of the AGN on the host galaxy molecular gas and star formation is a complex process. The high variability of AGN luminosity, the potential time delay between AGN activity and its effect, and the shared cold gas reservoir that fuels both the AGN and the star formation make it difficult to predict what trends, if any, would be seen in active galaxy populations \citep[e.g.][]{Hickox2014,Harrison2017,Luo2021}. Therefore, a potentially better way to test proposed AGN feedback models is to extract predictions of galaxy properties from cosmological simulations directly. 

In this study, we analyse three cosmological simulations using similar approaches to those taken by observers, described above, to establish the predicted relationships between AGN luminosity and star formation rates and molecular gas content. This approach enables us to: (1) determine if the observational results are indeed in tension with simulations where strong AGN feedback is present and (2) investigate how the different models of AGN feedback across the simulations result in different observational predictions. We will focus on two cosmological epochs in line with where observers have focussed their efforts: local galaxies at $z\simeq0$ (for which the data are easier to obtain) and cosmic noon, $z\simeq2$, when star formation and AGN activity peaked \citep{Madau2014}.

This paper is organised as follows: in Section \ref{methods} we explain our methodology and in Section~\ref{results} we present our analysis of the simulations following similar approaches to observational papers, namely looking at trends between galaxy properties and AGN luminosity (Section~\ref{corrs}), comparing AGN host galaxies to inactive galaxies on the {\fgas}$-M_\star$ plane (Section~\ref{Contours}), investigating the gas fraction distribution for the highest {\Lbol} systems (Section~\ref{high Lbol}), and comparing the gas-depleted and quenched fractions of AGN host galaxies with a mass-matched non-AGN sample (Section~\ref{gas depleted fractions}). We discuss the implications of our results in the context of the observations in Section~\ref{discussion} and offer suggestions for how to improve both observational and simulation-based studies, before presenting our conclusions in Section~\ref{conclusion}.

In this work, we assume a flat, $\Lambda$CDM cosmology, using values from the \cite{Planck2016cite}: $H_0 = 67.7 \rm{\ km s^{-1}\ Mpc^{-1}}$, $\Omega_m = 0.3$, $\Omega_\Lambda = 0.7$; in line with the cosmologies assumed by the three simulations we will study.

\section{Methods} \label{methods}

In this section we present our methods, including a brief introduction to the three cosmological simulations and the observational data used (Sections \ref{Simulations} \& \ref{Observational Samples}), the target galaxy quantities we will be considering (Section \ref{Galaxy quantities}), our AGN-selection and quenching definitions (Sections \ref{AGN Selection} \& \ref{Quenching Definition}) and finally our methods for calculating correlation coefficients (Section \ref{Correlation Coefficients}).

\subsection{Simulations} \label{Simulations}

\begin{figure}
    \centering
	\includegraphics[width=0.48\textwidth]{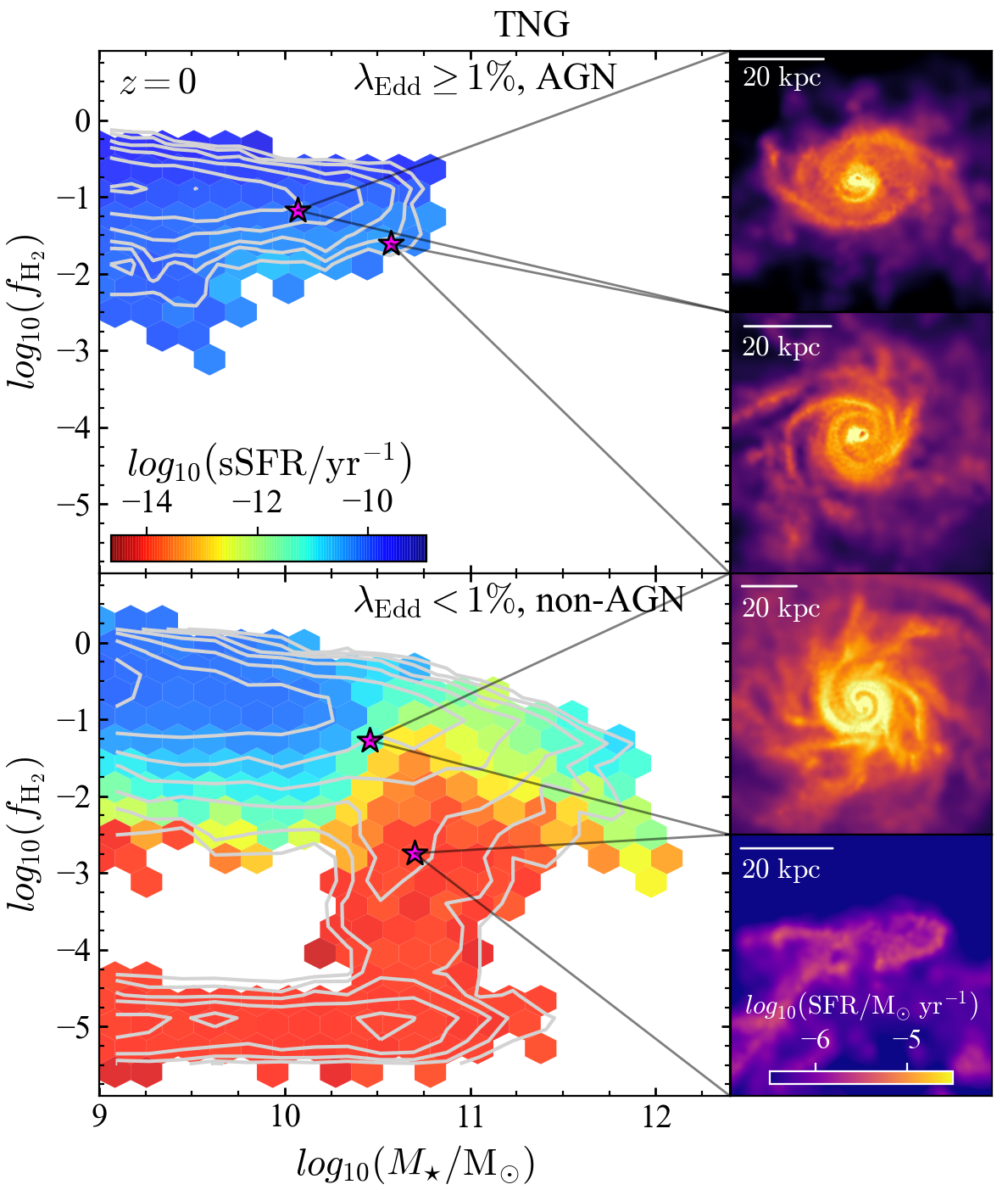}
    \caption{An example presentation of the simulation data we use in this work. The logarithmic number density contours of galaxies is shown in the $f_{H_2}-M_\star$ plane for TNG at $z=0$. Pixel bins are coloured by the mean sSFR of the galaxies enclosed. The upper panel shows an {\fedd}-selected AGN sample and the lower panel shows galaxies without an AGN. Star formation rate maps of four example galaxies are visualised in the right-hand columns. From the left panels we can see that AGN are hosted by galaxies with high {\fgas} and sSFR. The images also reveal that AGN host galaxies have similar morphology to other star-forming galaxies that do not host AGN.}
    \label{fig:summary}
\end{figure}

In this study, we have selected three of the current generation of hydrodynamic, cosmological simulations: {\sc IllustrisTNG} \citep{Springel2018,Pillepich2018b,Nelson2018,Marinacci2018,Naiman2018}, EAGLE \citep{Crain2015, Schaye2015} and SIMBA \citep{Dave2019}. These have comparable box sizes of $L \simeq 100 \rm{\ cMpc}$ (comoving Mpc) and trace dark matter, gas, stellar masses and SMBH properties, allowing us to search for trends between SMBH accretion rates, gas fractions, star formation and stellar masses within the context of a large host galaxy population ($\sim 10^4$ galaxies at $z=0$). As an illustration of the type of data available, Figure \ref{fig:summary} shows the galaxy population in {\sc IllustrisTNG} at $z=0$, displaying specific star formation rates as a function of molecular gas fraction and stellar mass for Eddington-ratio selected AGN and non-AGN samples (see Section~\ref{AGN Selection}) as well as visualisations of the star formation rate distribution in four representative galaxies from these populations. 
These example galaxies give us a first hint about our main finding: bright AGN tend to reside in extended, star-forming discs ( as seen in the first two panels on the right-hand side).

These simulations all model various crucial physical processes, including gas cooling, star formation, SMBH accretion, stellar feedback and AGN feedback. However, due to resolution limitations, these processes are not resolved from first principles, but rather modelled at a subgrid level. Of particular interest to this study are the prescriptions for AGN feedback included in each of the simulations. There are significant differences in how each simulation approaches feedback, which we summarise in Table \ref{Sim table}, show in Figure~\ref{fig:new_accretion}, and explain in more detail below.

To study the impact of AGN on the host galaxy, many observational studies estimate $H_2$ masses, using a tracer such as CO, and we would therefore like to test these observational results against the simulations' findings. However, due to computational limitations, the ISM is poorly resolved and the molecular gas phase is not directly traced in the simulations. Instead, we can estimate the molecular hydrogen by using `post-processed' models such as the one presented in \cite{Gnedin2011} who ran a set of high-resolution zoom-in simulations which followed the detailed chemical evolution of the gas and used these to derive fitting functions for the $H_2$ which can then be applied to cosmological simulations (a discussion of alternative post-processed models can be found in Appendix A of \citealt{Lagos2015} and in \citealt{Diemer2018}). These models have been applied in a similar way across the three simulations we consider in this work \citep{Lagos2015,Diemer2018,Diemer2019,Dave2019}, allowing us to consistently compare predictions for molecular hydrogen fractions between the simulations. Furthermore, the predictions from these models have been tested against observed galaxy populations and have been found to be in good agreement at $z=0$ \citep{Lagos2015,Diemer2019}. At $z=2$ there is a small discrepancy with the simulations predicting a lower molecular gas fraction than the observational constraints by a factor of 1.5 in EAGLE \citep{Lagos2015} and a factor of $2-3$ in {\sc IllustrisTNG} \citep{Popping2019}. SIMBA also likely shows such a discrepancy but it has not been extensively discussed in the literature. We describe our correction for this in Section~\ref{Quenching Definition}.

To ensure all galaxies in our simulation sample are sufficiently resolved, we take a stellar mass cut of {\MstarEq{\geq}{9}}. This is raised slightly to {\MstarEq{\geq}{9.5}} in SIMBA as black holes are only seeded once a galaxy has reached this size. We keep the lower cut for TNG and EAGLE to increase the sample size, although we note that raising this mass cut yielded no qualitative differences to our results.

We now provide an overview of the three simulations, focusing specifically on the implementation of SMBH accretion and AGN feedback, and their molecular gas estimates. For more details, the reader is referred to the release papers listed.

\subsubsection{IllustrisTNG}

\begingroup
\renewcommand{\arraystretch}{2.5}
\begin{table*}
\begin{tabular}{ccccccc}
\hline
\textbf{Simulation}              & \textbf{Mode Name}       & \textbf{Criteria}                                                            & \makecell{\textbf{Injection} \\ \textbf{Energy Type}}                                                     & \textbf{Direction} & \textbf{Description}     \\ \hline
\multirow{2}{*}{TNG}               & Thermal         & $\lambda_{\rm{Edd}} \geq \chi \left(M_{\rm{BH}} \right)$                                       & Thermal                                                         & Isotropic & \makecell{Continuous thermal energy injection \\ into small local environment} \\ \cline{3-7} & Kinetic         & $\lambda_{\rm{Edd}} < \chi\left(M_{\rm{BH}} \right)$                                          & Kinetic                                                         & \makecell{Random \\ (averages isotropic)}    & \makecell{Pulsed momentum kick \\ in random direction}     \\ \hline
EAGLE                                                & \makecell{Thermal \\ (single mode)}              & Always active                                                                    & Thermal                                                         & Isotropic & Pulsed thermal injection      \\ \hline
\multirow{3}{*}{SIMBA}             & Wind & Always active                                                                    & Kinetic                                                         & Bipolar   & Kinetic kick, $v\leq 1000 \rm{\ km\ s^{-1}}$         \\  \cline{3-7}
                                & Jet             & \makecell{$\lambda_{\rm{Edd}}<0.2$ \\ $M_{\rm{BH}}>10^{7.5}M_\odot$} & \makecell{Kinetic \\ (\& few \% thermal)}                                      & Bipolar   & \makecell{Kinetic kick, $v \leq 7000 \rm{\ km\ s^{-1}}$, \\ temperature raised to $T_{\rm{halo}}$}         \\  \cline{3-7}
                     & X-ray           & \makecell{$\lambda_{\rm{Edd}}<0.02$ \\ $f_{\rm{gas}}<0.2$}                         & \makecell{Thermal (non-ISM gas) or \\ thermal \& kinetic (ISM gas)} & Isotropic & Local thermal heating         \\ \hline
\end{tabular}
\caption{A summary of the subgrid implementations for AGN feedback in the three simulations. We note that, for simplicity, we neglect the X-ray mode in SIMBA as it has negligible impact on the initial quenching of galaxies \citep{Dave2019}. Refer to Section \ref{Simulations} for the definitions of $\chi \left( M_{\rm{BH}} \right)$ and $\lambda_{\rm{Edd}}$.}
\label{Sim table}
\end{table*}
\endgroup


{\sc IllustrisTNG}\footnote{\url{https://www.tng-project.org}} (\citealt{Springel2018,Pillepich2018b,Nelson2018,Marinacci2018,Naiman2018}, data release: \citealt{Nelson2019}, hereafter TNG) is a suite of simulations building on the original {\sc Illustris} simulation \citep{Vogelsberger2014b,Vogelsberger2014a} and using the moving-mesh code \textsc{arepo} \citep{Springel2010}. TNG has been run with three box sizes with approximate side lengths of 50, 100 and 300 cMpc. In this work we use the mid-sized run, TNG100 (side length $L=110.7 \rm{cMpc}$; baryonic mass resolution $m_{\rm{b}}=$\num{1.39e6}$\rm{\ M_\odot}$), as it provides the most direct comparison to EAGLE and SIMBA, although we also verify the convergence of our results by comparing them against the TNG300 run.

The calculated values for molecular gas masses are taken from the molecular and atomic hydrogen post-processed catalogues from \cite{Diemer2018,Diemer2019} using the prescription by \citealt{Gnedin2011}.

A key feature of the simulations is the black hole accretion model and feedback processes \citep{Weinberger2017,Weinberger2018}. The SMBH accretion rate is set according to the Bondi-Hoyle model \citep{Bondi1952} by performing a kernel-weighted average over nearest-neighbour gas cells around each SMBH particle. The accretion rate is also capped at the Eddington limit.

TNG features two modes of AGN feedback at low and high accretion rates which occur exclusively from each other. The mode a given SMBH is in at a particular time, is determined by its current Eddington ratio, {\fedd}, which is compared to a critical value, $\chi$, given by:

\begin{equation}
    \chi \left( M_{\rm{BH}} \right) = \text{min} \left[ 0.002 \left( \frac{M_{\textit{\rm{BH}}}}{10^8 M_\odot} \right)^2, 0.1 \right] ,
\end{equation}

\begin{figure*}
    \centering
    \includegraphics[width=0.95\textwidth]{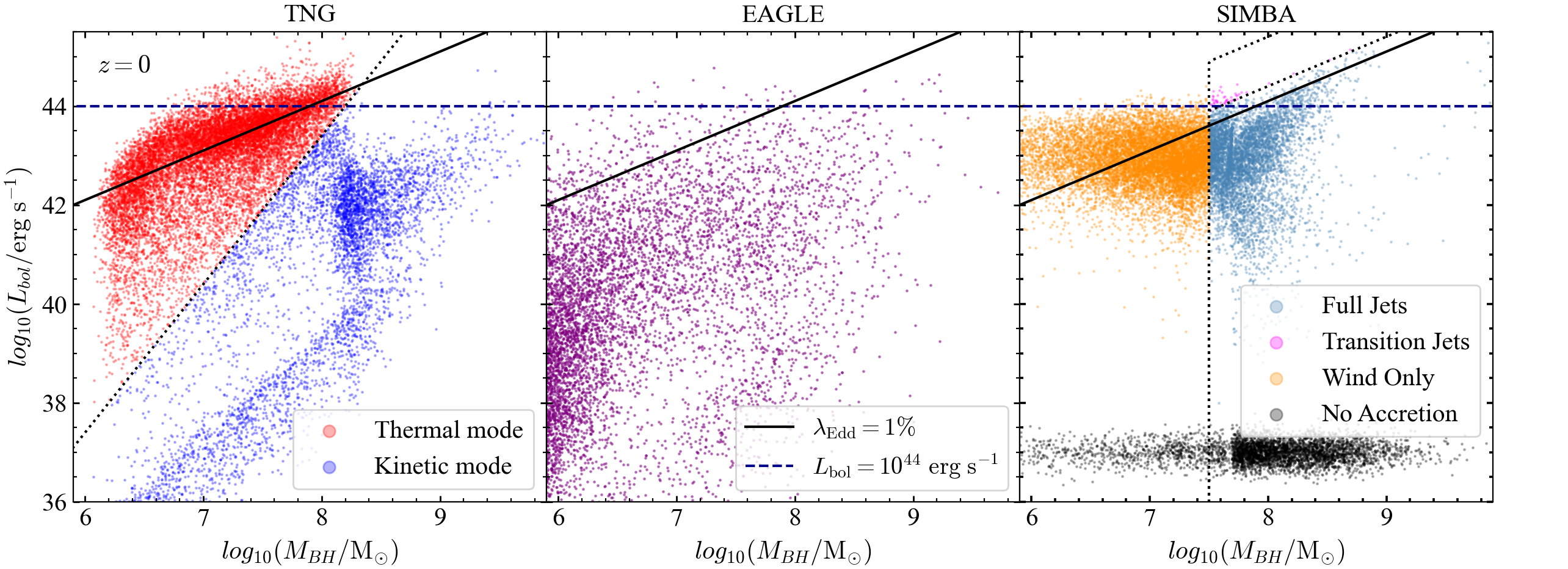}
    \caption{Bolometric luminosity (calculated from SMBH accretion rate; Equation~\ref{eq:Lbol}) versus black hole mass for the simulations. The dotted lines in TNG and SIMBA show the boundaries between the different feedback modes that can be active. We also show our two AGN definitions (Section \ref{AGN Selection}): the solid black line shows a constant Eddington ratio of {\fedd}$=1\%$ and the blue dashed line shows a high luminosity cut of {\Lbol}$=10^{44}\rm{\ erg s^{-1}}$.}
    \label{fig:new_accretion}
\end{figure*}

For $\lambda_{\rm{Edd}} \geq \chi$, the SMBH is in the high-accretion mode and for $\lambda_{\rm{Edd}} < \chi$, the SMBH is in the low-accretion regime. This split between the two modes can be seen in Figure \ref{fig:new_accretion}, in the accretion rate (scaled to bolometric luminosity; Equation~\ref{eq:Lbol}) versus black hole mass plane. 

The two modes are associated to two different feedback models, summarised here and also shown in Table \ref{Sim table}:

\begin{itemize}
    \item \textit{Thermal mode}: in this high-accretion mode, feedback energy is continually injected isotropically into the local environment in thermal form, heating the surrounding gas cells. SMBHs in this mode lie in the top left of Figure \ref{fig:new_accretion} and generally have low masses and high accretion rates. This mode is designed to reflect a `wind' or `quasar' mode \citep[e.g.][]{Sijacki2007}.
    \item \textit{Kinetic mode}: the energy is injected as momentum into the neighbouring gas cells of the black hole, directed  in a random orientation with a 0{\textdegree} opening angle, although this averages to an isotropic distribution after many episodes. A minimum energy for each kick is required, so the energy injected is stored up until this threshold is reached and then released in a pulse. These sources lie in the bottom right of Figure \ref{fig:new_accretion} and have high masses and low accretion rates \citep{Croton2006,Bower2006}.
\end{itemize}

Additionally, radiative feedback from the AGN is modelled by modifying the cooling function of gas cells in the vicinity of the SMBH which acts to suppress cooling \citep{Vogelsberger2013,Pillepich2018_methods}. However, this channel is only significant for sources with high Eddington ratios and is thus concurrent only with the thermal mode -- for simplicity we will not distinguish it from the thermal mode in this study.

\subsubsection{EAGLE}

EAGLE\footnote{\url{http://icc.dur.ac.uk/Eagle/index.php};\newline\url{https://eagle.strw.leidenuniv.nl}} (\citealt{Schaye2015,Crain2015}, data release: \citealt{McAlpine2016}) is a suite of cosmological simulations run with the smoothed particle hydrodynamics (SPH) code \textsc{gadget-3} \citep{Springel2005_gadget}. In this work, we use the largest box-size reference run, Ref-L100N1504, which has a side length of $L=100 \rm{\ cMpc}$ and a baryonic mass resolution of $m_{\rm{b}}=$\num{1.81e6}$\rm{\ M_\odot}$.

Molecular gas masses are calculated in post-processing \citep{Lagos2015} by first calculating the neutral gas fraction using the method of \cite{Rahmati2013} and then following the prescriptions of \cite{Gnedin2011} in a similar way to TNG. Previous results have noted that the SFRs in EAGLE are around 0.2 dex too low compared to observations \citep{Furlong2015,McAlpine2017} so, following these studies, we also scale the SFR up by this value.

In EAGLE, SMBH growth is also modelled via Bondi-Hoyle accretion, modified by a factor of the ratio between the Bondi and viscous time-scales \citep{Rosas-Guevara2016}, and is capped at the Eddington limit. The coupling of the released energy from the black hole to its nearest neighbour particles is achieved using a single mode of feedback \citep{Booth2009} that operates at any Eddington ratio, in contrast to the dual modes of TNG and SIMBA. Feedback energy is stored until it is sufficient to heat the surrounding particles by $\Delta T = 10^{7.5} \rm{K}$ and then stochastically injected as thermal energy. This `pulsed' nature of the thermal feedback prevents the energy being immediately radiated away and offsets cooling \citep{Booth2009}. This makes it more efficient at quenching the galaxy than the corresponding thermal mode in TNG which is how EAGLE can successfully reproduce the galaxy stellar mass function without utilising a kinetic feedback mechanism \citep{Schaye2015}.

\subsubsection{SIMBA}
SIMBA\footnote{\url{http://simba.roe.ac.uk}} \citep{Dave2019} is based on the earlier MUFASA simulation \citep{Dave2016} and features updated physics for the modelling of black hole growth and feedback. It is run using finite mass hydrodynamics from \textsc{gizmo} \citep{Hopkins2015}. In this paper, we will use the fiducial SIMBA run (m100n1024) which has a box side length of $L=147\ \textrm{cMpc}$ and a baryonic mass resolution of $m_{\rm{b}}=$\num{1.83e7}$\rm{\ M_\odot}$.

Unlike TNG and EAGLE, the molecular hydrogen content of the gas is calculated on-the-fly at each timestep (using the prescription of \citealt{Gnedin2011}), rather than calculated in post-processing, although we stress that this is still performed in subgrid fashion, as the molecular phase cannot be directly resolved. This allows SIMBA to utilise an $H_2$-based star formation model. In addition, the accretion of gas onto black holes depends on the gas phase: hot ($T>10^5\ \textrm{K}$) gas is accreted via Bondi-Hoyle accretion and cold ($T<10^5\ \rm{K}$) gas using a torque-limited accretion model \citep{Hopkins2011,Angles-Alcazar2017}, which sets the black hole accretion model of SIMBA apart from the other two simulations we employ.

The black hole feedback model in SIMBA features two primary modes, plus an additional X-ray feedback mode. Unlike in TNG, the feedback modes occur concurrently. The modes are summarised in Table \ref{Sim table} and briefly described here:

\begin{itemize}
    \item \textit{`Wind' mode:} black holes with high Eddington ratios ({\fedd}$>0.2$) provide a kinetic kick to nearby gas with an outflow velocity proportional to $\log{M_{\rm{BH}}}$ which can reach up to $1000 \textrm{\ km s}^{-1}$, based on the scaling relations of \cite{Fiore2017} and \cite{Ishibashi2018}. For both this mode and the `jet' mode (described below) the ejection is bipolar and parallel to the angular momentum vector of the inner disc used to calculate the SMBH accretion. Sources where only this mode is active are plotted in orange in Figure \ref{fig:new_accretion}.
    \item \textit{`Jet' mode:} for Eddington ratios $\lambda_{\rm{Edd}} < 0.2$, the `jet' mode switches on. Like the wind mode, this also ejects gas kinetically, although it can reach much higher velocities. Additionally, the temperature of this ejected gas is increased to the virial temperature of the halo although the thermal energy is typically only a few percent of the kinetic energy \citep{Dave2019}. The velocity of the outflow increases with decreasing Eddington ratio until it reaches a maximum speed of $7000\ \textrm{km s}^{-1}$ at  $\lambda_{\rm{Edd}} < 0.02$. The mass of the black hole must also exceed $10^{7.5} \textrm{M}_\odot$ for this mode to be activated. In Figure \ref{fig:new_accretion}, sources in pink have {\fedd}$=0.02-0.2$ and are labelled as `transition jets' as the maximum velocity has not yet been reached. Sources in blue have {\fedd}$<0.02$ and are labelled as `full jets' as the Eddington ratio is now low enough for the ejection velocity to meet its maximum. For consistency with previous papers, we keep the naming convention of `jet' for this mode, but we would like to stress that it does not seek to model a resolved jet structure as some high-resolution simulations have done (e.g. \citealt{Mukherjee2016,Mukherjee2018,Talbot2021,Bourne2021,Mandal2021}).
    \item \textit{`X-ray' feedback:} the final mode aims to model X-ray feedback from the accretion disc as a spherical, mostly thermal feedback mode. This requires full-speed jets (i.e. $\lambda_{\rm{Edd}} < 0.02$ and $M_{\rm{BH}}>10^{7.5} \textrm{M}_\odot$) and low molecular gas fractions ($f_{H_2}<0.2$) to be activated. The X-ray heating is applied to gas within the SMBH particle kernel and is either entirely thermal (for non-ISM gas) or half thermal and half kinetic (for ISM gas). \cite{Dave2019} found that this mode has a minimal effect on the galaxy mass function, but does help to fully quench the most massive galaxies \citep[][]{Cui2021,Appleby2021}. For simplicity, we will generally neglect this mode in our discussion as it has little effect on the initial quenching of galaxies and only becomes significant once the galaxy has already been mostly cleared of its gas.
\end{itemize}

\subsection{Observational Samples} \label{Observational Samples}

In this section, we present the observational data we use in our analysis. We select example observational samples with large numbers of sources with published information on {\em all of}: (1) AGN luminosities, (2) stellar masses, (3) molecular gas fractions and (4) star formation rates. Meeting these criteria, we selected two of the most recent samples at $z\simeq0$ \citep{Koss2021,Zhuang2021} and one at $z\simeq2$ \citep{Bischetti2021}.

\subsubsection{Koss et al. (2021)}

Our first low redshift sample is from the \textit{Swift}-BAT AGN Spectroscopic Survey\footnote{\url{https://www.bass-survey.com}} (BASS; \citealt{Koss2017}); an optical spectroscopic follow-up of AGN identified by the ultra-hard X-ray ($>10 \rm{\ keV}$) \textit{Swift}-BAT all-sky catalogue \citep{Baumgartner2013}. Stellar masses were calculated by combining near-IR data from 2MASS with mid-IR data from the AllWISE catalogue (see \citealt{Powell2018} for more details). Star formation rates were derived by decomposing spectral energy distributions (SEDs) of infrared data, mostly from \textit{Herschel} \citep{Shimizu2017,Ichikawa2019}. {\Lbol} is calculated by applying a bolometric correction to the intrinsic hard X-ray luminosity from the AGN \citep[][]{Marconi2004, Rosas-Guevara2016}, given by:

\begin{equation}
    \log{\left(\frac{L_{\rm{HX}}}{L_{\rm{bol}}}\right)} = -1.54 -0.24L - 0.012L^2 + 0.0015 L^3
\end{equation}

where $L = \log{\left(L_{\rm{bol}}/L_{\odot}\right)} -12$, which is then be solved for {\Lbol}.

In this study, we use a subsample of nearby ($0.01<z<0.05$) BAT AGN presented in \cite{Koss2021} for which there are additional CO(2-1) observations from JCMT and APEX, which are used by the authors to estimate measurements of molecular gas ($H_2$) masses. This gives us a sample of 213 AGN host galaxies with $H_2$ masses.

\subsubsection{Zhuang et al. (2021)}

Our other low redshift sample is presented in \cite{Zhuang2021} which is based on the catalogue formed by \cite{Liu2019} of Type I AGN from the Sloan Digital Sky Survey (SDSS;\citealt{York2000}, DR7: \citealt{Abazajian2009}) which were selected primarily based on a detection of broad-line $H \alpha$ emission. Galaxies in the redshift range $0.3<z<0.35$ were chosen, with the requirement of sufficient signal-to-noise to allow AGN selection from emission-line diagnostics. SFRs are calculated from [O II] $\lambda 3727$ and [O III] $\lambda 5007$ following the method of \cite{ZhuangHo2019}. Black hole masses are calculated from broad $\rm{H} \alpha$ (following \citealt{GreeneHo2005}), allowing inference of stellar masses based on the empirical scaling relations of \cite{Greene2020}. {\Lbol} is calculated by applying a bolometric correction to the optical continuum luminosity at $5100$ {\AA} \citep{McLure2004}.

To estimate the molecular gas masses of their galaxies, the method presented in \cite{Yesuf2019} is used which combines measurements of dust extinction, traced by the $\rm{H} \alpha/ \rm{H} \beta$ Balmer decrement, with gas-phase metallicities. This technique was calibrated against a sample of star-forming galaxies with CO measurements. \cite{Zhuang2021} show that this is a reliable method for Type I AGN as well being significantly less observationally intensive than CO measurements, resulting in estimated $H_2$ masses in a large sample of 453 AGN host galaxies to be calculated.

\subsubsection{Bischetti et al. (2021)} \label{Bisch}

For our high-redshift ($z\sim2$) sample, we use the data presented in \cite{Bischetti2021}, which is a compilation of the relevant available host galaxy data on AGN at this epoch. This combines a subsample from the WISSH QSOs project \citep{Bischetti2017}, with additional data from \cite{Perna2018}, X-ray selected QSOs from the SUPER survey \citep[][]{Circosta2021} and AGN-hosting SMGs \citep[][]{Bothwell2013}. Measurements of the IR luminosity were obtained from AGN-corrected SED fitting and used to derive the SFR \citep{Kennicutt1998}. Molecular gas masses were calculated from measurements of various CO or [C II] transitions (see \citealt{Bischetti2021} and references therin for details).

Due to the time-consuming observations required, there are only a small number of high-redshift AGN with measurements of all the quantities we require ({\Lbol}, $M_{H_2}$, SFR and $M_\star$). Therefore, to maximise the number of targets available, we select galaxies from this sample in the broad redshift range of $z=1-5$, yielding 20 sources with suitable data, with a mean redshift of $z=2.2$. However, we checked our results against the narrower redshift range of $z=2-3$ and found no qualitative differences. Additionally, the subsample we select has a slightly smaller range in {\fgas} than the parent sample. We discuss this further when we compare these observational results to the simulation data in Section~\ref{high Lbol}.

\begingroup
\renewcommand{\arraystretch}{1.2}

\begin{table*}
\begin{tabular}{ccccccccc}
                                               \hline
                                               & \multicolumn{3}{c}{$z=0$} & & \multicolumn{3}{c}{$z=2$} \\
                                               & TNG     & EAGLE   & SIMBA &  & TNG     & EAGLE  & SIMBA  \\ \hline
All galaxies                                   & 18991   & 12355   & 22389&  & 11807   & 7856   & 5133   \\
$\lambda_{\rm{Edd}} \geq 1\%$                  & 6280    & 245     & 5163  & & 11078   & 717    & 4601   \\
$L_{\rm{bol}} \geq 10^{44} \rm{\ erg\ s^{-1}}$ & 1423    & 27      & 386&    & 4361    & 228    & 2829   \\
Combined selection                         & 1263    & 18      & 170    && 4285    & 217    & 2729  \\\hline
\end{tabular}

\caption{Sample sizes for our various AGN selection criteria at $z=0$ and $z=2$. `All galaxies' shows the number of sources at each redshift (after applying stellar mass cuts) and the `Combined selection' requires both a high {\Lbol} and high {\fedd}.}
\label{AGN table}
\end{table*}

\endgroup

\subsection{Key Quantities} \label{Galaxy quantities}

In this study, we focus on two key global galaxy quantities and investigate how they depend on the luminosity of the AGN. The galaxy properties are the specific star formation rate;

\begin{equation}
    \rm{sSFR} = \frac{\rm{SFR}}{\mathit{M_\star}},
\end{equation}

where $M_\star$ is the stellar mass of the galaxy, and the molecular gas fraction;

\begin{equation}
    f_{H_2} = \frac{M_{H_2}}{M_\star},
\end{equation}

where $M_{H_2}$ is the mass of molecular hydrogen gas.

These quantities are commonly investigated in observational studies to probe the effect the AGN has on the molecular gas and how that influences the galaxy's star formation (e.g. \citealt{Harrison2017,Shangguan2018,Rosario2018,Scholtz2018,Zhuang2021}). Additionally, sSFR is frequently used to determine whether a galaxy is star-forming or quiescent (see Section \ref{Quenching Definition} for our quenching/gas-depletion definitions). 

Due to the resolution limits of the simulations, very low star formation rates and molecular gas masses values are not resolved (this is partly resolution-dependent and is also affected by the SFR averaging timescales, see Appendix A of \citealt{Donnari2019}). To track these galaxies, we artificially plots these sources in the figures with arbitrarily low values of sSFR$=10^{-14} \rm{\ yr^{-1}}$ and {\fgas}$=10^{-5}$, respectively. For presentation reasons we scatter these points around these values with a standard deviation of 0.2. (see \citealt{Weinberger2018} who follow a similar method). We note that these exact values are not used in any calculation in this study and they are only used to visually represent them as a quenched and gas-depleted populations in the figures. These sources represent around $10-20\%$ of all galaxies across the three simulations in our sample at $z=0$.

For each simulation, we chose an aperture of either $30 \rm{\ kpc}$ or twice the stellar half-mass radius depending on the default choice of the original simulation teams. As the majority of the star formation of a galaxy falls well within both these definitions, there is a negligible difference between them. For a more detailed discussion of the effect of aperture choice, see \cite{Donnari2019} and Appendix C in \cite{Weinberger2018}. 

Another factor of consideration is the timescale over which we calculate the star formation rate in the simulations. \cite{Donnari2019} investigated the effect of changing the SFR averaging timescale, from instantaneous to $1000 \rm{\ Myr}$. By calculating the resulting star formation main sequence for each timescale, they showed that varying the timescale makes negligible difference to the slope of this sequence, although there is an offset of around $0.1-0.2 \rm{\ dex}$ between the shortest ($10 \rm{\ Myr}$) and longest ($1000 \rm{\ Myr}$) timescales. These differences are smaller than the range of quenching definitions we use which are the dominant source of uncertainty in our results for correlation coefficients and quenched fraction (see Section~\ref{Quenching Definition}). Therefore, we chose to use only the instantaneous SFRs in this study.

A final choice is whether to include both centrals and satellites in our simulation galaxy sample, with different approaches being taken in previous studies (\citealt{Donnari2019} make no distinction, whereas \citealt{Weinberger2018} only selects centrals). We include both centrals and satellites, with central galaxies making up $60-70\%$ of our sample across the three simulations. We also test our results on a centrals-only selection and find no qualitative difference, except for a slight reduction in quenched and gas-depleted fractions in lower-$M_\star$ galaxies, e.g. galaxies with {\MstarEq{\lesssim}{10}} have a global gas-depleted fraction of $7-13$ percentage points lower when we select only for centrals. We find that this change is even lower in AGN hosts, changing the gas-depleted fraction by $\lesssim 5$ percentage points.

\subsection{AGN Selection Criteria} \label{AGN Selection}

Our goal is to broadly emulate the approach taken by observational studies to investigate the relationships between AGN activity and host galaxy properties (see Section~\ref{sec:intro}). Therefore, we design two AGN-selection methods that roughly emulate current observational limits in separating luminous AGN hosting galaxies from non-AGN galaxies; one based on the bolometric luminosity of the AGN ({\Lbol}) and the other on the Eddington ratio ({\fedd}).

In the simulations, the bolometric luminosity is calculated from the instantaneous accretion rate of the central SMBH in each galaxy:

\begin{equation}
    L_{\rm{bol}} = \epsilon_r \dot{M}_{\rm{accr}} c^2 , \label{eq:Lbol}
\end{equation}
    
where $\dot{M}_{\rm{accr}}$ is the mass accretion rate flowing onto the black hole and $\epsilon_r$ is the radiative efficiency of the AGN. EAGLE and SIMBA use the canonical value of $\epsilon_r=0.1$ \citep{Crain2015,Dave2019} and TNG uses $\epsilon_r=0.2$ \citep{Weinberger2017}. A reasonable observational cut-off for studying high-luminosity AGN is {\LbolEq{\geq}{44}} which we use as our luminosity-based selection. This corresponds to the lower end of the bolometric luminosities in the \cite{Koss2021} comparison sample (see Figure~\ref{fig:hist_cont}).

However, this simple relation between the accretion rate and {\Lbol} may only be valid for high-Eddington ratio, Shakura-Sunyaev discs (SSDs; \citealt{Shakura1973}). In low-{\fedd} systems, the disc becomes geometrically thick and radiatively inefficient. The impact of the accretion efficiency on {\Lbol} motivates a second AGN definition based on the Eddington ratio. Following \cite{Rosas-Guevara2016}, we assume that sources with $\lambda_{\text{Edd}} \geq 1\%$ are radiatively efficient, thin discs, with high X-ray luminosities and we define these as our high-{\fedd} selection. We also define a `combined' selection where we require the AGN to satisfy both criteria. There are also other methods to account for inefficient AGN accretion. For instance, in \cite{Habouzit2021,Habouzit2022}, {\Lbol} is scaled down proportionally to the Eddington ratio if $\lambda_{\rm{Edd}}<10\%$. We tested our results using this model and found it yielded qualitatively similar results to our combined AGN definition, showing our conclusions are insensitive to the exact model used for inefficient AGN accretion.

Table \ref{AGN table} shows the number of sources for each of these selection criteria at $z=0$ and $z=2$, across the three simulations. We note that, especially at $z=0$, EAGLE shows far fewer high-{\Lbol} sources than the other two simulations. This can also be seen by the scarcity of points above the blue dashed line in Figure \ref{fig:new_accretion}. Previous work has noted the lower AGN luminosity distribution function in EAGLE compared to TNG and SIMBA \citep[][]{Habouzit2022} and it has been suggested that this is caused by both a lower BH mass function in the high-$M_{\rm{BH}}$ regime in EAGLE \citep[][]{Habouzit2021} as well as lower accretion rates in these massive BHs, possibly due to strong supernovae feedback, the efficient thermal feedback model and the modified Bondi-Hoyle accretion model \citep[][]{Rosas-Guevara2016,Habouzit2022}. This discrepancy between the simulations does not effect our ability to investigate the trends across the three simulations; however, we do take care to account for the lower statistics of AGN host galaxies in EAGLE in our discussion of the results.

\subsection{Quenching \& Gas-Depletion Definitions} \label{Quenching Definition}

There are various ways to define a galaxy as having quenched star formation, including the distance from the main sequence \citep{Weinberger2018}, a cut in sSFR \citep{Donnari2019} or based on a colour selection (e.g. UVJ diagrams, \citealt{Donnari2019,Akins2021}). In this study, we use a main sequence-based quenching definition. We prefer this over a colour-based definition as this would require assumptions and modelling of the dust content of galaxies \citep[e.g.][]{Trayford2016,Akins2021}. We use the main-sequence model presented in \cite{Weinberger2018} for their analysis of the TNG simulation. This is a modification of the \cite{Ellison15} model, with an additional redshift dependence based on \cite{Schreiber2015}. We note that we repeated our analysis using other main sequence definitions \citep{Speagle2014} including a non-linear model \citep{Whitaker2014} and found it made a negligible difference to the results compared to the systematic uncertainties outlined below. For the molecular gas fraction, we use a similar method, utilising the scaling relations presented in \cite{Tacconi2018} as our `main-sequence' model for this quantity. Both of these models are redshift-dependent, as the normalisation of both the star-forming and gas-fraction main sequences are higher at $z=2$ than $z=0$. As noted before, the simulations tend to under-predict {\fgas} at $z=2$ by a factor of $1.5-3$ \citep{Lagos2015,Popping2019}, therefore we shift the scaling relation in \cite{Tacconi2018} down by $0.3\ \rm{dex}$ to better match the peak of the {\fgas}$-M_\star$ plane in the simulations. 

As illustrated in Figure \ref{fig:res_method}, we then define a galaxy as `quenched' if it lies more than a certain distance, $\Delta_{\rm{MS}}$, below the star-forming main sequence \citep{Weinberger2018}. Likewise, a galaxy is defined as `gas-depleted' if it lies $>\Delta_{\rm{MS}}$ below the gas fraction scaling relation \citep{Tacconi2018}. To check the sensitivity of our results to this cut, we vary this value between $-1\ \rm{dex}\ \leq \Delta_{\rm{MS}}\leq-2\ \rm{dex}$; a range similar to what has been presented in previous works. We show this range for sSFR on Figure \ref{fig:res_method} by the red shaded region. Where relevant, the uncertainty this adds to our results is represented by shaded violin plots, showing the posterior probability density for changing values of $\Delta_{\rm{MS}}$.

\subsection{Correlation Coefficients} \label{Correlation Coefficients}

\begin{figure*}
    \centering
    \includegraphics[width=0.9\textwidth]{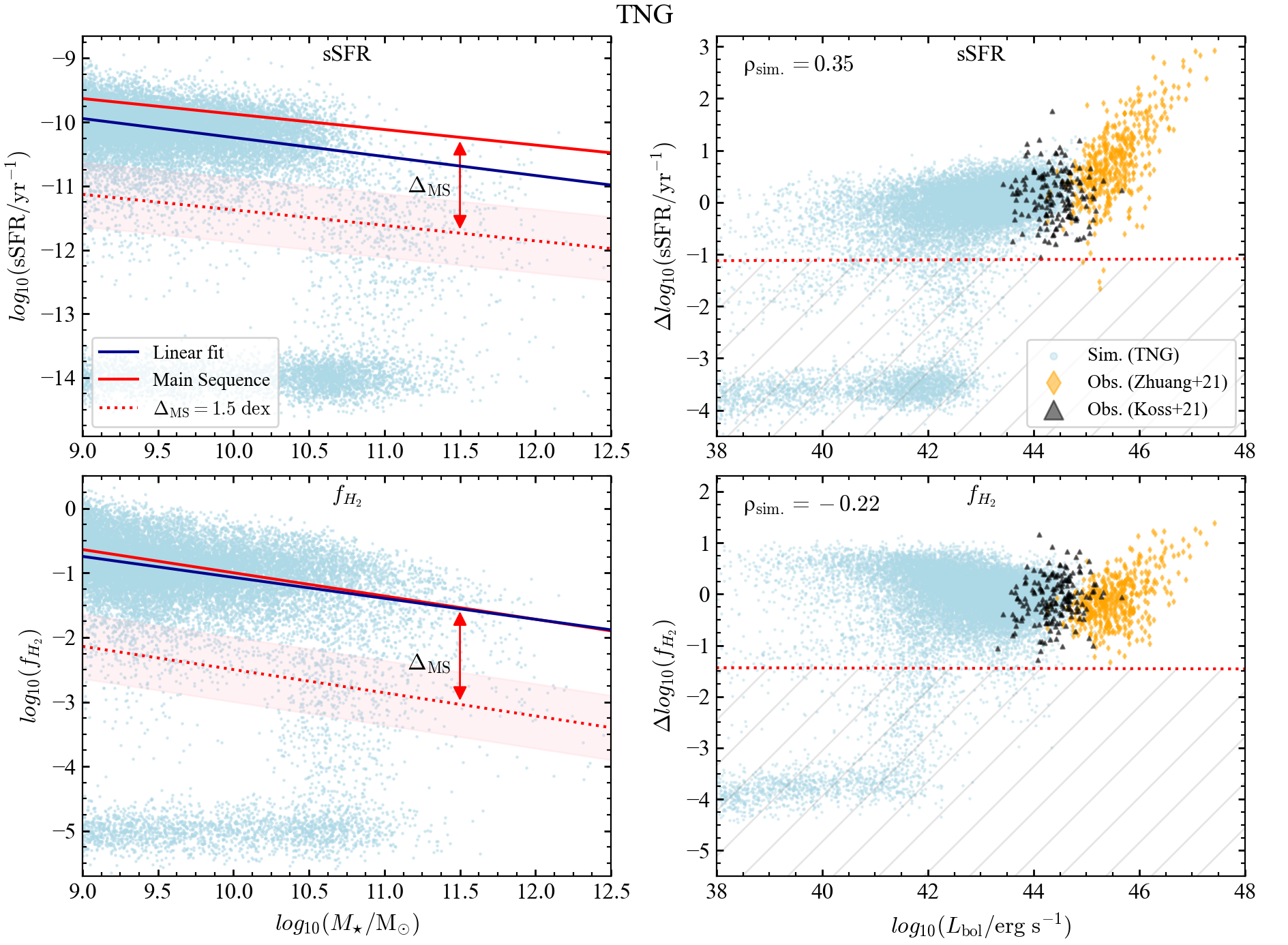}
    \caption{Demonstration of how the $M_\star$ dependency is removed from the quantities of interest and the correlation coefficient between the resulting residuals and {\Lbol} is calculated, using TNG at $z=0$ as an example. {\em Left Column:} sSFR (upper panel) and {\fgas} (lower panel) against $M_\star$. The red line shows the main sequence scaling relationship adopted for each quantity. Our quenched/gas-depleted definition is galaxies that lie some distance ($\Delta_{\rm{MS}}$) below these relations, which is allowed to vary between $-1\ \rm{dex}\ \leq \Delta_{\rm{MS}}\leq-2\ \rm{dex}$, as shown by the red shaded region. The blue line shows a linear regression to the remaining points, after excluding the quenched/gas-depleted sources from the fit.  {\em Right Column:} the resulting residuals' relationship with {\Lbol}, after removing the $M_\star$ dependence including comparison observational data. Calculated Spearman correlation coefficients are shown in the top left corner of the panels, revealing a weak positive correlation for sSFR and a weak negative correlation for $f_{H_2}$. The hatched region shows the quenched/gas-depleted sources that have been excluded from the correlation calculation.}
    \label{fig:res_method}
\end{figure*}

Part of our analysis involves quantifying the trends seen between galaxy properties and AGN luminosity by calculating correlation coefficients between these quantities. However, as our galaxy properties of interest have a known dependence on stellar mass, which itself causes a correlation with $L_{\rm{bol}}$ \citep[e.g.][]{Stanley2017,Scholtz2018,Tacconi2020}, we first need to correct for the effect of $M_\star$. To do this, we use a residual method which is represented in Figure \ref{fig:res_method}, using the TNG simulation as an example. The panels in the left-hand column show the sSFR- and {\fgas}-$M_\star$ planes. 

For this correlation coefficient analysis we only consider `star-forming' galaxies which are all galaxies that are not defined as quenched (for the sSFR analyses) or gas-depleted (for the {\fgas} analyses). We discuss quenched and gas-depleted fractions in detail in Section \ref{gas depleted fractions}, but, for a rough guide, the `star-forming' fraction for these simulations at $z=0$ is $69-80\%$ for sSFR and $79-86\%$ for {\fgas}. Once we have selected our star-forming sample, we characterise the stellar mass dependence. This approach avoids including the unresolved SFR and molecular gas mass values (see Section \ref{Galaxy quantities}) in the correlation analysis which would have to be arbitrarily placed at some low value. Furthermore, this is well motivated because the majority of AGN host galaxies fall within the `star forming' population (see Section \ref{high Lbol} for a discussion on the quenched and gas-depleted fractions for all AGN-selected galaxies). However, we also tested the robustness of our results by including the quenched and unresolved systems and found it made a negligible difference to the qualitative trends.

\begin{figure*}
    \centering
	\includegraphics[width=0.98\textwidth]{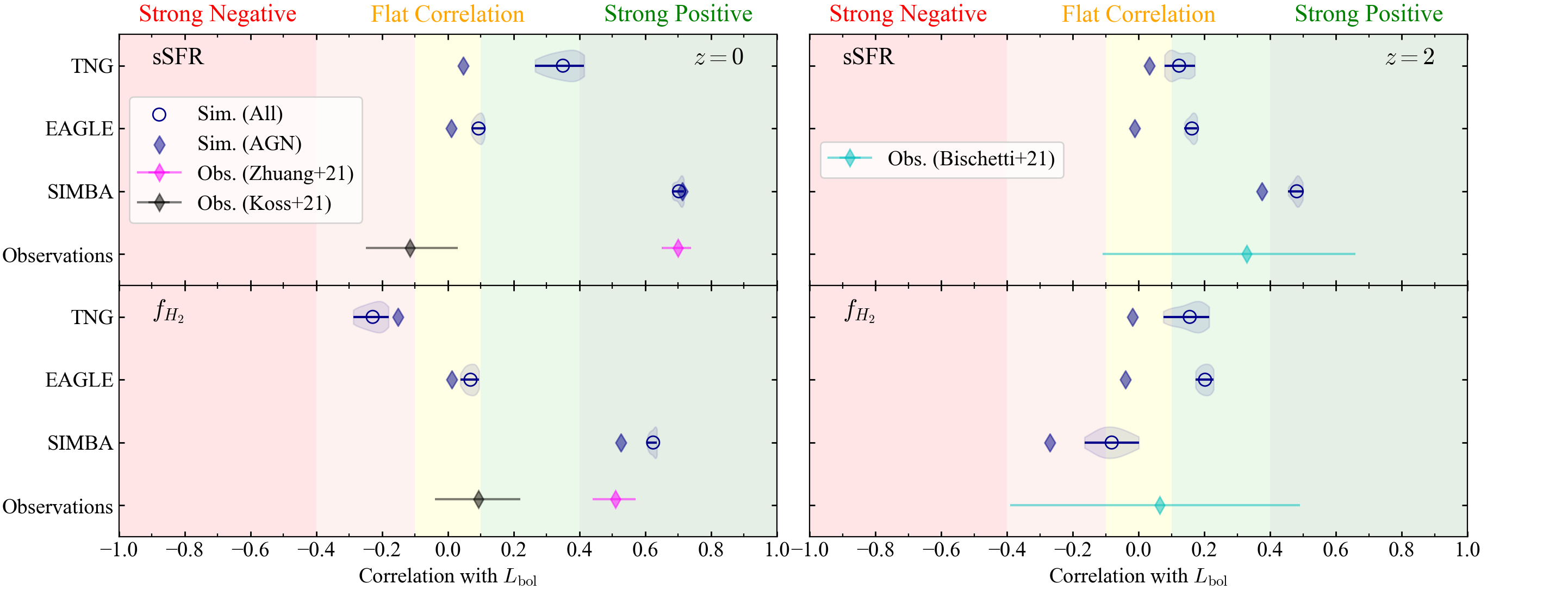}
    \caption{Correlation coefficients between {\Lbol} and host galaxy properties (upper panel: sSFR; lower panel: {\fgas}) after removing the dependence on $M_\star$ for the simulations and observational data investigated. The left column shows the results for $z=0$ and the right column shows $z=2$. For the simulation data, the mean result for star-forming galaxies is shown by the blue circles, with the shaded regions showing the probability densities for the uncertainty on this value, based on varying the quenching or gas-depletion definition (see Section \ref{Correlation Coefficients}). The filled diamonds show the correlation of only the AGN (high-{\fedd} selected sources). We also show the data from the observations as diamond-shaped points, with black points for \protect\cite{Koss2021} and purple for \protect\cite{Zhuang2021}, where the error bars on these values show the $95 \%$ confidence interval on the calculation of the correlation. At $z=2$, the cyan points show the sample from \protect\cite{Bischetti2021}. Although there are clear differences in the predictions, it can be seen that none of the simulations predict strong negative correlations ($\rho<-0.4$) between {\fgas} or sSFR with {\Lbol}, qualitatively consistent with observational results.}
    \label{fig:correlations}
\end{figure*}

After we have excluded the quenched or gas-depleted galaxies, we perform a linear regression on the remaining data points in the sSFR or {\fgas}-$M_\star$ plane and calculate the residual of this fit for each point. This is plotted against {\Lbol} as can be seen in the top right panel of Figure \ref{fig:res_method}. Points in the red hatched region show the quenched/depleted and thus excluded galaxies for an example value of $\Delta_{\rm{MS}}=-1.5$. The Spearman correlation coefficient, $\rho$, is then calculated between the residuals and {\Lbol}. This can take values between -1 and 1, where -1 is a strong negative correlation and 1 is a strong positive correlation with values around 0 showing no correlation between the quantities. In this study we consider the strength of the correlation as `strong' for $|\rho| > 0.4$, `weak' for $0.1 \leq |\rho| < 0.4$ and `flat' for $|\rho| < 0.1$. This method is equivalent to computing the semi-partial correlation between sSFR and $M_\star$ (this approach is similar to that in \citealt{Zhuang2021}, although they control for $M_{H_2}$ rather than $M_\star$). Figure \ref{fig:res_method} shows an example of a weak positive (sSFR, \textit{top right}) and weak negative ($f_{H_2}$, \textit{bottom right}) correlation in the TNG simulation, with the correlation coefficient values shown in the corner of each panel. The uncertainty caused by the varying quenching/depletion definition is propagated through to give a probability density region of correlation coefficients. These can be seen as the blue shaded `violin' plots on Figure \ref{fig:correlations}. We discuss this figure and present the results of this correlation analysis in Section~\ref{corrs}.

\section{Results} \label{results}

Here we present our results from an investigation of the relationships between AGN bolometric luminosities and host galaxies' specific star formation rates (sSFR) and molecular gas fractions ({\fgas}), as predicted from three state-of-the-art cosmological simulations (TNG, EAGLE and SIMBA). We follow the broad approaches taken in many observational papers, which search for evidence of the impact of AGN on their host galaxies by investigating: (1) correlations of sSFR and {\fgas} with {\Lbol} (Section \ref{corrs}); (2) a comparison between AGN and non-AGN host galaxies in the {\fgas}$-M_\star$ or ${\rm sSFR}-M_\star$ plane (Section \ref{Contours}); and (3) gas fraction distributions for the highest {\Lbol}  sources (Section \ref{high Lbol}) as well as the fraction of AGN which are quenched or gas-depleted (Section \ref{gas depleted fractions}). As motivated in Section \ref{Observational Samples} we investigate this at both $z=0$ and $z=2$. In Sections \ref{Contours}, \ref{high Lbol} \& \ref{gas depleted fractions}, we note that we obtain qualitatively similar results for sSFR as we do for {\fgas}. Therefore, for brevity, we mostly show the results for {\fgas} in the main body of the paper, referring interested readers to the equivalent plots for sSFR in Appendix \ref{ap: ssfr}.

\subsection{Correlations with $L_{\rm{bol}}$} \label{corrs}

For each of TNG, EAGLE and SIMBA, we compute the strength and sign of the Spearman correlation coefficient, $\rho$, between $L_{\rm{bol}}$ and our key galaxy properties of interest: sSFR and {\fgas} (see Section \ref{Galaxy quantities}). For this analysis we focus only on `star-forming' galaxies, which we define as those which remain after excluding quenched galaxies (for sSFR analyses) or gas-depleted galaxies (for {\fgas} analyses) following Section~\ref{Quenching Definition}. For this star-forming population, we then remove the intrinsic correlation with stellar mass following Section \ref{Correlation Coefficients}. We note that we return to consider the quenched and gas-depleted galaxies in Sections \ref{high Lbol} \& \ref{gas depleted fractions}.

The results from the correlation analysis are presented in Figure~\ref{fig:correlations}, which shows the calculated Spearman correlation coefficients between {\Lbol} and sSFR (\textit{top row}) and  between {\Lbol} and {\fgas} (\textit{bottom row}) at both $z=0$ (\textit{left column}) and $z=2$ (\textit{right column}). The open circles show the star-forming galaxy population and the diamonds show the {\fedd}-selected AGN. Several observational papers have investigated the trends between star formation rates or molecular gas fractions and {\Lbol} as evidence for AGN feedback \citep[][]{Lutz2010,Mainieri2011,Page2012,Harrison2012,Stanley2015,Kakkad2017,Shangguan2019,Ramasawmy2019,Jarvis2020,Circosta2021}. However, we find that the simulations do not predict any strong negative correlations between {\Lbol} and host galaxy properties.

Out of the 12 coefficients calculated, eight show positive trends (populating the right hand side of Figure \ref{fig:correlations}), three show flat trends and only one is (weakly) negatively correlated, with $\rho=-0.23$. Even for this negative trend, which is seen in TNG, there is only a reduction in the mean gas fraction of around 1 dex, over $\sim 6$ orders of magnitude in AGN luminosity (see lower left panel of Figure \ref{fig:res_method}).

Looking at Figure \ref{fig:correlations} in more detail, there are clear differences between the three simulations for their predictions of the trends between sSFR or {\fgas} and {\Lbol}. If we first consider the sSFR of the entire star-forming galaxy population with no AGN selection (open circles) at $z=0$ (\textit{top left}): TNG is weakly positive with $\rho=0.35$, EAGLE shows a negligible correlation (i.e. flat trend) of $\rho = 0.09$ and SIMBA shows a  strong positive relation of $\rho = 0.70$. Looking at {\fgas} at $z=0$, we see that SIMBA is again strongly positively correlated ($\rho=0.62$) and EAGLE again shows negligible correlation ($\rho=0.07$), whilst TNG shows a weak negative correlation ($\rho=-0.23$). If we also consider the results at $z=2$ (\textit{right column}), there is generally more agreement across the three simulations. However, there are clearly different predictions of the evolution of these relationships. At $z=2$, TNG now shows weakly positive correlations for both quantities, whereas there is a weak negative correlation at $z$=0 for gas fraction. EAGLE shows very little evolution across the two epochs, although there is a weakly positive trend with gas fraction at $z=2$ ($\rho = 0.20$). SIMBA is still strongly correlated in sSFR at $z=2$, however, it shows a flat trend with gas fraction ($\rho = -0.08$), which is a significant change from the strong positive trend at $z=0$.

We next consider the correlation coefficients for only star-forming galaxies in the simulations that are also selected to be AGN based on their Eddington ratios ({\fedd}$>1\%$, see Section \ref{AGN Selection} and shown as filled diamonds here). We find that selecting only high-{\fedd} AGN for this correlation analysis has either little effect or reduces the strength of the trends observed, i.e. $\rho$ values remain unchanged, or in most cases move closer to zero. This latter effect is likely due to a smaller dynamic range in {\Lbol} being sampled. In no case does only considering high-{\fedd} AGN convert a positive correlation to a negative one, showing that they generally behave in a similar way to the overall star-forming galaxy population in the simulations. The only exception here is in {\fgas} in SIMBA at $z=2$, where the trend moves from uncorrelated ($\rho = -0.05$) to weakly negative ($\rho = -0.27$).

The comparison observational data, as described in Section \ref{Observational Samples}, are shown at the bottom of each panel of Figure \ref{fig:correlations}. Of our two primary comparison samples at $z=0$, the \cite{Koss2021} sample overlaps more with the {\Lbol} parameter space of the simulations (see Figure~\ref{fig:res_method}). This shows a mostly flat trend which is qualitatively consistent with the results of EAGLE and TNG (for sSFR) for AGN-selected samples. In contrast, the \cite{Zhuang2021} data predicts very strong positive trends. This tension between the two observational samples could be caused by the different selection criteria used by the two studies, or be because the sources in \cite{Zhuang2021} probe higher values of {\Lbol}, possibly corroborating the work of \cite{Rosario2012} who found that positive correlations are only seen once the AGN luminosity exceeds {\LbolEq{>}{44.7}}.

At $z=2$, the observational data from \cite{Bischetti2021} suggest a positive correlation for sSFR and a flat correlation for {\fgas}. Due to the small number of sources, the uncertainties are large, but we note that these error bars  cover a similar range to the spread we find between the simulations. 

It is beyond the scope of this work to perform a more quantitative assessment to compare the observations to the simulations, which would require careful considerations of sample selection effects and controlling for uncertainties in conversion factors between observables and derived quantities, etc.. Instead, we use these results to demonstrate in a {\em qualitative} sense that the broad behaviour seen in the simulations in these trends is consistent with observations, in that no strong negative correlations are observed between gas fractions and star formation rates.

Although there is agreement between the simulations in not predicting strong negative correlations, we have shown that the differences in their predictions are still quite stark. However, by simply looking at trends or correlation coefficients, it is difficult to understand the details of the investigated relationships and to infer whether this is due to their differing AGN feedback subgrid models or because of other factors. Therefore, in Sections \ref{Contours}, \ref{high Lbol} \& \ref{gas depleted fractions} we consider the {\em distribution} of galaxy properties, also considering the different feedback modes used in the simulations, to help understand the differences seen here.

\subsection{AGN and non-AGN in the {\fgas}$-M_\star$ plane} \label{Contours}

\begin{figure*}
    \centering
    \includegraphics[width=0.95\textwidth]{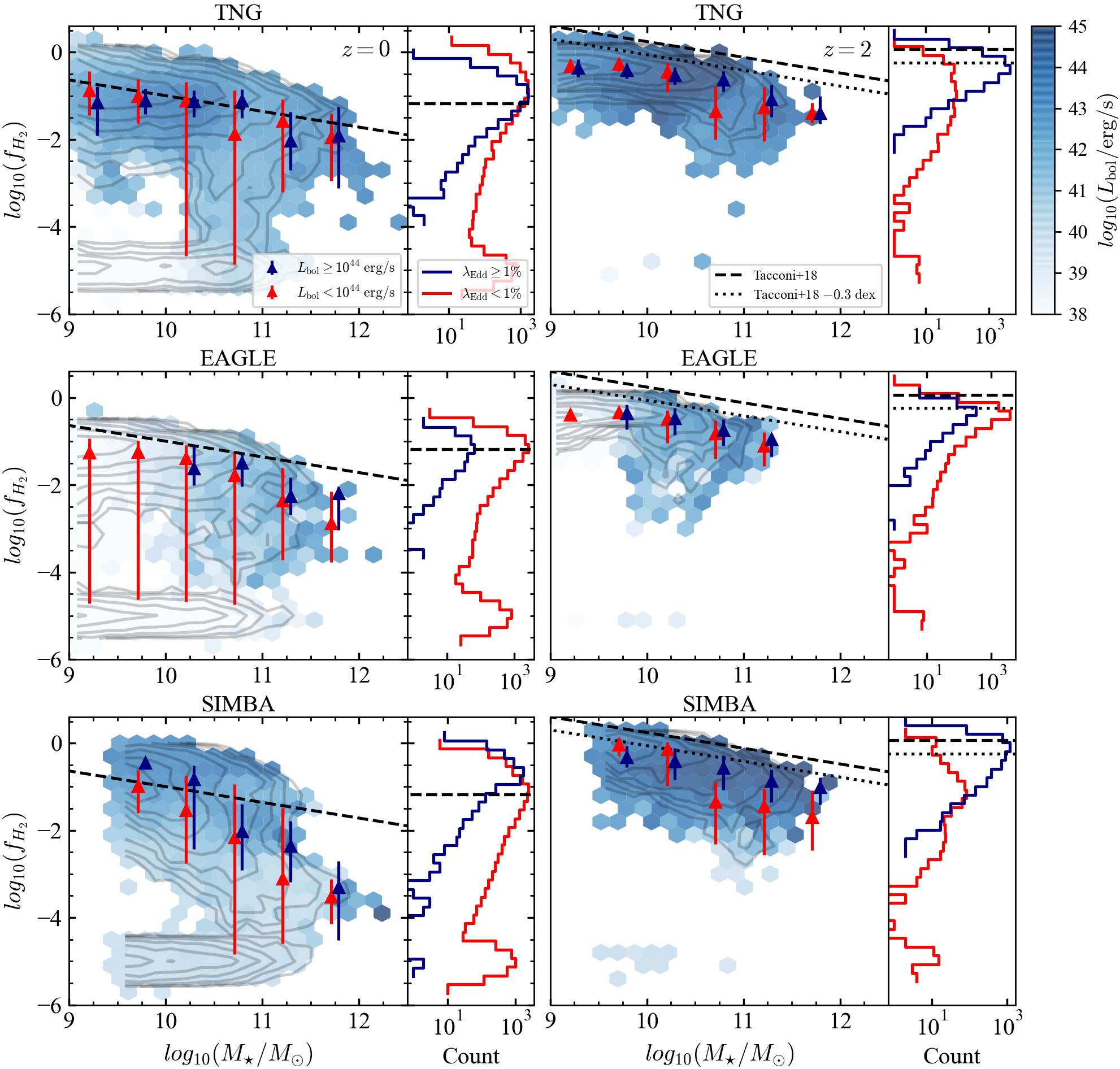}
    \caption{The molecular gas fraction against stellar mass at $z=0$ (left column) and $z=2$ (right column) for the three simulations. Contour lines show logarithmic number density and the colouring of the pixels shows the mean bolometric luminosity of the sources within each bin. Galaxies with unresolved $f_{H_2}$ have been scattered around $f_{H_2}=10^{-5}$. The dashed line shows the observed {\fgas} main sequence from \protect\cite{Tacconi2018}. At $z=2$, we shift this line down by $0.3 \rm{\ dex}$, shown as the dotted line. The triangular points in the left-hand subpanels show the median and 16\ts{th}-84\ts{th} percentiles for AGN (blue) and non-AGN (red) galaxies based on an {\Lbol} selection, grouped in stellar mass bins of 0.5 dex.The histograms in the right-hand subpanels show the logarithm of the number of sources classified as AGN or non-AGN based on an {\fedd} definition, with the dotted line showing the expected gas fraction for a fiducial $M_{\star}=10^{10.5}\ \rm{M_{\odot}}$ galaxy on the main sequence. We can see that both the {\Lbol}- and {\fedd}-selected sample of AGN reside preferentially in gas-rich galaxies.}
    \label{fig:contours}
\end{figure*}

Another common method that observers have used to look for evidence of the effect of AGN feedback, is to compare the host galaxy properties of active galaxies to those without known AGN \citep{Rosario2018,Bischetti2021,Circosta2021}. This is often done by comparing the location of AGN host galaxies with mass-matched galaxies without AGN, or to established scaling relations of sSFR or {\fgas} with stellar mass for `main sequence' star-forming galaxies. In this section, we aim to emulate this method using the simulations, by considering our two ways of categorising AGN, bolometric luminosities and Eddington ratios (Section \ref{AGN Selection}), and comparing their host galaxy's {\fgas} to that of inactive galaxies in the simulations and also to the gas fraction scaling relations of \cite{Tacconi2018}. Although we focus on {\fgas}, the equivalent figures for sSFR are shown in Appendix \ref{ap: ssfr}, for which we obtain broadly consistent conclusions. Unlike for the correlation coefficient analysis, which only considered `star-forming galaxies', for the following analysis we now consider the whole galaxy sample.

Figure \ref{fig:contours} presents the molecular gas fractions as a function of stellar mass at $z=0$ ({\em left column}) and $z=2$ ({\em right column}) for galaxies above $M_\star \geq 10^{9} \rm{\ M_\odot}$ in the three simulations. The contour lines in the main panels show logarithmic number density of sources and the galaxies have also been grouped into hexagonal bins, which are then coloured by the mean bolometric luminosity of the AGN within. We also plot the median and 16\ts{th}-84\ts{th} percentiles for a luminosity-selected ({\LbolEq{\geq}{44}}) sample of AGN (blue triangles) and for the corresponding non-AGN sample (red), grouped in stellar mass bins of 0.5 dex. The histograms to the side show logarithmic number counts of sources with high {\fedd} ratios ($>1\%$) in blue and with low {\fedd} in red. As an observational comparison, we show the molecular gas scaling relation from \cite{Tacconi2018} as our `main-sequence' by the black dashed line in the {\fgas}$-M_{\star}$ plane. On the histograms, this is represented by another black line showing the main-sequence gas fraction for a fiducial galaxy of $M_{\star}=10^{10.5}\ \rm{M_{\odot}}$.

Considering the distribution of all galaxies in this {\fgas}$-M_{\star}$ plane (grey contours), we can see that all three simulations reproduce the observed galaxy bimodality with star forming systems clustered around the main sequence at {\fgas}$\sim 10^{-1}$ and a large gas-depleted fraction below (i.e. $15-25\%$ of the galaxies lie $>1$ dex below the gas fraction main sequence at $z=0$). At $z=0$, the simulations are generally successful at reproducing the gradient and normalisation of the observed gas scaling relations \citep{Lagos2015,Diemer2018,Diemer2019,Dave2019}. At $z=2$, all three simulations slightly under-predict the gas fraction, as previously noted in \cite{Lagos2015} and \cite{Popping2019}. Figure \ref{fig4sSFR} shows a similar result for the sSFR.

In Figure \ref{fig:contours}, the darker blue pixels show the brightest AGN (highest accretion rates) and the lighter pixels show the lowest accretion rate systems. In TNG at $z=0$, we can see that the most luminous AGN lie around the gas fraction main sequence and at lower stellar masses. We see a strong downwards `plume' of quenching galaxies all with similar luminosities of {\Lbol}$=10^{42}\rm{\ erg\ s^{-1}}$ (i.e. these low luminosity systems would not be detected as `AGN' in most observational work). As we further discuss in Section \ref{sim differences} this dramatic plume of low-{\fgas} galaxies over a narrow mass range ($M_\star = 10^{10.2-11} \rm{\ M_\odot}$) is associated with a switch to the low-accretion (`kinetic') feedback mode in TNG \citep{Weinberger2018,Terrazas2020}. At $z=2$, there are very few gas-depleted systems and the mean {\Lbol} is high all along the main sequence, peaking at $M_\star \simeq 10^{10.5} \rm{\ M_\odot}$.

The brightest {\Lbol} sources in EAGLE are in high-mass galaxies ({\MstarEq{\gtrsim}{10.5}}) on the main sequence and the mean {\Lbol} increases smoothly from low- to high-$M_\star$; a trend seen at both redshifts.

The mean {\Lbol} in SIMBA is high ({\LbolEq{\gtrsim}{43}}) around the main sequence line, but at around $1-2$ dex below this, the mean {\Lbol} drops significantly in both low- and high-$M_{\star}$ galaxies (although there are a few high-{\Lbol} sources here as we explore in Section \ref{high Lbol}). Like the other two simulations, there are very few galaxies with unresolved {\fgas} at $z=2$, although the mean {\Lbol} of the galaxies towards the bottom of the main-sequence grouping is much lower than similar galaxies in the other two simulations. 

Considering a luminosity selection of {\LbolEq{\geq}{44}}, the triangular points, representing the median gas fraction of AGN and non-AGN across different mass bins, show that {\Lbol}-selected AGN reside on or near the peak of the gas-rich galaxies and the observational main sequence in all three simulations. The median gas fraction of the AGN are always close to the non-AGN points across the three simulations, except they show a much narrower range: the 16\ts{th}-84\ts{th} percentiles shown in Figure \ref{fig:contours} never extend down to the lowest `unresolved' gas fractions.

As motivated in Section \ref{AGN Selection}, we can also make an AGN selection based on the Eddington ratio. Looking at the right-hand subplots, we can see that in all three simulations the highly-accreting AGN (shown by the blue line) reside in high-{\fgas} galaxies on the main sequence with the peak of the AGN population matching or even exceeding the peak of the non-AGN population (in red). We explore this further in Section \ref{high Lbol} by considering the gas-depletion fractions of these high-{\fedd} systems.

In Figure \ref{fig4sSFR} we present the same analysis for the ${\rm sSFR}-M_\star$ plane. The results are generally consistent with those for {\fgas}, but there are some small differences, especially in TNG. For example, at $z=2$ in TNG there is a larger population of quenched galaxies than there is gas-depleted. These quenched galaxies are mostly associated with the kinetic feedback mode and lie in the narrow stellar mass range $M_\star = 10^{10.2-11} \rm{\ M_\odot}$ (we discuss this further in Section \ref{sim differences}). This suggests that, in the TNG model, a galaxy does not need to be fully depleted of molecular gas before the star formation has quenched at $z=2$. We also see that the non-AGN sample in this mass range at $z=0$ has a much lower median sSFR compared to the {\Lbol}-selected AGN sample. This is in contrast to the result for the molecular gas, where the two samples have similar median values for {\fgas} (Figure~\ref{fig:contours}). Additionally, the highest mass bin ($M_\star \geq 10^{11.5} \rm{\ M_\odot}$) in SIMBA at $z=0$ is entirely quenched, whereas the galaxies had non-negligible gas fractions in the {\fgas} plane. Nevertheless, the median value for the AGN sample is still higher than the non-AGN for both sSFR and {\fgas} in the high-mass ($M_\star \geq 10^{11} \rm{\ M_\odot}$) regime in SIMBA.

From Figure \ref{fig:contours} (and \ref{fig4sSFR}), we can conclude that the simulations predict that AGN preferentially live in high-sSFR, high-{\fgas} galaxies. This result holds for both an {\Lbol}- and {\fedd}-based AGN selection, showing that the simulations do not predict large differences between active and inactive galaxy populations, despite the implementation of strong AGN feedback models. This is in qualitative agreement with observational work at $z=0$ \citep[e.g.][]{Rosario2018,Jarvis2020}. However, the observations have less of a coherent picture at $z\simeq 2$ where some studies find AGN might tend to live in lower {\fgas} sources \citep[e.g.][]{Kakkad2017,Perna2018,Circosta2021}, potentially in tension with these results from the simulations.

Despite a broadly consistent picture across the three simulations, we find that there are differences in the predictions between the three simulations, especially in the exact location of the brightest AGN in the {\fgas}$-M_{\star}$ plane. In Section \ref{high Lbol} we explore these brightest systems and investigate the effects of the feedback and accretion modes.

\subsection{{\fgas} Distributions in AGN} \label{high Lbol}

\begin{figure}
    \centering
    \includegraphics[width=0.45\textwidth]{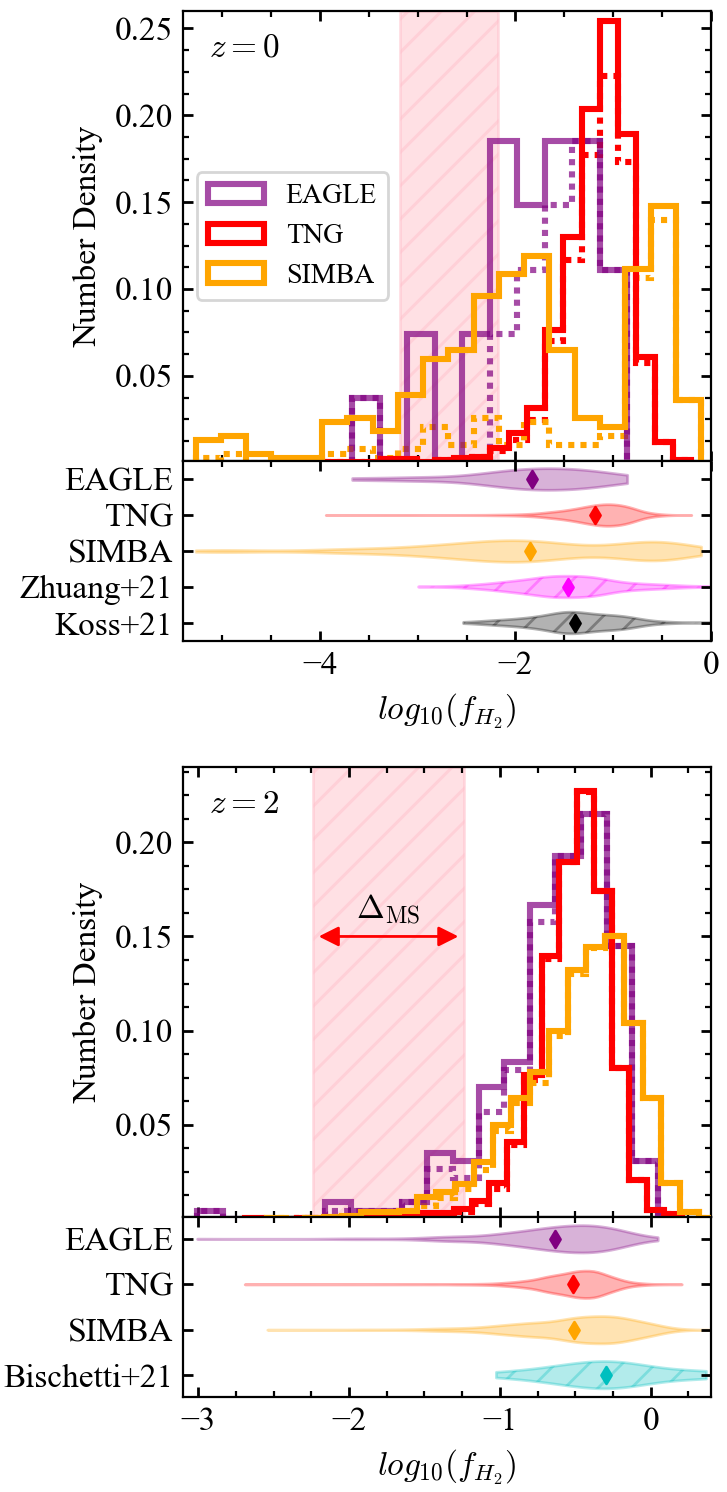}
    \caption{Gas fraction distributions for bright ({\LbolEq{\geq}{44}}) AGN. The top part of each panel shows the binned density distribution in $f_{H_2}$ for the three simulations. The red shaded region to the left shows a range of definitions for gas-depletion. The dotted lines show the result if the additional AGN criteria that {\fedd}$\geq1\%$ is required. The lower part shows violin plots of the density distributions for both the simulation data and observational samples, where the mean value of $f_{H_2}$ is shown by the diamond-shaped point. Although the simulations are in good agreement at $z=2$, the predictions at $z=0$ are starkly different for our {\Lbol}-selection.}
    \label{fig:hist_highLbol}
\end{figure}

To investigate the effect of powerful AGN on their host galaxies, many observational studies look for evidence of feedback in the brightest ({\Lbol}$\gtrsim 10^{44} \rm{\ erg s^{-1}}$) AGN \citep[e.g.][]{Schulze2019,Scholtz2021,Zhuang2021,Bischetti2021}. In this section, we look at the simulation predictions for the distribution of {\fgas} among the brightest AGN in the simulations.

The results for the gas fraction distribution in {\LbolEq{\geq}{44}} AGN hosts are shown in Figure \ref{fig:hist_highLbol} for both $z=0$ \textit{(top panel)} and $z=2$ \textit{(bottom panel)}. The top part of each panel shows the number density distributions in {\fgas} for the three simulations and the bottom part shows violin density plots, with the observations as a comparison. Gas-depleted systems lie to the left of the plot.

At $z=2$ (\textit{bottom panel} of Figure \ref{fig:hist_highLbol}), all three simulations show very similar behaviour. There is a roughly Gaussian distribution at a gas fraction of {\fgasEq{\simeq}{-0.5}} with a slightly longer tail extending into the depleted region to the left. The shapes of the distributions compare quite well with the observations from the \cite{Bischetti2021} sample as can be seen in the violin plots. We note that the full \cite{Bischetti2021} sample does contain some sources with lower {\fgas} (down to {\fgasEq{\simeq}{-1.4}}), but as we require values for all of {\fgas}, $M_\star$, SFR and {\Lbol} we had to select a subsample from their catalogue (see Section~\ref{Bisch}). However, the mean {\fgas} value from their full sample is comparable to the mean of our subsample, even if our range in {\fgas} is slightly narrower.

However, at $z=0$, there is much less agreement between the simulations. TNG still shows a similar shape to $z=2$ with a peak at just below {\fgasEq{\simeq}{-1}} and a tail extending down towards the gas-depleted region. SIMBA, however, shows a bimodal structure with a higher peak at {\fgasEq{\simeq}{-0.7}} and a second peak at {\fgasEq{\simeq}{-2}} with a tail which extends significantly into the depleted region. This includes a few sources with `zero' gas fractions which have been artificially scattered at {\fgasEq{\simeq}{-3.5}}. However, this lower peak almost disappears if we include the additional requirement of {\fedd}$\geq 1\%$, shown by the orange dotted line. This bimodality may be caused by the low-accretion feedback mode in SIMBA, which we discuss more in the Section~\ref{sim differences}. Due to the poorer statistics of EAGLE (see Section \ref{AGN Selection}), it is difficult to precisely describe the distribution, but there appears to be a broad peak at {\fgasEq{\simeq}{-2}} and some sources in the gas-depleted region.

If we consider the observational samples at $z=0$, we can see they both show a roughly symmetrical distribution around {\fgasEq{\simeq}{-1.5}} and are broadly consistent with these high-{\Lbol} AGN residing in gas-rich galaxies in the simulations. A similar result is found if we look at sSFR, as seen in Figure \ref{fig5sSFR} in the Appendix. The observational data at $z=0$ seems to be consistent with a single high-{\fgas} peak, although we caution that the samples may not be complete for low-{\fgas} and there is limited overlap in {\Lbol} with the sample from \cite{Zhuang2021} (see Section~\ref{sim differences}).

\subsection{Gas-Depleted Fractions of AGN} \label{gas depleted fractions}

\begin{figure}
    \centering
    \includegraphics[width=0.48\textwidth]{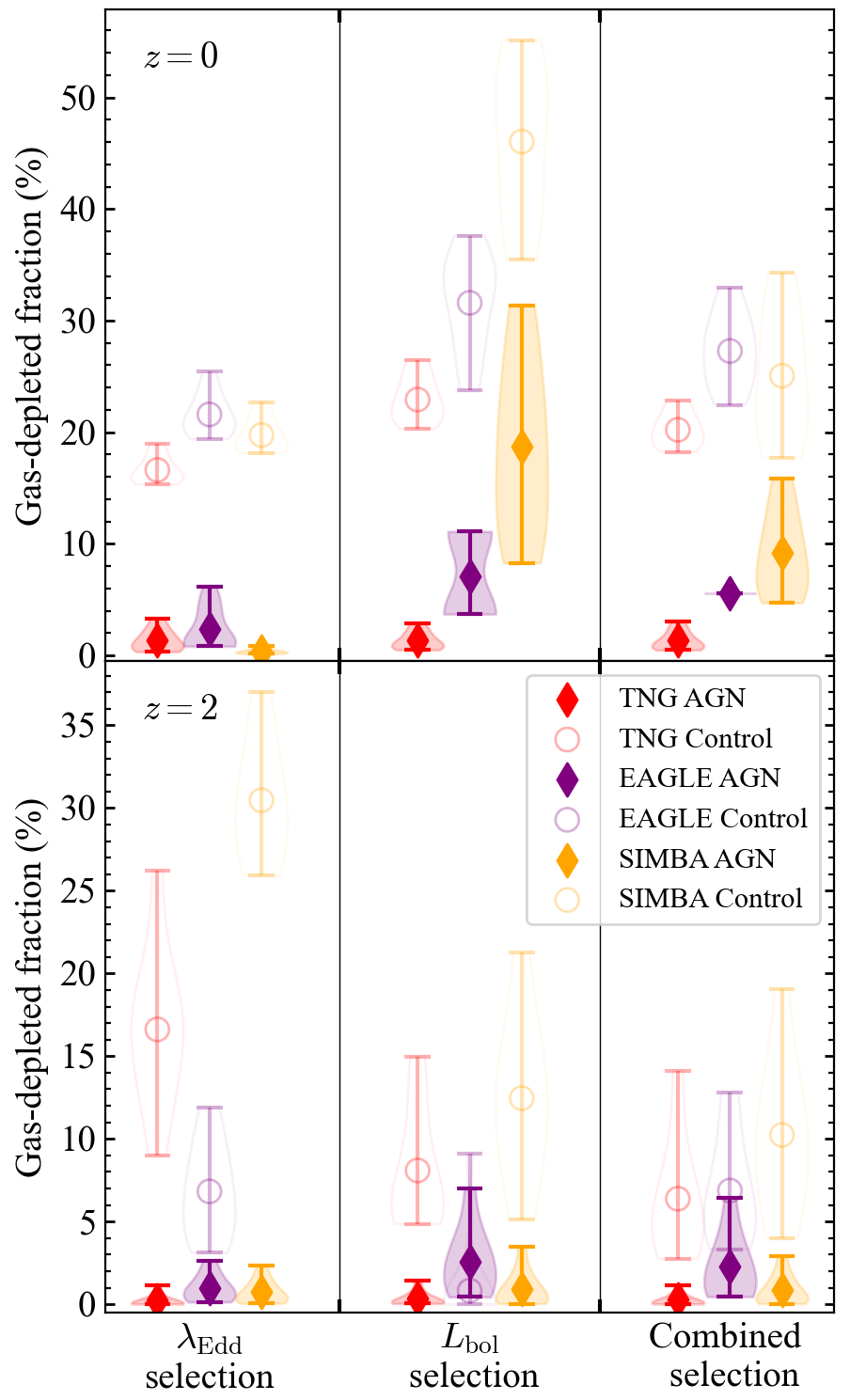}
    \caption{Gas-depleted fraction ($f_{\rm{GD}}$) for three AGN selection criteria: an Eddington ratio selection ({\fedd}$>1\%$), a luminosity cut ({\Lbol}$>10^{44} \rm{\ erg s^{-1}}$), and the result if we combine the two selections. Each selection is also compared to a stellar-mass-matched control sample of non-AGN from each simulation, shown as the fainter points. The uncertainty regions are calculated from varying the offset from the main sequence ($\Delta_{\rm{MS}}$) for the gas-depleted definition (Section \ref{Quenching Definition}). The points represent the mean value for $f_{\rm{GD}}$. At $z=0$, all three simulations show very low gas depleted fractions compared to the control galaxies for the high-{\fedd} selection, but SIMBA has a significantly higher depleted fraction for the high-{\Lbol} selection.}
    \label{fig:f_Q}
\end{figure}

To quantify the distributions seen in Figure~\ref{fig:hist_highLbol} in more detail, we calculate the gas-depleted fraction, $f_{\rm{GD}}$, of the AGN host galaxies across the whole galaxy sample based on the offset to the gas fraction main sequence in \cite{Tacconi2018}. A galaxy is defined as quenched if it lies more than some distance, $\Delta_{\rm{MS}}$, below this main sequence which is both $M_\star$- and $z$-dependent (see Section \ref{Quenching Definition}). We vary this value in the range $\Delta_{\rm{MS}} \in [1,2]$ dex to test the sensitivity of our results to the quenching definition. The red shaded region in Figure \ref{fig:hist_highLbol} shows the variation in this quenching cut for an example galaxy of $M_\star=10^{10.5} \rm{\ M_\odot}$. 

We present the results of this analysis in Figure \ref{fig:f_Q}. The range of gas-depletion definitions from $\Delta_{\rm{MS}}$ is represented by violin plots showing the probability density and the diamond points show the mean value. The depleted fraction is investigated for three AGN selections: an Eddington ratio selection ({\fedd}$>1\%$), a luminosity cut ({\Lbol}$>10^{44} \rm{\ erg s^{-1}}$), and finally the result if we combine the two selections (high-{\Lbol} and high-{\fedd}). We also create a control sample for each AGN selection criteria by randomly sampling from the non-AGN population to match the stellar-mass distribution of AGN. This is repeated for each simulation with each AGN selection and this mass-matched control sample is plotted as the faint circular point in Figure \ref{fig:f_Q} to provide a reference sample.

At $z=0$ we can see clear differences between the simulations. Taking a high-{\fedd} cut, all three simulations show low gas-depleted fractions for the AGN ({\fGD}$\lesssim 6\%$) which are always much lower than the control galaxy samples ($18\%<${\fGD}$<24\%$). This is also demonstrated by the blue histograms in the right-hand panels of Figure \ref{fig:contours} which show that very few high-{\fedd} AGN reside in gas-depleted galaxies. However, for the high-{\Lbol} selection, we see a higher depletion fraction in SIMBA ($8\%<${\fGD}$< 31 \%$) -- due to the lower peak of the bimodal distribution seen in Figure \ref{fig:f_Q} -- and much lower depleted fractions in TNG and EAGLE. Combining these two selections has the effect of reducing the depletion fraction in SIMBA to $4\%<${\fGD}$<16\%$ (although this is still the highest of the simulations) but has little impact on EAGLE or TNG. The reasons for this dramatic change in the distribution in SIMBA are discussed further in Section \ref{sim differences}.

At $z=2$ there are fewer differences between the simulations with all three showing {\fGD}$< 7\%$ regardless of AGN selection. This is lower than the control samples although these show large variation between the simulations, especially in the {\fedd}-selection criteria. This is because in SIMBA and TNG, the high-{\fedd} sources are more heavily biased towards lower $M_\star$ galaxies which have a higher quenched fraction at $z=2$.

The results for sSFR (Figure~\ref{fig6sSFR}) are again similar, with generally low quenched fractions across all the simulations regardless of AGN selection, except for the high fraction at $z=0$ for {\Lbol}-selected AGN in SIMBA which is also seen in {\fgas}. Again, the AGN-selected galaxies always have significantly lower quenched fractions than their mass-matched non-AGN counterparts.

In this section, we have shown that gas-depleted and quenched fractions in the simulations are lower for AGN than for a mass-matched control sample of galaxies, showing that AGN are preferentially found in gas-rich and star-forming galaxies. However, we have also shown that at $z=0$ there are differences between the simulations in their predictions for the gas fraction distributions when selecting using a luminosity cut ({\LbolEq{\geq}{44}}). In the Section~\ref{sim differences}, we will explore whether these differences are linked to the subgrid SMBH feedback models.

\section{Discussion} \label{discussion}

Many observational studies seek evidence of AGN feedback, as invoked by simulations, by examining trends between star formation rates or molecular gas content and the AGN luminosity, or by comparing active to inactive galaxies \citep[e.g.][]{Page2012,Harrison2012,Scholtz2018,Rosario2018,Schulze2019,Shangguan2019,Florez2020,Circosta2021,Zhuang2021,Koss2021,Ji2022,Scholtz2021}. However, these observational results have not led to a clear picture of negative feedback by AGN. Therefore, to help understand this, we have presented results on the relationship between AGN activity and host galaxy molecular gas fractions and specific star formation rates as directly predicted by three state-of-the-art cosmological simulations (TNG, EAGLE and SIMBA). Here we discuss these results in the context of the observations and possible future directions to test the theoretical framework of AGN feedback.

\subsection{Cosmological models predict AGN preferentially live in gas rich, star-forming galaxies} \label{main discussion}

The overall picture we have shown is that the simulations do not predict strong negative correlations between AGN luminosity and host galaxy properties of {\fgas} or sSFR (Figure \ref{fig:correlations}) in local galaxies ($z=0$) or at cosmic noon ($z=2$). Furthermore, by considering either an {\Lbol}- or a {\fedd}-based AGN definition, we found that AGN are predicted to almost exclusively reside in gas-rich, star-forming galaxies (Figure \ref{fig:contours} and Figure \ref{fig4sSFR}). More quantitatively, we found that the simulations generally show low gas-depleted fractions (mean fraction, {\fGD}$<10\%$ at $z=0$ for a combined AGN selection) in AGN hosts compared to a mass-matched control sample (Figure \ref{fig:hist_highLbol}; Figure \ref{fig:f_Q}). None of these tests have shown what might be considered a clear sign of immediate quenching by luminous AGN.  

We note that a small number of observational studies, mostly of high redshift AGN, have suggested luminous AGN may have lower gas fractions than matched non-active galaxies \citep[e.g.][]{Perna2018,Circosta2021,Bischetti2021}. At face value, the simulation predictions would not agree with these observations. However, throughout this study we have not attempted to account for such effects as the selection criteria of individual studies (e.g. using different wavelengths to select AGN), or different methods to calculate SFRs. These are all important for robust quantitative comparisons and may introduce some systematic differences, \citep[e.g.][]{Harrison2017,Ji2022}, which may also explain the different correlation coefficients seen in the observational samples investigated here (see Figure \ref{fig:correlations}). A thorough investigation is beyond the scope of this work. Nonetheless, in a broad qualitative sense, the simulations agree with most observational studies that find flat or positive correlations between {\Lbol} and sSFR and {\fgas}, and that AGN are preferentially hosted in gas-rich \citep{Rosario2018,Kirkpatrick2019,Jarvis2020,Shangguan2020b,Zhuang2021,Koss2021,Valentino2021,Salvestrini2022} and star-forming \citep{Rosario2012,Zhuang2019,Ji2022,Kim2022} galaxies. This result is also in agreement with previous studies of these simulations which mostly investigated star formation rates \citep[e.g.][]{McAlpine2017,Scholtz2018}.

These results show that the observational result of finding luminous AGN located preferentially in star-forming or gas-rich galaxies is not in tension with negative AGN feedback, as implemented in these cosmological simulations \citep[see also e.g.][]{Thacker2014,Scholtz2018,Jackson2020}. This is likely not surprising when noting that AGN and host galaxy star formation both require a gas reservoir. Indeed, if the gas is heated, destroyed or ejected by the feedback mechanism, this will not only affect the star formation, causing the galaxy to quench, but will also reduce the gas available to be accreted by the SMBH, limiting the accretion rate and the luminosity. This suggests that a realistic feedback process is one in which an observable impact on the galaxy-wide molecular gas and star formation properties is not observed in-situ with a highly luminous phase of AGN activity or, alternatively, a feedback process powered by low accretion rates, not associated with bolometrically luminous AGN \citep[][]{Harrison2012,Scholtz2018,Florez2020,Luo2021}. Indeed, the simulations do invoke different prescriptions of AGN feedback (see Table~\ref{Sim table}) and observationally testing how realistic these are is an important step as we discuss in Section~\ref{sim differences}. Additionally, this does not rule out a more localised immediate effect of the AGN, as we discuss in Section~\ref{future work}.

Furthermore, even if there is no direct, rapid impact on cold, dense gas, outflows still deposit large amounts of energy into diffuse halo gas \citep[e.g.][]{Costa2014,Costa2018b} or the CGM \citep[e.g.][]{Zinger2020}, increasing the cooling time and preventing gas accretion back onto the galaxy, preventing future star formation. This effect operates over Gyr timescales so will be shaped by the cumulative energy injected by many outflows.

Another factor to consider when observationally searching for the impact of luminous AGN on their host galaxies, is the high variability of accretion rate and corresponding bolometric luminosity. The observed luminosity of an AGN -- whether through optical, IR or X-ray measurements -- is the key way we categorise galaxies into active or inactive. However, it only captures the \textit{instantaneous} accretion rate of the SMBH and therefore is only a weak indicator of the cumulative energy injected into the host galaxy during feedback. It has been shown both observationally and in simulations that AGN accretion rates can vary by multiple orders of magnitude over timescales much shorter than typical star formation episodes \citep[e.g.][]{Novak2011,Hickox2014,Schawinski2015}. Consequently, very little information on the total energy input by a SMBH on timescales relevant to star formation can be inferred from a single measurement of {\Lbol} \citep{Harrison2017}. 

Rather than looking at the instantaneous quantity of {\Lbol}, we could use the mass of the black hole ($M_{\rm{BH}}$) as an integrated quantity of the SMBH's accretion history. For example, \cite{Piotrowska2021} study the predictive power of different galaxy properties on whether or not a galaxy is quenched, in the EAGLE, Illustris and TNG simulations, and in observations. They find that it is the mass of the SMBH, not its accretion rate, that is the strongest predictor for quenching. This corroborates the study of \cite{Terrazas2017}, who found that quiescence correlates strongly with black hole mass both in an observational sample and the Illustris simulation, finding that the sSFR was a smoothly declining function of the specific black hole mass ($M_{\rm{BH}}/M_{\star}$), and the conclusion of \cite{Thomas2019} who found black hole mass to be a key indicator of quenching in the SIMBA simulation. Therefore, these authors conclude that black hole mass is a good indicator of quiescence as it provides a robust, model-independent feedback tracer which is insensitive to the mode of feedback. However, this still tells us little about the details of how or when the energy was deposited. 

To test our understanding of the physical processes behind AGN feedback, we need to look for {\em different} predictions across the simulations, which all have different subgrid implementations of feedback mechanisms (Section~\ref{Simulations}; Table~\ref{Sim table}). Indeed our results have shown different predicted behaviours across the three simulations of galaxy properties as a function of AGN luminosity (Figures~\ref{fig:correlations}, \ref{fig:contours}, \ref{fig:hist_highLbol} and \ref{fig:f_Q}). We now discuss how these differences may relate to the subgrid feedback models and whether this could allow us to evaluate how realistic the implementations of AGN feedback are.

\subsection{Different feedback modes lead to different simulation predictions at $z=0$} \label{sim differences}

\begin{figure*}
    \centering
    \includegraphics[width=0.95\textwidth]{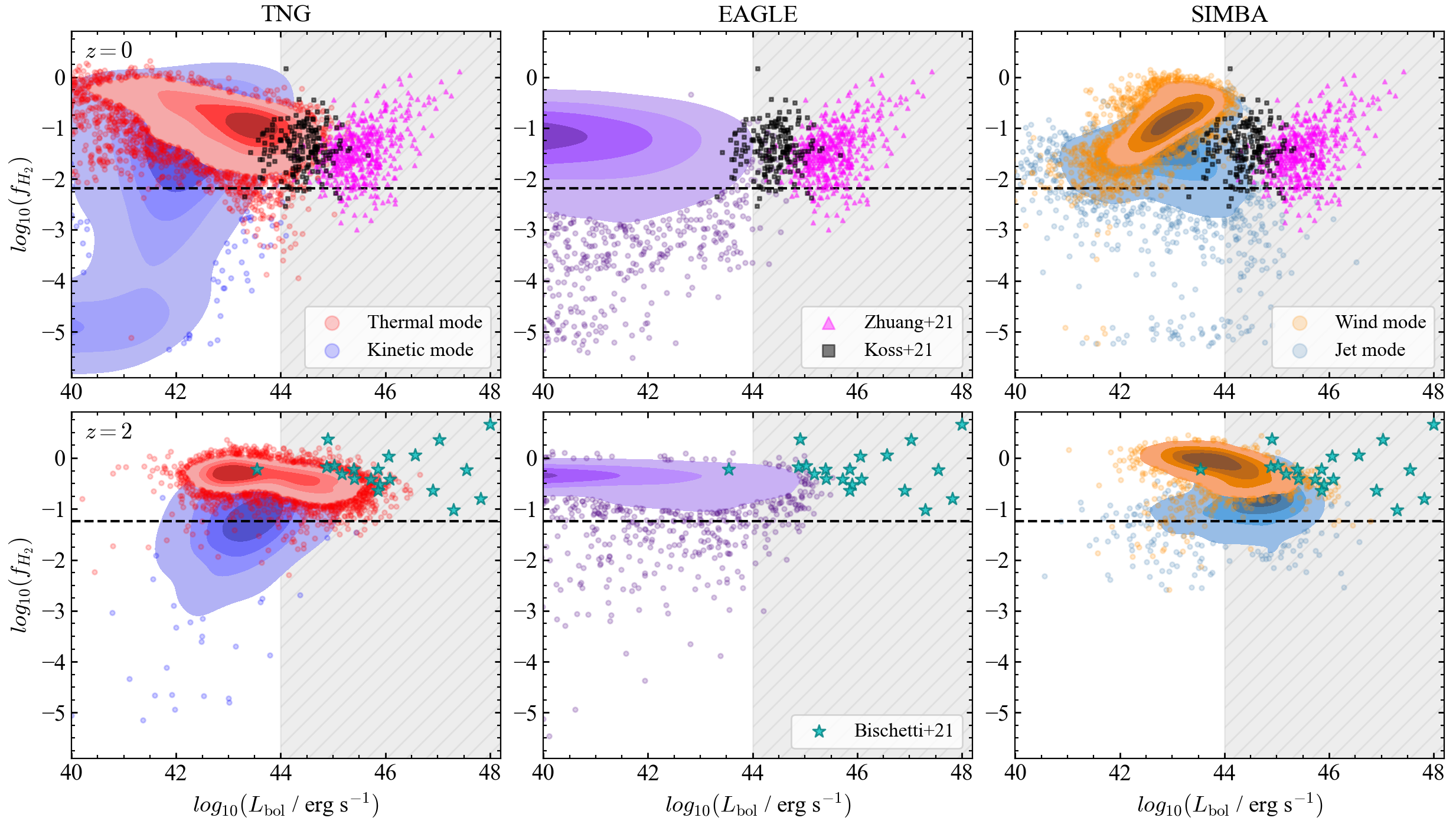}
    \caption{The distribution of gas fraction with $L_{\rm{bol}}$ for each of the three simulations at $z=0$ (\textit{top row}) and $z=2$ (\textit{bottom row}). Shaded contours show the number density containing 90\% of the subhalos and  the extreme systems that lie outside are plotted as single points. The observational samples are also shown as coloured points. The hatched grey region to the right corresponds to a cut of {\Lbol}$>10^{44} \rm{erg s^{-1}}$ and the black dashed line shows a rough gas-depletion definition for a galaxy with {\MstarEq{=}{10.5}} to guide the eye. Galaxies with unresolved gas fraction are scatted around $f_{H_2}=10^{-5}$. The simulation data is coloured according to the accretion mode the AGN is in. In contrast to the other two simulations, at $z=0$ SIMBA predicts a significant fraction of high-{\Lbol} systems with depleted gas reservoirs, which are all in the `jet mode' (grey data points).}
    \label{fig:hist_cont}
\end{figure*}

The main differences between the simulations predictions are: (1) the strength of the correlations between sSFR and {\fgas} with {\Lbol} (Figure \ref{fig:correlations}); (2) the distribution of AGN in the {\fgas} and sSFR vs $M_\star$ plane (Figures~\ref{fig:contours} and \ref{fig4sSFR}) and; (3) the varying gas-depleted and quenched fractions for the highest {\Lbol} and {\fedd} AGN (Figures~\ref{fig:hist_highLbol} \& \ref{fig:f_Q} and \ref{fig5sSFR} \& \ref{fig6sSFR}). Interestingly, these different predictions show much greater variation at $z=0$ than at $z=2$, which might seem surprising because free parameters in cosmological simulations are generally calibrated on the local universe \citep[][]{Schaye2015,Springel2018,Dave2019}.

To explore these differences further, Figure \ref{fig:hist_cont} shows the gas fraction against the bolometric luminosity for the whole galaxy sample in the three simulations. We also split the data from TNG and SIMBA into their two primary feedback modes. The grey hatched region to the right of each panel shows our {\LbolEq{\geq}{44}} selection. To guide the eye, the black dotted line shows a rough gas-depleted definition, assuming a fiducial {\MstarEq{=}{10.5}} galaxy (typical of our AGN hosts) and a tight cut of $\Delta_{\rm{MS}}=-1$. The equivalent plot for sSFR can be found in the appendix (Figure~\ref{fig7sSFR}).

At $z=0$, almost all TNG galaxies in the thermal mode are found above the dashed line in Figure \ref{fig:hist_cont}, i.e. they are not gas depleted, although they do seem to show a slight negative trend with {\Lbol} (see also Figure~\ref{fig:correlations}). In contrast, gas depleted galaxies are almost exclusively found in the kinetic mode. There are very few high-{\Lbol}, low-{\fgas} sources and they are heavily outnumbered by the large number of bright quasars at {\fgasEq{\simeq}{-1}}, causing very low gas-depleted fractions, regardless of the AGN selection (Figure \ref{fig:f_Q}). We can see this switch in accretion mode more clearly in Figure \ref{fig:new_accretion}: low mass SMBHs generally have high accretion rates and are in the thermal mode, growing in size until they reach the invoked accretion mode boundary (black dotted line) at which point they drop in luminosity and become kinetic mode sources. This creates a large population of kinetic mode sources at a black hole mass of $M_{\rm{BH}}\simeq 10^{8.1-8.4} \rm{\ M_{\odot}}$ and luminosity of {\LbolEq{\simeq}{41-43}}. This can also be seen in the downward `plume' in Figure \ref{fig:contours} at a stellar mass of $M_\star \simeq 10^{10.5-11} \rm{\ M_\odot}$ and is linked to the onset of quenching in these massive galaxies. 

Our findings for TNG corroborate the work of \cite{Terrazas2020} and \cite{Zinger2020} who both find a critical black hole mass threshold for quiescence of $M_{\rm{BH}}\simeq 10^{8.2} \rm{\ M_{\odot}}$, above which the switch to the kinetic feedback mode efficiently pushed cold gas out of the galaxy, causing a sharp decrease in the amount of material available to cool and form stars \citep[see also,][]{Weinberger2018}. From Figure \ref{fig:hist_cont} we can see that this gas-depleted population in the kinetic mode is already beginning to be established at $z=2$, although the relatively low gas-depleted fraction in the overall galaxy population at this redshift suggests it has not had time to fully establish a large gas-depleted population. Interestingly, Figure~\ref{fig7sSFR} shows that, for sSFR, there is a quenched population in the kinetic mode at $z=2$, although these sources make up a much lower proportion of the total galaxy population than at $z=0$.

For both redshifts investigated, the results for EAGLE in Figure \ref{fig:hist_cont} show a similar picture; a flat trend in {\fgas} with {\Lbol}, as also seen in Figure \ref{fig:correlations}. This similarity between the epochs is likely to be due to the single feedback mode; the dominant feedback mode is the same at both $z=0$ and $z=2$ and there isn't a low-{\fedd} mode kicking in at later times unlike in TNG and SIMBA. The consistently flat trends seen in EAGLE suggest that the instantaneous {\Lbol} is not driving an immediate gas-depleting impact. This could be due to the fact that the energy input is `pulsed' \citep[following][]{Booth2009}, which separates the accretion rate at any given time from its later cumulative impact on the galaxy. This is in contrast to the otherwise similar thermal mode in TNG where the energy input is continuous. Therefore, this may explain the weak negative correlations between {\fgas} and {\Lbol} seen in Figure~\ref{fig:hist_cont} for the thermal mode in TNG, but which are flat in EAGLE.

If we consider the correlations seen in SIMBA (Figure \ref{fig:correlations}), we notice that the direction of the trend between {\fgas} and {\Lbol} reverses between $z=2$ and $z=0$. Figure \ref{fig:hist_cont} suggests that this is driven by the `wind'-mode sources (in orange) which dominate the correlations. However, the biggest difference from TNG is in the location of the `jet'-mode sources (in grey) which can have high accretion rates, and thus be bolometrically luminous, despite having relatively low Eddington ratios (see Figure~\ref{fig:new_accretion}). Because the `jet' mode is more efficient at removing gas from the host galaxy, due to more powerful kinetic outflows and thermal heating \citep[][]{Thomas2019}, we thus find high-{\Lbol} AGN in low-{\fgas} galaxies in SIMBA which we do not see in EAGLE or TNG. This population is the source of the lower peak seen in Figure \ref{fig:hist_highLbol} and causes the high quenched fraction in Figure \ref{fig:f_Q} when we select AGN based on their {\Lbol}. This also explains why this peak and the gas depleted fraction is reduced when we select for both luminosity and Eddington ratio -- these sources making up this depleted population have low {\fedd}, despite their high {\Lbol}. This could present an observational challenge as low-{\fedd} AGN may be accreting inefficiently and thus be less bolometrically luminous than their high accretion rate may imply (see Section \ref{AGN Selection}). 

We have shown that the different predicted trends with bolometric luminosity (Figure~\ref{fig:correlations}) and the different distribution of galaxy's gas fractions and sSFRs as a function of AGN luminosity (Figure~\ref{fig:contours} and \ref{fig:hist_highLbol}) can be attributed, at least in part, to the different feedback modes invoked by the three different simulations. These differences are greater at lower redshifts, possibly because the low-accretion modes in TNG and SIMBA only become dominant at late cosmic times and it is these modes that are designed to have the greatest impact in galaxy quenching \citep[][]{Weinberger2017}. EAGLE does not have such a mode-switch and thus is is the most similar across the two epochs. This suggests that any observational work attempting to test these simulations with the types of experiments discussed would have more diagnostic power if performed on more local galaxies. However there may be other factors that act from $z=2$ to $z=0$ apart from AGN feedback that create such differences, for example, the switch from cold to hot halo gas that happens around cosmic noon \citep[][]{Dekel2006}.

The varied subgrid implementations of feedback do produce some quantifiable differences between the simulations which may help to constrain the models, for example, the narrow {\Lbol} and stellar mass range of the quenching kinetic-mode systems in TNG (Figures~\ref{fig:contours} and \ref{fig:hist_cont}) around a characteristic black hole mass of $M_{\rm{BH}}\simeq 10^{8.1-8.4} \rm{\ M_{\odot}}$ stands in contrast to EAGLE where there is no preferred {\Lbol} for galaxies undergoing quenching. Additionally, SIMBA is the only simulation to predict quenched galaxies with high SMBH accretion rates. 

\subsection{Limitations and Outlook} \label{future work}


One of the limitations of this study is the lack of comparable high-{\Lbol} AGN in these cosmological simulations with respect to the observational samples. As we can see in Figure \ref{fig:hist_cont}, at $z=0$ there is only a small overlap with the data from \cite{Koss2021} ({\LbolEq{\simeq}{43.5-45}}) and almost no overlap with the data from \cite{Zhuang2021} ({\LbolEq{\gtrsim}{44.5}}). This is problematic, as some studies have suggested that quasar feedback is only effective in this extremely luminous regime \citep[][]{Costa2018b,Valentino2021}. Such systems may exist in these simulations, but they are too short-lived to be captured in the limited box-size and coarsely time-sampled output snap-shot data of the current generation of cosmological simulations. To improve the sampling of high-{\Lbol} AGN, we could increase the volume of these big-box cosmological simulations, or run zoom-in simulations on massive halos where we can specify high accretion rates to test the effectiveness of quasar feedback in this regime. The larger volume of TNG300 may offer some improvements here; it contains 273 galaxies at $z=0$ with an AGN luminosity of {\LbolEq{\geq}{45}} compared with only 10 in TNG100, but the brightest AGN is only {\LbolEq{=}{46.3}} which still leads to the extremely luminous end of the parameter space being too sparsely populated.

Additionally, improvements to the numerical treatment of the molecular phase of the ISM are required for a more realistic comparison to observations. This demands a combined treatment of complex chemistry and non-equilibrium low-temperature cooling \citep[e.g.][]{Richings2014a,Richings2014b}, dust, and radiative transfer \citep[e.g.][]{Rosdahl2015}, which have not been modelled in the current generation of cosmological simulations. Although the post-processing models used in this work are useful for estimating the molecular gas mass in a galaxy, by not directly simulating the ISM we may be missing important feedback channels related to the direct effect of AGN on molecular gas \citep[e.g.][]{Bieri2017,Costa2018b,Costa2020}. Furthermore, applying next-generation models of quasar winds \citep{Costa2020}, jets \citep[][]{Mukherjee2016,Mukherjee2018,Talbot2021,Bourne2021,Mandal2021,Tanner2022} and radiation pressure \citep[][]{Bieri2017,Costa2018a,Costa2018b,Ishibashi2018} to cosmological simulations will make it possible to test more realistic AGN physical mechanisms by comparing to observations and consequently provide new insights into the true physical mechanisms that quench massive galaxies.

In this study, we have only looked at global galaxy properties and broadly found no connection between AGN luminosity and reduced gas content or star formation rates. However, building on the improved spatial resolution offered by modern instruments, many observers are looking at individual regions of galaxies to see where quenching or star formation might be occurring through high spatial-resolution and multi-wavelength observations \citep[e.g.][]{Yazeedi2021,Girdhar2022}. Some studies have found that central regions are gas-depleted which could be linked to AGN activity \citep{Ellison2021}, although others have found no difference in the central molecular gas content of active and inactive galaxies \citep{Rosario2018}. There have also been studies looking at the radial profiles of galaxies in the simulations which found that quenching galaxies in EAGLE and Illustris have too centrally-concentrated star formation \citep[][]{Starkenburg2019}, although \cite{Appleby2020} found that SIMBA did sufficiently suppress central star formation to match their observational sample. Investigating the timescale on which such central suppression occurs and seeing if this could be linked to a specific AGN feedback event could be an interesting avenue for future analysis of these simulations. 

Finally, we note that there are several interesting populations of active galaxies, beyond just simply focusing on the most luminous, that may be interesting for understanding feedback and for testing simulations. In particular, galaxies undergoing the transition between star-forming and quiescent could display some unique properties in how the AGN is interacting with the host. Post-Starburst Galaxies (PSBs; also known as E+A galaxies) have been proposed as recently quenched galaxies, with evidence of AGN winds in some cases \citep[][]{Wild2016,Baron2018,Baron2022}. Another population is Red Geyser galaxies \citep[][]{Cheung2016,Roy2021}. These host low-luminosity AGN, but exhibit powerful outflows which could be responsible for their quiescence. Such systems could share many properties with the quenching population seen in TNG, where the SMBH has low accretion rates ({\LbolEq{\simeq}{42}}) but through the kinetic mode feedback is able to launch outflows that act to quench the galaxy. Studying these complex and nuanced systems may provide another way of furthering our understanding of the intricate interaction between the AGN and host galaxies, allowing us to include ever more physically-motivated models in our theories of galaxy evolution.

\section{Conclusions} \label{conclusion}

We have investigated the predicted relationships between AGN activity and the host galaxy properties of molecular gas fraction ({\fgas}) and specific star formation rate (sSFR) in three contemporary cosmological simulations (TNG, EAGLE and SIMBA) for epochs of $z=0$ and $z=2$. We were motivated by observational studies that have searched for signatures of AGN feedback by looking at trends between {\Lbol} and galaxy properties, or that compared AGN to non-AGN. Several observational studies have suggested that the lack of negative trends between sSFR or {\fgas} and AGN luminosity could indicate that AGN feedback may be ineffective. To investigate this, we followed the broad approaches taken in observational studies by looking at: (1) the correlation coefficients between the AGN luminosity and the host galaxy properties, after controlling for stellar mass; (2) the location of the AGN host galaxies compared to non-AGN hosts in the sSFR$-$ and {\fgas}$-M_\star$ plane; and (3) the gas-depleted and quenched fractions in AGN host galaxies. For all of these methods, we explore the effect of two different AGN selection definitions: a luminosity-based cut of {\LbolEq{\geq}{44}}, and an Eddington ratio criterion of {\fedd}$\geq 1\%$. Our main results are:

\begin{enumerate}
    \item In all three simulations we see no strong negative correlations between {\Lbol} and sSFR or {\fgas} at $z=0$ or $z=2$ for star-forming galaxies (Figure \ref{fig:correlations}). Out of the 12 correlations investigated, only one shows even a weak negative correlation with the rest either being flat or positively correlated. This result is consistent whether we consider the whole star-forming population or only those with an Eddington ratio-selected AGN.
    \item AGN are predicted to preferentially reside in gas-rich, high-sSFR galaxies in all three simulations at both redshifts investigated. This result holds whether we consider an {\Lbol}- or an {\fedd}-based AGN selection (Figure \ref{fig:contours}). 
    \item Overall, the distributions in {\fgas} for the highest {\Lbol} and {\fedd} sources (Figure~\ref{fig:hist_highLbol}) show a low ($\lesssim 7\%$) gas depleted fraction (Figure~\ref{fig:f_Q}) which is always lower than for a mass-matched control sample of inactive galaxies. However, unlike the other simulations, SIMBA shows a bimodal distribution with a unique population of high-{\Lbol} galaxies located in gas-depleted and quenched galaxies.
\end{enumerate}

All three simulations are heavily reliant on negative AGN feedback in order to reproduce realistic galaxies and yet we have shown that: (1) despite this negative feedback, we do not see any predictions of strong negative correlations between {\Lbol} and sSFR or {\fgas} and (2) AGN host galaxies are not predicted to contain significantly lower gas fractions or star formation activity compared to mass-matched non-AGN host galaxies. Therefore, this demonstrates that the result of finding bright AGN in star-forming, gas rich galaxies is not necessarily in contradiction with the presence of negative feedback. This could be due to various reasons, including timescale effects and the relative contribution of high or low accretion rate AGN to causing the quenching (Section~\ref{main discussion}).

However, the simulations do produce some different predictions for the trends of global galaxy properties with {\Lbol} (Figure~\ref{fig:correlations}) and the distribution of galaxies in the {\Lbol}$-${\fgas} plane (Figure~\ref{fig:hist_cont}). These can be attributed to the different subgrid accretion and feedback models implemented (Figures \ref{fig:new_accretion} \& \ref{fig:hist_cont}) which produce unique populations; for example, the highly-accreting, but low-{\fedd} population in SIMBA, or the downwards `plume' of kinetic-mode galaxies at low-{\Lbol} in TNG, which are both associated with gas-depleted galaxies. We also note that these differences are greater at $z=0$ compared to $z=2$, perhaps suggesting that observations in local Universe will provide more diagnostic power for testing the different simulations.

Based on the results from this work, we suggest that focusing on population trends of only the most luminous AGN does not provide strong diagnostic power for testing AGN feedback. Instead, we recommend that observational efforts focus on testing some of the unusual quenching/quenched populations seen in the simulations and try to link these to observational samples. Complementary to this is observationally investigating AGN host galaxies in detail to understand the spatially-resolved physical processes of AGN feedback. Together, these kinds of studies could help to constrain which feedback models are the most realistic and help to inform future theoretical models of galaxy evolution.

\section*{Acknowledgements}

We thank the referee for their constructive report, which helped improve the clarity of the manuscript. We also express our gratitude to Stuart McAlpine for his assistance with the EAGLE simulation data, as well as to Luis Ho, Manuela Bischetti, Michael Koss and Chiara Circosta for help collating the observational samples. We would also like to thank Miranda Jarvis for her early work on this project and Volker Springel, Stephen Molyneux and Aishwarya Girdhar for their helpful comments on the paper. SRW acknowledges funding from the Deutsche Forschungsgemeinschaft (DFG, German Research Foundation) under Germany's Excellence Strategy: EXC-2094-390783311. CMH acknowledges funding from a United Kingdom Research and Innovation grant (project: MR/V022830/1).

\section*{Data Availability}

The simulation data used in this work can be accessed via their respective websites:

\begin{itemize}
    \item {\sc IllustrisTNG}: \url{https://www.tng-project.org}
    \item EAGLE: \url{http://icc.dur.ac.uk/Eagle/index.php}
    \item SIMBA: \url{http://simba.roe.ac.uk}
\end{itemize}

The molecular gas masses are included in every snapshot in SIMBA; separate catalogues for five snapshots (including $z=0$ and $z=2$) are available on the TNG webpage; and catalogues for EAGLE are available on request.

Data for the observational samples used are available from the respective release papers: \citealt{Koss2021,Zhuang2021,Bischetti2021}.


\bibliographystyle{mnras}
\bibliography{lib.bib}

\begin{thebibliography}{}
\makeatletter
\relax
\def\mn@urlcharsother{\let\do\@makeother \do\$\do\&\do\#\do\^\do\_\do\%\do\~}
\def\mn@doi{\begingroup\mn@urlcharsother \@ifnextchar [ {\mn@doi@}
  {\mn@doi@[]}}
\def\mn@doi@[#1]#2{\def\@tempa{#1}\ifx\@tempa\@empty \href
  {http://dx.doi.org/#2} {doi:#2}\else \href {http://dx.doi.org/#2} {#1}\fi
  \endgroup}
\def\mn@eprint#1#2{\mn@eprint@#1:#2::\@nil}
\def\mn@eprint@arXiv#1{\href {http://arxiv.org/abs/#1} {{\tt arXiv:#1}}}
\def\mn@eprint@dblp#1{\href {http://dblp.uni-trier.de/rec/bibtex/#1.xml}
  {dblp:#1}}
\def\mn@eprint@#1:#2:#3:#4\@nil{\def\@tempa {#1}\def\@tempb {#2}\def\@tempc
  {#3}\ifx \@tempc \@empty \let \@tempc \@tempb \let \@tempb \@tempa \fi \ifx
  \@tempb \@empty \def\@tempb {arXiv}\fi \@ifundefined
  {mn@eprint@\@tempb}{\@tempb:\@tempc}{\expandafter \expandafter \csname
  mn@eprint@\@tempb\endcsname \expandafter{\@tempc}}}

\bibitem[\protect\citeauthoryear{Abazajian et~al.,}{Abazajian
  et~al.}{2009}]{Abazajian2009}
Abazajian K.~N.,  et~al., 2009, \mn@doi [\apjs] {10.1088/0067-0049/182/2/543},
  182, 543

\bibitem[\protect\citeauthoryear{Akins, Narayanan, Whitaker, Davé, Lower,
  Bezanson, Feldmann  \& Kriek}{Akins et~al.}{2021}]{Akins2021}
Akins H.~B.,  Narayanan D.,  Whitaker K.~E.,  Davé R.,  Lower S.,  Bezanson
  R.,  Feldmann R.,   Kriek M.,  2021, arXiv e-prints, p. arXiv:2105.12748

\bibitem[\protect\citeauthoryear{{Al Yazeedi}, {Katkov}, {Gelfand},
  {Wylezalek}, {Zakamska}  \& {Liu}}{{Al Yazeedi} et~al.}{2021}]{Yazeedi2021}
{Al Yazeedi} A.,  {Katkov} I.~Y.,  {Gelfand} J.~D.,  {Wylezalek} D.,
  {Zakamska} N.~L.,   {Liu} W.,  2021, \mn@doi [\apj]
  {10.3847/1538-4357/abf5e1}, \href
  {https://ui.adsabs.harvard.edu/abs/2021ApJ...916..102A} {916, 102}

\bibitem[\protect\citeauthoryear{{Angl{\'e}s-Alc{\'a}zar}, {Dav{\'e}},
  {Faucher-Gigu{\`e}re}, {{\"O}zel}  \& {Hopkins}}{{Angl{\'e}s-Alc{\'a}zar}
  et~al.}{2017}]{Angles-Alcazar2017}
{Angl{\'e}s-Alc{\'a}zar} D.,  {Dav{\'e}} R.,  {Faucher-Gigu{\`e}re} C.-A.,
  {{\"O}zel} F.,   {Hopkins} P.~F.,  2017, \mn@doi [\mnras]
  {10.1093/mnras/stw2565}, \href
  {https://ui.adsabs.harvard.edu/abs/2017MNRAS.464.2840A} {464, 2840}

\bibitem[\protect\citeauthoryear{{Appleby}, {Dav{\'e}}, {Kraljic},
  {Angl{\'e}s-Alc{\'a}zar}  \& {Narayanan}}{{Appleby}
  et~al.}{2020}]{Appleby2020}
{Appleby} S.,  {Dav{\'e}} R.,  {Kraljic} K.,  {Angl{\'e}s-Alc{\'a}zar} D.,
  {Narayanan} D.,  2020, \mn@doi [\mnras] {10.1093/mnras/staa1169}, \href
  {https://ui.adsabs.harvard.edu/abs/2020MNRAS.494.6053A} {494, 6053}

\bibitem[\protect\citeauthoryear{{Appleby}, {Dav{\'e}}, {Sorini},
  {Storey-Fisher}  \& {Smith}}{{Appleby} et~al.}{2021}]{Appleby2021}
{Appleby} S.,  {Dav{\'e}} R.,  {Sorini} D.,  {Storey-Fisher} K.,   {Smith} B.,
  2021, \mn@doi [\mnras] {10.1093/mnras/stab2310}, \href
  {https://ui.adsabs.harvard.edu/abs/2021MNRAS.507.2383A} {507, 2383}

\bibitem[\protect\citeauthoryear{{Azadi} et~al.,}{{Azadi}
  et~al.}{2015}]{Azadi2015}
{Azadi} M.,  et~al., 2015, \mn@doi [\apj] {10.1088/0004-637X/806/2/187}, \href
  {https://ui.adsabs.harvard.edu/abs/2015ApJ...806..187A} {806, 187}

\bibitem[\protect\citeauthoryear{Azadi et~al.,}{Azadi et~al.}{2017}]{Azadi2017}
Azadi M.,  et~al., 2017, \mn@doi [\apj] {10.3847/1538-4357/835/1/27}, 835, 27

\bibitem[\protect\citeauthoryear{{Baldry}, {Glazebrook}, {Brinkmann},
  {Ivezi{\'c}}, {Lupton}, {Nichol}  \& {Szalay}}{{Baldry}
  et~al.}{2004}]{Baldry2004}
{Baldry} I.~K.,  {Glazebrook} K.,  {Brinkmann} J.,  {Ivezi{\'c}} {\v{Z}}.,
  {Lupton} R.~H.,  {Nichol} R.~C.,   {Szalay} A.~S.,  2004, \mn@doi [\apj]
  {10.1086/380092}, \href
  {https://ui.adsabs.harvard.edu/abs/2004ApJ...600..681B} {600, 681}

\bibitem[\protect\citeauthoryear{Baron \& Netzer}{Baron \&
  Netzer}{2019}]{Baron2019}
Baron D.,  Netzer H.,  2019, \mn@doi [\mnras] {10.1093/MNRAS/STZ1070}, 486,
  4290

\bibitem[\protect\citeauthoryear{{Baron} et~al.,}{{Baron}
  et~al.}{2018}]{Baron2018}
{Baron} D.,  et~al., 2018, \mn@doi [\mnras] {10.1093/mnras/sty2113}, \href
  {https://ui.adsabs.harvard.edu/abs/2018MNRAS.480.3993B} {480, 3993}

\bibitem[\protect\citeauthoryear{{Baron}, {Netzer}, {Lutz}, {Prochaska}  \&
  {Davies}}{{Baron} et~al.}{2022}]{Baron2022}
{Baron} D.,  {Netzer} H.,  {Lutz} D.,  {Prochaska} J.~X.,   {Davies} R.~I.,
  2022, \mn@doi [\mnras] {10.1093/mnras/stab3232}, \href
  {https://ui.adsabs.harvard.edu/abs/2022MNRAS.509.4457B} {509, 4457}

\bibitem[\protect\citeauthoryear{Baumgartner, Tueller, Markwardt, Skinner,
  Barthelmy, Mushotzky, Evans  \& Gehrels}{Baumgartner
  et~al.}{2013}]{Baumgartner2013}
Baumgartner W.~H.,  Tueller J.,  Markwardt C.~B.,  Skinner G.~K.,  Barthelmy
  S.,  Mushotzky R.~F.,  Evans P.~A.,   Gehrels N.,  2013, \mn@doi [\apjs]
  {10.1088/0067-0049/207/2/19}, 207, 19

\bibitem[\protect\citeauthoryear{{Beckmann} et~al.,}{{Beckmann}
  et~al.}{2017}]{Beckmann2017}
{Beckmann} R.~S.,  et~al., 2017, \mn@doi [\mnras] {10.1093/mnras/stx1831},
  \href {https://ui.adsabs.harvard.edu/abs/2017MNRAS.472..949B} {472, 949}

\bibitem[\protect\citeauthoryear{{Bernhard}, {Mullaney}, {Daddi}, {Ciesla}  \&
  {Schreiber}}{{Bernhard} et~al.}{2016}]{Bernhard2016}
{Bernhard} E.,  {Mullaney} J.~R.,  {Daddi} E.,  {Ciesla} L.,   {Schreiber} C.,
  2016, \mn@doi [\mnras] {10.1093/mnras/stw973}, \href
  {https://ui.adsabs.harvard.edu/abs/2016MNRAS.460..902B} {460, 902}

\bibitem[\protect\citeauthoryear{{Bieri}, {Dubois}, {Rosdahl}, {Wagner}, {Silk}
   \& {Mamon}}{{Bieri} et~al.}{2017}]{Bieri2017}
{Bieri} R.,  {Dubois} Y.,  {Rosdahl} J.,  {Wagner} A.,  {Silk} J.,   {Mamon}
  G.~A.,  2017, \mn@doi [\mnras] {10.1093/mnras/stw2380}, \href
  {https://ui.adsabs.harvard.edu/abs/2017MNRAS.464.1854B} {464, 1854}

\bibitem[\protect\citeauthoryear{Bischetti et~al.,}{Bischetti
  et~al.}{2017}]{Bischetti2017}
Bischetti M.,  et~al., 2017, \mn@doi [\aap] {10.1051/0004-6361/201629301},
  \href {https://ui.adsabs.harvard.edu/abs/2017A&A...598A.122B} {598, A122}

\bibitem[\protect\citeauthoryear{Bischetti et~al.,}{Bischetti
  et~al.}{2021}]{Bischetti2021}
Bischetti M.,  et~al., 2021, \mn@doi [A\&A] {10.1051/0004-6361/202039057}, 645,
  A33

\bibitem[\protect\citeauthoryear{{Bluck} et~al.,}{{Bluck}
  et~al.}{2020}]{Bluck2020}
{Bluck} A. F.~L.,  et~al., 2020, \mn@doi [\mnras] {10.1093/mnras/staa2806},
  \href {https://ui.adsabs.harvard.edu/abs/2020MNRAS.499..230B} {499, 230}

\bibitem[\protect\citeauthoryear{{Bondi}}{{Bondi}}{1952}]{Bondi1952}
{Bondi} H.,  1952, \mn@doi [\mnras] {10.1093/mnras/112.2.195}, \href
  {https://ui.adsabs.harvard.edu/abs/1952MNRAS.112..195B} {112, 195}

\bibitem[\protect\citeauthoryear{{Booth} \& {Schaye}}{{Booth} \&
  {Schaye}}{2009}]{Booth2009}
{Booth} C.~M.,  {Schaye} J.,  2009, \mn@doi [\mnras]
  {10.1111/j.1365-2966.2009.15043.x}, \href
  {https://ui.adsabs.harvard.edu/abs/2009MNRAS.398...53B} {398, 53}

\bibitem[\protect\citeauthoryear{{Bothwell} et~al.,}{{Bothwell}
  et~al.}{2013}]{Bothwell2013}
{Bothwell} M.~S.,  et~al., 2013, \mn@doi [\mnras] {10.1093/mnras/sts562}, \href
  {https://ui.adsabs.harvard.edu/abs/2013MNRAS.429.3047B} {429, 3047}

\bibitem[\protect\citeauthoryear{{Bourne} \& {Sijacki}}{{Bourne} \&
  {Sijacki}}{2021}]{Bourne2021}
{Bourne} M.~A.,  {Sijacki} D.,  2021, \mn@doi [\mnras]
  {10.1093/mnras/stab1662}, \href
  {https://ui.adsabs.harvard.edu/abs/2021MNRAS.506..488B} {506, 488}

\bibitem[\protect\citeauthoryear{{Bower}, {Benson}, {Malbon}, {Helly}, {Frenk},
  {Baugh}, {Cole}  \& {Lacey}}{{Bower} et~al.}{2006}]{Bower2006}
{Bower} R.~G.,  {Benson} A.~J.,  {Malbon} R.,  {Helly} J.~C.,  {Frenk} C.~S.,
  {Baugh} C.~M.,  {Cole} S.,   {Lacey} C.~G.,  2006, \mn@doi [\mnras]
  {10.1111/j.1365-2966.2006.10519.x}, \href
  {https://ui.adsabs.harvard.edu/abs/2006MNRAS.370..645B} {370, 645}

\bibitem[\protect\citeauthoryear{{Cattaneo} et~al.,}{{Cattaneo}
  et~al.}{2009}]{Cattaneo2009}
{Cattaneo} A.,  et~al., 2009, \mn@doi [\nat] {10.1038/nature08135}, \href
  {https://ui.adsabs.harvard.edu/abs/2009Natur.460..213C} {460, 213}

\bibitem[\protect\citeauthoryear{{Chen}, {He}, {Ho}, {Gu}, {Wang}, {Zhuang},
  {Liu}  \& {Wang}}{{Chen} et~al.}{2022}]{Chen2022}
{Chen} Z.,  {He} Z.,  {Ho} L.~C.,  {Gu} Q.,  {Wang} T.,  {Zhuang} M.,  {Liu}
  G.,   {Wang} Z.,  2022, \mn@doi [Nature Astronomy]
  {10.1038/s41550-021-01561-3}, \href
  {https://ui.adsabs.harvard.edu/abs/2022NatAs.tmp....5C} {}

\bibitem[\protect\citeauthoryear{{Cheung} et~al.,}{{Cheung}
  et~al.}{2016}]{Cheung2016}
{Cheung} E.,  et~al., 2016, \mn@doi [\nat] {10.1038/nature18006}, \href
  {https://ui.adsabs.harvard.edu/abs/2016Natur.533..504C} {533, 504}

\bibitem[\protect\citeauthoryear{{Choi}, {Somerville}, {Ostriker}, {Naab}  \&
  {Hirschmann}}{{Choi} et~al.}{2018}]{Choi2018}
{Choi} E.,  {Somerville} R.~S.,  {Ostriker} J.~P.,  {Naab} T.,   {Hirschmann}
  M.,  2018, \mn@doi [\apj] {10.3847/1538-4357/aae076}, \href
  {https://ui.adsabs.harvard.edu/abs/2018ApJ...866...91C} {866, 91}

\bibitem[\protect\citeauthoryear{{Cicone}, {Brusa}, {Ramos Almeida}, {Cresci},
  {Husemann}  \& {Mainieri}}{{Cicone} et~al.}{2018}]{Cicone2018}
{Cicone} C.,  {Brusa} M.,  {Ramos Almeida} C.,  {Cresci} G.,  {Husemann} B.,
  {Mainieri} V.,  2018, \mn@doi [Nature Astronomy] {10.1038/s41550-018-0406-3},
  \href {https://ui.adsabs.harvard.edu/abs/2018NatAs...2..176C} {2, 176}

\bibitem[\protect\citeauthoryear{{Circosta} et~al.,}{{Circosta}
  et~al.}{2021}]{Circosta2021}
{Circosta} C.,  et~al., 2021, \mn@doi [\aap] {10.1051/0004-6361/202039270},
  \href {https://ui.adsabs.harvard.edu/abs/2021A&A...646A..96C} {646, A96}

\bibitem[\protect\citeauthoryear{{Costa}, {Sijacki}  \& {Haehnelt}}{{Costa}
  et~al.}{2014}]{Costa2014}
{Costa} T.,  {Sijacki} D.,   {Haehnelt} M.~G.,  2014, \mn@doi [\mnras]
  {10.1093/mnras/stu1632}, \href
  {https://ui.adsabs.harvard.edu/abs/2014MNRAS.444.2355C} {444, 2355}

\bibitem[\protect\citeauthoryear{{Costa}, {Rosdahl}, {Sijacki}  \&
  {Haehnelt}}{{Costa} et~al.}{2018a}]{Costa2018a}
{Costa} T.,  {Rosdahl} J.,  {Sijacki} D.,   {Haehnelt} M.~G.,  2018a, \mn@doi
  [\mnras] {10.1093/mnras/stx2598}, \href
  {https://ui.adsabs.harvard.edu/abs/2018MNRAS.473.4197C} {473, 4197}

\bibitem[\protect\citeauthoryear{{Costa}, {Rosdahl}, {Sijacki}  \&
  {Haehnelt}}{{Costa} et~al.}{2018b}]{Costa2018b}
{Costa} T.,  {Rosdahl} J.,  {Sijacki} D.,   {Haehnelt} M.~G.,  2018b, \mn@doi
  [\mnras] {10.1093/mnras/sty1514}, \href
  {https://ui.adsabs.harvard.edu/abs/2018MNRAS.479.2079C} {479, 2079}

\bibitem[\protect\citeauthoryear{Costa, Pakmor  \& Springel}{Costa
  et~al.}{2020}]{Costa2020}
Costa T.,  Pakmor R.,   Springel V.,  2020, \mn@doi [\mnras]
  {10.1093/mnras/staa2321}, 497, 5229

\bibitem[\protect\citeauthoryear{Crain et~al.,}{Crain et~al.}{2015}]{Crain2015}
Crain R.~A.,  et~al., 2015, \mn@doi [\mnras] {10.1093/mnras/stv725}, 450, 1937

\bibitem[\protect\citeauthoryear{{Croton} et~al.,}{{Croton}
  et~al.}{2006}]{Croton2006}
{Croton} D.~J.,  et~al., 2006, \mn@doi [\mnras]
  {10.1111/j.1365-2966.2005.09675.x}, \href
  {https://ui.adsabs.harvard.edu/abs/2006MNRAS.365...11C} {365, 11}

\bibitem[\protect\citeauthoryear{{Cui}, {Dav{\'e}}, {Peacock},
  {Angl{\'e}s-Alc{\'a}zar}  \& {Yang}}{{Cui} et~al.}{2021}]{Cui2021}
{Cui} W.,  {Dav{\'e}} R.,  {Peacock} J.~A.,  {Angl{\'e}s-Alc{\'a}zar} D.,
  {Yang} X.,  2021, \mn@doi [Nature Astronomy] {10.1038/s41550-021-01404-1},
  \href {https://ui.adsabs.harvard.edu/abs/2021NatAs...5.1069C} {5, 1069}

\bibitem[\protect\citeauthoryear{{Dav{\'e}}, {Thompson}  \&
  {Hopkins}}{{Dav{\'e}} et~al.}{2016}]{Dave2016}
{Dav{\'e}} R.,  {Thompson} R.,   {Hopkins} P.~F.,  2016, \mn@doi [\mnras]
  {10.1093/mnras/stw1862}, 462, 3265

\bibitem[\protect\citeauthoryear{Davé, Anglés-Alcázar, Narayanan, Li,
  Rafieferantsoa  \& Appleby}{Davé et~al.}{2019}]{Dave2019}
Davé R.,  Anglés-Alcázar D.,  Narayanan D.,  Li Q.,  Rafieferantsoa M.~H.,
  Appleby S.,  2019, \mn@doi [MNRAS] {10.1093/mnras/stz937}, 486, 2827

\bibitem[\protect\citeauthoryear{{Dekel} \& {Birnboim}}{{Dekel} \&
  {Birnboim}}{2006}]{Dekel2006}
{Dekel} A.,  {Birnboim} Y.,  2006, \mn@doi [\mnras]
  {10.1111/j.1365-2966.2006.10145.x}, \href
  {https://ui.adsabs.harvard.edu/abs/2006MNRAS.368....2D} {368, 2}

\bibitem[\protect\citeauthoryear{Diemer et~al.,}{Diemer
  et~al.}{2018}]{Diemer2018}
Diemer B.,  et~al., 2018, \mn@doi [\apjs] {10.3847/1538-4365/aae387}, 238, 33

\bibitem[\protect\citeauthoryear{Diemer et~al.,}{Diemer
  et~al.}{2019}]{Diemer2019}
Diemer B.,  et~al., 2019, \mn@doi [\mnras] {10.1093/mnras/stz1323}, 487, 1529

\bibitem[\protect\citeauthoryear{Donnari et~al.,}{Donnari
  et~al.}{2019}]{Donnari2019}
Donnari M.,  et~al., 2019, \mn@doi [\mnras] {10.1093/mnras/stz712}, 485, 4817

\bibitem[\protect\citeauthoryear{{Dubois}, {Peirani}, {Pichon}, {Devriendt},
  {Gavazzi}, {Welker}  \& {Volonteri}}{{Dubois} et~al.}{2016}]{Dubois2016}
{Dubois} Y.,  {Peirani} S.,  {Pichon} C.,  {Devriendt} J.,  {Gavazzi} R.,
  {Welker} C.,   {Volonteri} M.,  2016, \mn@doi [\mnras]
  {10.1093/mnras/stw2265}, \href
  {https://ui.adsabs.harvard.edu/abs/2016MNRAS.463.3948D} {463, 3948}

\bibitem[\protect\citeauthoryear{{Ellison}, {Patton}  \& {Hickox}}{{Ellison}
  et~al.}{2015}]{Ellison15}
{Ellison} S.~L.,  {Patton} D.~R.,   {Hickox} R.~C.,  2015, \mn@doi [\mnras]
  {10.1093/mnrasl/slv061}, \href
  {https://ui.adsabs.harvard.edu/abs/2015MNRAS.451L..35E} {451, L35}

\bibitem[\protect\citeauthoryear{{Ellison} et~al.,}{{Ellison}
  et~al.}{2021}]{Ellison2021}
{Ellison} S.~L.,  et~al., 2021, \mn@doi [\mnras] {10.1093/mnrasl/slab047},
  \href {https://ui.adsabs.harvard.edu/abs/2021MNRAS.505L..46E} {505, L46}

\bibitem[\protect\citeauthoryear{Fabian}{Fabian}{2012}]{Fabian2012}
Fabian A.~C.,  2012, \mn@doi [\araa] {10.1146/annurev-astro-081811-125521}, 50,
  455

\bibitem[\protect\citeauthoryear{Fiore et~al.,}{Fiore et~al.}{2017}]{Fiore2017}
Fiore F.,  et~al., 2017, \mn@doi [\aap] {10.1051/0004-6361/201629478}, 601, 143

\bibitem[\protect\citeauthoryear{Florez et~al.,}{Florez
  et~al.}{2020}]{Florez2020}
Florez J.,  et~al., 2020, \mn@doi [MNRAS] {10.1093/mnras/staa2200}, 497, 3273

\bibitem[\protect\citeauthoryear{Furlong et~al.,}{Furlong
  et~al.}{2015}]{Furlong2015}
Furlong M.,  et~al., 2015, \mn@doi [MNRAS] {10.1093/mnras/stv852}, 450, 4486

\bibitem[\protect\citeauthoryear{{Girdhar} et~al.,}{{Girdhar}
  et~al.}{2022}]{Girdhar2022}
{Girdhar} A.,  et~al., 2022, \mn@doi [\mnras] {10.1093/mnras/stac073}, \href
  {https://ui.adsabs.harvard.edu/abs/2022MNRAS.tmp..170G} {}

\bibitem[\protect\citeauthoryear{Gnedin \& Kravtsov}{Gnedin \&
  Kravtsov}{2011}]{Gnedin2011}
Gnedin N.~Y.,  Kravtsov A.~V.,  2011, \mn@doi [\apj]
  {10.1088/0004-637X/728/2/88}, 728, 88

\bibitem[\protect\citeauthoryear{Greene \& Ho}{Greene \&
  Ho}{2005}]{GreeneHo2005}
Greene J.~E.,  Ho L.~C.,  2005, \mn@doi [\apj] {10.1086/431897}, 630, 122

\bibitem[\protect\citeauthoryear{Greene, Strader  \& Ho}{Greene
  et~al.}{2020}]{Greene2020}
Greene J.~E.,  Strader J.,   Ho L.~C.,  2020, \mn@doi [\araa]
  {10.1146/annurev-astro-032620-021835}, 58, 257

\bibitem[\protect\citeauthoryear{{G{\"u}rkan} et~al.,}{{G{\"u}rkan}
  et~al.}{2015}]{Gurkan2015}
{G{\"u}rkan} G.,  et~al., 2015, \mn@doi [\mnras] {10.1093/mnras/stv1502}, \href
  {https://ui.adsabs.harvard.edu/abs/2015MNRAS.452.3776G} {452, 3776}

\bibitem[\protect\citeauthoryear{{Habouzit} et~al.,}{{Habouzit}
  et~al.}{2021}]{Habouzit2021}
{Habouzit} M.,  et~al., 2021, \mn@doi [\mnras] {10.1093/mnras/stab496}, \href
  {https://ui.adsabs.harvard.edu/abs/2021MNRAS.503.1940H} {503, 1940}

\bibitem[\protect\citeauthoryear{{Habouzit} et~al.,}{{Habouzit}
  et~al.}{2022}]{Habouzit2022}
{Habouzit} M.,  et~al., 2022, \mn@doi [\mnras] {10.1093/mnras/stab3147}, \href
  {https://ui.adsabs.harvard.edu/abs/2022MNRAS.509.3015H} {509, 3015}

\bibitem[\protect\citeauthoryear{Harrison}{Harrison}{2017}]{Harrison2017}
Harrison C.~M.,  2017, \mn@doi [Nature Astronomy] {10.1038/s41550-017-0165}, 1,
  0165

\bibitem[\protect\citeauthoryear{{Harrison} et~al.,}{{Harrison}
  et~al.}{2012}]{Harrison2012}
{Harrison} C.~M.,  et~al., 2012, \mn@doi [\apjl] {10.1088/2041-8205/760/1/L15},
  \href {https://ui.adsabs.harvard.edu/abs/2012ApJ...760L..15H} {760, L15}

\bibitem[\protect\citeauthoryear{{Hickox}, {Mullaney}, {Alexander}, {Chen},
  {Civano}, {Goulding}  \& {Hainline}}{{Hickox} et~al.}{2014}]{Hickox2014}
{Hickox} R.~C.,  {Mullaney} J.~R.,  {Alexander} D.~M.,  {Chen} C.-T.~J.,
  {Civano} F.~M.,  {Goulding} A.~D.,   {Hainline} K.~N.,  2014, \mn@doi [\apj]
  {10.1088/0004-637X/782/1/9}, \href
  {https://ui.adsabs.harvard.edu/abs/2014ApJ...782....9H} {782, 9}

\bibitem[\protect\citeauthoryear{{Hopkins}}{{Hopkins}}{2015}]{Hopkins2015}
{Hopkins} P.~F.,  2015, \mn@doi [\mnras] {10.1093/mnras/stv195}, \href
  {https://ui.adsabs.harvard.edu/abs/2015MNRAS.450...53H} {450, 53}

\bibitem[\protect\citeauthoryear{{Hopkins} \& {Quataert}}{{Hopkins} \&
  {Quataert}}{2011}]{Hopkins2011}
{Hopkins} P.~F.,  {Quataert} E.,  2011, \mn@doi [\mnras]
  {10.1111/j.1365-2966.2011.18542.x}, \href
  {https://ui.adsabs.harvard.edu/abs/2011MNRAS.415.1027H} {415, 1027}

\bibitem[\protect\citeauthoryear{{Husemann}, {Davis}, {Jahnke}, {Dannerbauer},
  {Urrutia}  \& {Hodge}}{{Husemann} et~al.}{2017}]{Husemann2017}
{Husemann} B.,  {Davis} T.~A.,  {Jahnke} K.,  {Dannerbauer} H.,  {Urrutia} T.,
   {Hodge} J.,  2017, \mn@doi [\mnras] {10.1093/mnras/stx1123}, \href
  {https://ui.adsabs.harvard.edu/abs/2017MNRAS.470.1570H} {470, 1570}

\bibitem[\protect\citeauthoryear{Ichikawa et~al.,}{Ichikawa
  et~al.}{2019}]{Ichikawa2019}
Ichikawa K.,  et~al., 2019, \mn@doi [\apj] {10.3847/1538-4357/aaef8f}, 870, 31

\bibitem[\protect\citeauthoryear{{Irodotou} et~al.,}{{Irodotou}
  et~al.}{2021}]{Irodotou2021}
{Irodotou} D.,  et~al., 2021, arXiv e-prints, \href
  {https://ui.adsabs.harvard.edu/abs/2021arXiv211011368I} {p. arXiv:2110.11368}

\bibitem[\protect\citeauthoryear{Ishibashi, Fabian  \& Maiolino}{Ishibashi
  et~al.}{2018}]{Ishibashi2018}
Ishibashi W.,  Fabian A.~C.,   Maiolino R.,  2018, \mn@doi [\mnras]
  {10.1093/mnras/sty236}, 476, 512

\bibitem[\protect\citeauthoryear{Jackson, Rosario, Alexander, Scholtz, McAlpine
   \& Bower}{Jackson et~al.}{2020}]{Jackson2020}
Jackson T.~M.,  Rosario D.~J.,  Alexander D.~M.,  Scholtz J.,  McAlpine S.,
  Bower R.~G.,  2020, \mn@doi [\mnras] {10.1093/mnras/staa2414}, 498, 2323

\bibitem[\protect\citeauthoryear{Jarvis et~al.,}{Jarvis
  et~al.}{2020}]{Jarvis2020}
Jarvis M.~E.,  et~al., 2020, \mn@doi [\mnras] {10.1093/mnras/staa2196}, 498,
  1560

\bibitem[\protect\citeauthoryear{{Ji}, {Giavalisco}, {Kirkpatrick}, {Kocevski},
  {Daddi}, {Delvecchio}  \& {Hatcher}}{{Ji} et~al.}{2022}]{Ji2022}
{Ji} Z.,  {Giavalisco} M.,  {Kirkpatrick} A.,  {Kocevski} D.,  {Daddi} E.,
  {Delvecchio} I.,   {Hatcher} C.,  2022, \mn@doi [\apj]
  {10.3847/1538-4357/ac3837}, \href
  {https://ui.adsabs.harvard.edu/abs/2022ApJ...925...74J} {925, 74}

\bibitem[\protect\citeauthoryear{{Kakkad} et~al.,}{{Kakkad}
  et~al.}{2017}]{Kakkad2017}
{Kakkad} D.,  et~al., 2017, \mn@doi [\mnras] {10.1093/mnras/stx726}, \href
  {https://ui.adsabs.harvard.edu/abs/2017MNRAS.468.4205K} {468, 4205}

\bibitem[\protect\citeauthoryear{Kennicutt}{Kennicutt}{1998}]{Kennicutt1998}
Kennicutt R.~C.,  1998, \mn@doi [\araa] {10.1146/annurev.astro.36.1.189}, 36,
  189

\bibitem[\protect\citeauthoryear{{Khandai}, {Di Matteo}, {Croft}, {Wilkins},
  {Feng}, {Tucker}, {DeGraf}  \& {Liu}}{{Khandai} et~al.}{2015}]{Khandai2015}
{Khandai} N.,  {Di Matteo} T.,  {Croft} R.,  {Wilkins} S.,  {Feng} Y.,
  {Tucker} E.,  {DeGraf} C.,   {Liu} M.-S.,  2015, \mn@doi [\mnras]
  {10.1093/mnras/stv627}, \href
  {https://ui.adsabs.harvard.edu/abs/2015MNRAS.450.1349K} {450, 1349}

\bibitem[\protect\citeauthoryear{{Kim} et~al.,}{{Kim} et~al.}{2022}]{Kim2022}
{Kim} C.,  et~al., 2022, \mn@doi [\apj] {10.3847/1538-4357/ac5407}, \href
  {https://ui.adsabs.harvard.edu/abs/2022ApJ...928...73K} {928, 73}

\bibitem[\protect\citeauthoryear{{King}}{{King}}{2003}]{King2003}
{King} A.,  2003, \mn@doi [\apjl] {10.1086/379143}, \href
  {https://ui.adsabs.harvard.edu/abs/2003ApJ...596L..27K} {596, L27}

\bibitem[\protect\citeauthoryear{{Kirkpatrick}, {Sharon}, {Keller}  \&
  {Pope}}{{Kirkpatrick} et~al.}{2019}]{Kirkpatrick2019}
{Kirkpatrick} A.,  {Sharon} C.,  {Keller} E.,   {Pope} A.,  2019, \mn@doi
  [\apj] {10.3847/1538-4357/ab223a}, \href
  {https://ui.adsabs.harvard.edu/abs/2019ApJ...879...41K} {879, 41}

\bibitem[\protect\citeauthoryear{{Kormendy} \& {Ho}}{{Kormendy} \&
  {Ho}}{2013}]{Kormendy2013}
{Kormendy} J.,  {Ho} L.~C.,  2013, \mn@doi [\araa]
  {10.1146/annurev-astro-082708-101811}, \href
  {https://ui.adsabs.harvard.edu/abs/2013ARA&A..51..511K} {51, 511}

\bibitem[\protect\citeauthoryear{{Koss} et~al.,}{{Koss}
  et~al.}{2017}]{Koss2017}
{Koss} M.,  et~al., 2017, \mn@doi [\apj] {10.3847/1538-4357/aa8ec9}, \href
  {https://ui.adsabs.harvard.edu/abs/2017ApJ...850...74K} {850, 74}

\bibitem[\protect\citeauthoryear{Koss et~al.,}{Koss et~al.}{2021}]{Koss2021}
Koss M.~J.,  et~al., 2021, \mn@doi [\apjs] {10.3847/1538-4365/ABCBFE}, 252, 29

\bibitem[\protect\citeauthoryear{Lagos et~al.,}{Lagos et~al.}{2015}]{Lagos2015}
Lagos C. D.~P.,  et~al., 2015, \mn@doi [\mnras] {10.1093/mnras/stv1488}, 452,
  3815

\bibitem[\protect\citeauthoryear{Liu, Liu, Dong, Zhou, Wang, Lu  \& Yuan}{Liu
  et~al.}{2019}]{Liu2019}
Liu H.-Y.,  Liu W.-J.,  Dong X.-B.,  Zhou H.,  Wang T.,  Lu H.,   Yuan W.,
  2019, \mn@doi [\apjs] {10.3847/1538-4365/ab298b}, 243, 21

\bibitem[\protect\citeauthoryear{{Luo} et~al.,}{{Luo} et~al.}{2021}]{Luo2021}
{Luo} R.,  et~al., 2021, \mn@doi [\apj] {10.3847/1538-4357/abd5ac}, \href
  {https://ui.adsabs.harvard.edu/abs/2021ApJ...908..221L} {908, 221}

\bibitem[\protect\citeauthoryear{{Lutz} et~al.,}{{Lutz}
  et~al.}{2010}]{Lutz2010}
{Lutz} D.,  et~al., 2010, \mn@doi [\apj] {10.1088/0004-637X/712/2/1287}, \href
  {https://ui.adsabs.harvard.edu/abs/2010ApJ...712.1287L} {712, 1287}

\bibitem[\protect\citeauthoryear{{Madau} \& {Dickinson}}{{Madau} \&
  {Dickinson}}{2014}]{Madau2014}
{Madau} P.,  {Dickinson} M.,  2014, \mn@doi [\araa]
  {10.1146/annurev-astro-081811-125615}, \href
  {https://ui.adsabs.harvard.edu/abs/2014ARA&A..52..415M} {52, 415}

\bibitem[\protect\citeauthoryear{{Mainieri} et~al.,}{{Mainieri}
  et~al.}{2011}]{Mainieri2011}
{Mainieri} V.,  et~al., 2011, \mn@doi [\aap] {10.1051/0004-6361/201117259},
  \href {https://ui.adsabs.harvard.edu/abs/2011A&A...535A..80M} {535, A80}

\bibitem[\protect\citeauthoryear{{Mandal}, {Mukherjee}, {Federrath},
  {Nesvadba}, {Bicknell}, {Wagner}  \& {Meenakshi}}{{Mandal}
  et~al.}{2021}]{Mandal2021}
{Mandal} A.,  {Mukherjee} D.,  {Federrath} C.,  {Nesvadba} N. P.~H.,
  {Bicknell} G.~V.,  {Wagner} A.~Y.,   {Meenakshi} M.,  2021, \mn@doi [\mnras]
  {10.1093/mnras/stab2822}, \href
  {https://ui.adsabs.harvard.edu/abs/2021MNRAS.508.4738M} {508, 4738}

\bibitem[\protect\citeauthoryear{Marconi, Risaliti, Gilli, Hunt, Maiolino  \&
  Salvati}{Marconi et~al.}{2004}]{Marconi2004}
Marconi A.,  Risaliti G.,  Gilli R.,  Hunt L.~K.,  Maiolino R.,   Salvati M.,
  2004, \mn@doi [\mnras] {10.1111/j.1365-2966.2004.07765.x}, 351, 169

\bibitem[\protect\citeauthoryear{{Marinacci} et~al.,}{{Marinacci}
  et~al.}{2018}]{Marinacci2018}
{Marinacci} F.,  et~al., 2018, \mn@doi [\mnras] {10.1093/mnras/sty2206}, \href
  {https://ui.adsabs.harvard.edu/abs/2018MNRAS.480.5113M} {480, 5113}

\bibitem[\protect\citeauthoryear{McAlpine et~al.,}{McAlpine
  et~al.}{2016}]{McAlpine2016}
McAlpine S.,  et~al., 2016, \mn@doi [Astronomy and Computing]
  {10.1016/j.ascom.2016.02.004}, 15, 72

\bibitem[\protect\citeauthoryear{McAlpine, Bower, Harrison, Crain, Schaller,
  Schaye  \& Theuns}{McAlpine et~al.}{2017}]{McAlpine2017}
McAlpine S.,  Bower R.~G.,  Harrison C.~M.,  Crain R.~A.,  Schaller M.,  Schaye
  J.,   Theuns T.,  2017, \mn@doi [\mnras] {10.1093/mnras/stx658}, 468, 3395

\bibitem[\protect\citeauthoryear{McCarthy, Schaye, Bower, Ponman, Booth,
  Vecchia  \& Springel}{McCarthy et~al.}{2011}]{McCarthy2011}
McCarthy I.~G.,  Schaye J.,  Bower R.~G.,  Ponman T.~J.,  Booth C.~M.,  Vecchia
  C.~D.,   Springel V.,  2011, \mn@doi [\mnras]
  {10.1111/j.1365-2966.2010.18033.x}, 412, 1965

\bibitem[\protect\citeauthoryear{McLure \& Dunlop}{McLure \&
  Dunlop}{2004}]{McLure2004}
McLure R.~J.,  Dunlop J.~S.,  2004, \mnras, 352, 1390

\bibitem[\protect\citeauthoryear{Mukherjee, Bicknell, Sutherland  \&
  Wagner}{Mukherjee et~al.}{2016}]{Mukherjee2016}
Mukherjee D.,  Bicknell G.~V.,  Sutherland R.,   Wagner A.,  2016, \mn@doi
  [\mnras] {10.1093/mnras/stw1368}, 461, 967

\bibitem[\protect\citeauthoryear{Mukherjee, Bicknell, Wagner, Sutherland  \&
  Silk}{Mukherjee et~al.}{2018}]{Mukherjee2018}
Mukherjee D.,  Bicknell G.~V.,  Wagner A.~Y.,  Sutherland R.~S.,   Silk J.,
  2018, \mn@doi [\mnras] {10.1093/mnras/sty1776}, 479, 5544

\bibitem[\protect\citeauthoryear{{Mullaney} et~al.,}{{Mullaney}
  et~al.}{2012}]{Mullaney2012}
{Mullaney} J.~R.,  et~al., 2012, \mn@doi [\apjl] {10.1088/2041-8205/753/2/L30},
  \href {https://ui.adsabs.harvard.edu/abs/2012ApJ...753L..30M} {753, L30}

\bibitem[\protect\citeauthoryear{{Naiman} et~al.,}{{Naiman}
  et~al.}{2018}]{Naiman2018}
{Naiman} J.~P.,  et~al., 2018, \mn@doi [\mnras] {10.1093/mnras/sty618}, \href
  {https://ui.adsabs.harvard.edu/abs/2018MNRAS.477.1206N} {477, 1206}

\bibitem[\protect\citeauthoryear{{Nelson} et~al.,}{{Nelson}
  et~al.}{2018}]{Nelson2018}
{Nelson} D.,  et~al., 2018, \mn@doi [\mnras] {10.1093/mnras/stx3040}, \href
  {https://ui.adsabs.harvard.edu/abs/2018MNRAS.475..624N} {475, 624}

\bibitem[\protect\citeauthoryear{Nelson et~al.,}{Nelson
  et~al.}{2019}]{Nelson2019}
Nelson D.,  et~al., 2019, \mn@doi [Computational Astrophysics and Cosmology]
  {10.1186/s40668-019-0028-x}, 6, 2

\bibitem[\protect\citeauthoryear{{Novak}, {Ostriker}  \& {Ciotti}}{{Novak}
  et~al.}{2011}]{Novak2011}
{Novak} G.~S.,  {Ostriker} J.~P.,   {Ciotti} L.,  2011, \mn@doi [\apj]
  {10.1088/0004-637X/737/1/26}, \href
  {https://ui.adsabs.harvard.edu/abs/2011ApJ...737...26N} {737, 26}

\bibitem[\protect\citeauthoryear{{Page} et~al.,}{{Page}
  et~al.}{2012}]{Page2012}
{Page} M.~J.,  et~al., 2012, \mn@doi [\nat] {10.1038/nature11096}, \href
  {https://ui.adsabs.harvard.edu/abs/2012Natur.485..213P} {485, 213}

\bibitem[\protect\citeauthoryear{Perna et~al.,}{Perna et~al.}{2018}]{Perna2018}
Perna M.,  et~al., 2018, \mn@doi [\aap] {10.1051/0004-6361/201833040}, 619, 90

\bibitem[\protect\citeauthoryear{Pillepich et~al.,}{Pillepich
  et~al.}{2017}]{Pillepich2018_methods}
Pillepich A.,  et~al., 2017, \mn@doi [\mnras] {10.1093/mnras/stx2656}, 473,
  4077

\bibitem[\protect\citeauthoryear{{Pillepich} et~al.,}{{Pillepich}
  et~al.}{2018}]{Pillepich2018b}
{Pillepich} A.,  et~al., 2018, \mn@doi [\mnras] {10.1093/mnras/stx3112}, \href
  {https://ui.adsabs.harvard.edu/abs/2018MNRAS.475..648P} {475, 648}

\bibitem[\protect\citeauthoryear{{Piotrowska}, {Bluck}, {Maiolino}  \&
  {Peng}}{{Piotrowska} et~al.}{2021}]{Piotrowska2021}
{Piotrowska} J.~M.,  {Bluck} A. F.~L.,  {Maiolino} R.,   {Peng} Y.,  2021,
  \mn@doi [\mnras] {10.1093/mnras/stab3673}, \href
  {https://ui.adsabs.harvard.edu/abs/2021MNRAS.tmp.3414P} {}

\bibitem[\protect\citeauthoryear{{Planck Collaboration XIII}}{{Planck
  Collaboration XIII}}{2016}]{Planck2016cite}
{Planck Collaboration XIII} 2016, \mn@doi [\aap] {10.1051/0004-6361/201525830},
  \href {https://ui.adsabs.harvard.edu/abs/2016A&A...594A..13P} {594, A13}

\bibitem[\protect\citeauthoryear{Popping et~al.,}{Popping
  et~al.}{2019}]{Popping2019}
Popping G.,  et~al., 2019, \mn@doi [\apj] {10.3847/1538-4357/ab30f2}, 882, 137

\bibitem[\protect\citeauthoryear{Powell et~al.,}{Powell
  et~al.}{2018}]{Powell2018}
Powell M.~C.,  et~al., 2018, \mn@doi [\apj] {10.3847/1538-4357/aabd7f}, 858,
  110

\bibitem[\protect\citeauthoryear{Rahmati, Pawlik, Raičevì  \& Schaye}{Rahmati
  et~al.}{2013}]{Rahmati2013}
Rahmati A.,  Pawlik A.~H.,  Raičevì M.,   Schaye J.,  2013, \mn@doi [MNRAS]
  {10.1093/mnras/stt066}, 430, 2427

\bibitem[\protect\citeauthoryear{{Ramasawmy}, {Stevens}, {Martin}  \&
  {Geach}}{{Ramasawmy} et~al.}{2019}]{Ramasawmy2019}
{Ramasawmy} J.,  {Stevens} J.,  {Martin} G.,   {Geach} J.~E.,  2019, \mn@doi
  [\mnras] {10.1093/mnras/stz1093}, \href
  {https://ui.adsabs.harvard.edu/abs/2019MNRAS.486.4320R} {486, 4320}

\bibitem[\protect\citeauthoryear{{Ramos Almeida} et~al.,}{{Ramos Almeida}
  et~al.}{2021}]{RamosAlmeida2022}
{Ramos Almeida} C.,  et~al., 2021, arXiv e-prints, \href
  {https://ui.adsabs.harvard.edu/abs/2021arXiv211113578R} {p. arXiv:2111.13578}

\bibitem[\protect\citeauthoryear{{Richings}, {Schaye}  \&
  {Oppenheimer}}{{Richings} et~al.}{2014a}]{Richings2014a}
{Richings} A.~J.,  {Schaye} J.,   {Oppenheimer} B.~D.,  2014a, \mn@doi [\mnras]
  {10.1093/mnras/stu525}, \href
  {https://ui.adsabs.harvard.edu/abs/2014MNRAS.440.3349R} {440, 3349}

\bibitem[\protect\citeauthoryear{{Richings}, {Schaye}  \&
  {Oppenheimer}}{{Richings} et~al.}{2014b}]{Richings2014b}
{Richings} A.~J.,  {Schaye} J.,   {Oppenheimer} B.~D.,  2014b, \mn@doi [\mnras]
  {10.1093/mnras/stu1046}, \href
  {https://ui.adsabs.harvard.edu/abs/2014MNRAS.442.2780R} {442, 2780}

\bibitem[\protect\citeauthoryear{Rosario et~al.,}{Rosario
  et~al.}{2012}]{Rosario2012}
Rosario D.~J.,  et~al., 2012, \mn@doi [A&A] {10.1051/0004-6361/201219258}, 545,
  45

\bibitem[\protect\citeauthoryear{Rosario et~al.,}{Rosario
  et~al.}{2013}]{Rosario2013aa}
Rosario D.~J.,  et~al., 2013, \mn@doi [A&A] {10.1051/0004-6361/201322196}, 560,
  72

\bibitem[\protect\citeauthoryear{Rosario et~al.,}{Rosario
  et~al.}{2018}]{Rosario2018}
Rosario D.~J.,  et~al., 2018, \mn@doi [MNRAS] {10.1093/mnras/stx2670}, 473,
  5658

\bibitem[\protect\citeauthoryear{Rosas-Guevara, Bower, Schaye, McAlpine,
  Vecchia, Frenk, Schaller  \& Theuns}{Rosas-Guevara
  et~al.}{2016}]{Rosas-Guevara2016}
Rosas-Guevara Y.,  Bower R.~G.,  Schaye J.,  McAlpine S.,  Vecchia C.~D.,
  Frenk C.~S.,  Schaller M.,   Theuns T.,  2016, \mn@doi [MNRAS]
  {10.1093/mnras/stw1679}, 462, 190

\bibitem[\protect\citeauthoryear{{Rosdahl} \& {Teyssier}}{{Rosdahl} \&
  {Teyssier}}{2015}]{Rosdahl2015}
{Rosdahl} J.,  {Teyssier} R.,  2015, \mn@doi [\mnras] {10.1093/mnras/stv567},
  \href {https://ui.adsabs.harvard.edu/abs/2015MNRAS.449.4380R} {449, 4380}

\bibitem[\protect\citeauthoryear{{Roy} et~al.,}{{Roy} et~al.}{2021}]{Roy2021}
{Roy} N.,  et~al., 2021, \mn@doi [\apj] {10.3847/1538-4357/abf1e6}, \href
  {https://ui.adsabs.harvard.edu/abs/2021ApJ...913...33R} {913, 33}

\bibitem[\protect\citeauthoryear{{Salvestrini} et~al.,}{{Salvestrini}
  et~al.}{2022}]{Salvestrini2022}
{Salvestrini} F.,  et~al., 2022, arXiv e-prints, \href
  {https://ui.adsabs.harvard.edu/abs/2022arXiv220315825S} {p. arXiv:2203.15825}

\bibitem[\protect\citeauthoryear{{Schawinski} et~al.,}{{Schawinski}
  et~al.}{2014}]{Schawinski2014}
{Schawinski} K.,  et~al., 2014, \mn@doi [\mnras] {10.1093/mnras/stu327}, \href
  {https://ui.adsabs.harvard.edu/abs/2014MNRAS.440..889S} {440, 889}

\bibitem[\protect\citeauthoryear{{Schawinski}, {Koss}, {Berney}  \&
  {Sartori}}{{Schawinski} et~al.}{2015}]{Schawinski2015}
{Schawinski} K.,  {Koss} M.,  {Berney} S.,   {Sartori} L.~F.,  2015, \mn@doi
  [\mnras] {10.1093/mnras/stv1136}, \href
  {https://ui.adsabs.harvard.edu/abs/2015MNRAS.451.2517S} {451, 2517}

\bibitem[\protect\citeauthoryear{Schaye et~al.,}{Schaye
  et~al.}{2015}]{Schaye2015}
Schaye J.,  et~al., 2015, \mn@doi [\mnras] {10.1093/mnras/stu2058}, 446, 521

\bibitem[\protect\citeauthoryear{Scholtz et~al.,}{Scholtz
  et~al.}{2018}]{Scholtz2018}
Scholtz J.,  et~al., 2018, \mn@doi [\mnras] {10.1093/mnras/stx3177}, 475, 1288

\bibitem[\protect\citeauthoryear{Scholtz et~al.,}{Scholtz
  et~al.}{2021}]{Scholtz2021}
Scholtz J.,  et~al., 2021, \mn@doi [\mnras] {10.1093/MNRAS/STAB1631}, 505, 5469

\bibitem[\protect\citeauthoryear{Schreiber et~al.,}{Schreiber
  et~al.}{2015}]{Schreiber2015}
Schreiber C.,  et~al., 2015, \mn@doi [\aap] {10.1051/0004-6361/201425017}, 575

\bibitem[\protect\citeauthoryear{{Schulze} et~al.,}{{Schulze}
  et~al.}{2019}]{Schulze2019}
{Schulze} A.,  et~al., 2019, \mn@doi [\mnras] {10.1093/mnras/stz1746}, \href
  {https://ui.adsabs.harvard.edu/abs/2019MNRAS.488.1180S} {488, 1180}

\bibitem[\protect\citeauthoryear{Shakura \& Sunyaev}{Shakura \&
  Sunyaev}{1973}]{Shakura1973}
Shakura N.~I.,  Sunyaev R.~A.,  1973, \aap, 500, 33

\bibitem[\protect\citeauthoryear{Shangguan \& Ho}{Shangguan \&
  Ho}{2019}]{Shangguan2019}
Shangguan J.,  Ho L.~C.,  2019, \mn@doi [\apj] {10.3847/1538-4357/ab0555}, 873,
  90

\bibitem[\protect\citeauthoryear{Shangguan, Ho  \& Xie}{Shangguan
  et~al.}{2018}]{Shangguan2018}
Shangguan J.,  Ho L.~C.,   Xie Y.,  2018, \mn@doi [\apj]
  {10.3847/1538-4357/aaa9be}, 854, 158

\bibitem[\protect\citeauthoryear{Shangguan, Ho, Bauer, Wang  \&
  Treister}{Shangguan et~al.}{2020}]{Shangguan2020b}
Shangguan J.,  Ho L.~C.,  Bauer F.~E.,  Wang R.,   Treister E.,  2020, \mn@doi
  [\apj] {10.3847/1538-4357/aba8a1}, 899, 112

\bibitem[\protect\citeauthoryear{{Shimizu}, {Mushotzky}, {Mel{\'e}ndez},
  {Koss}, {Barger}  \& {Cowie}}{{Shimizu} et~al.}{2017}]{Shimizu2017}
{Shimizu} T.~T.,  {Mushotzky} R.~F.,  {Mel{\'e}ndez} M.,  {Koss} M.~J.,
  {Barger} A.~J.,   {Cowie} L.~L.,  2017, \mn@doi [\mnras]
  {10.1093/mnras/stw3268}, \href
  {https://ui.adsabs.harvard.edu/abs/2017MNRAS.466.3161S} {466, 3161}

\bibitem[\protect\citeauthoryear{{Sijacki}, {Springel}, {Di Matteo}  \&
  {Hernquist}}{{Sijacki} et~al.}{2007}]{Sijacki2007}
{Sijacki} D.,  {Springel} V.,  {Di Matteo} T.,   {Hernquist} L.,  2007, \mn@doi
  [\mnras] {10.1111/j.1365-2966.2007.12153.x}, \href
  {https://ui.adsabs.harvard.edu/abs/2007MNRAS.380..877S} {380, 877}

\bibitem[\protect\citeauthoryear{Silk \& Rees}{Silk \& Rees}{1998}]{Silk1998}
Silk J.,  Rees M.~J.,  1998, Astron. Astrophys, 331, 1

\bibitem[\protect\citeauthoryear{{Smirnova-Pinchukova}
  et~al.,}{{Smirnova-Pinchukova} et~al.}{2021}]{Smirnova-Pinchukova2021}
{Smirnova-Pinchukova} I.,  et~al., 2021, arXiv e-prints, \href
  {https://ui.adsabs.harvard.edu/abs/2021arXiv211110419S} {p. arXiv:2111.10419}

\bibitem[\protect\citeauthoryear{{Somerville}, {Hopkins}, {Cox}, {Robertson}
  \& {Hernquist}}{{Somerville} et~al.}{2008}]{Somerville2008}
{Somerville} R.~S.,  {Hopkins} P.~F.,  {Cox} T.~J.,  {Robertson} B.~E.,
  {Hernquist} L.,  2008, \mn@doi [\mnras] {10.1111/j.1365-2966.2008.13805.x},
  \href {https://ui.adsabs.harvard.edu/abs/2008MNRAS.391..481S} {391, 481}

\bibitem[\protect\citeauthoryear{Speagle, Steinhardt, Capak  \&
  Silverman}{Speagle et~al.}{2014}]{Speagle2014}
Speagle J.~S.,  Steinhardt C.~L.,  Capak P.~L.,   Silverman J.~D.,  2014,
  \mn@doi [\apjs] {10.1088/0067-0049/214/2/15}, 214, 15

\bibitem[\protect\citeauthoryear{{Springel}}{{Springel}}{2005}]{Springel2005_gadget}
{Springel} V.,  2005, \mn@doi [\mnras] {10.1111/j.1365-2966.2005.09655.x},
  \href {https://ui.adsabs.harvard.edu/abs/2005MNRAS.364.1105S} {364, 1105}

\bibitem[\protect\citeauthoryear{{Springel}}{{Springel}}{2010}]{Springel2010}
{Springel} V.,  2010, \mn@doi [\mnras] {10.1111/j.1365-2966.2009.15715.x},
  \href {https://ui.adsabs.harvard.edu/abs/2010MNRAS.401..791S} {401, 791}

\bibitem[\protect\citeauthoryear{{Springel} et~al.,}{{Springel}
  et~al.}{2018}]{Springel2018}
{Springel} V.,  et~al., 2018, \mn@doi [\mnras] {10.1093/mnras/stx3304}, \href
  {https://ui.adsabs.harvard.edu/abs/2018MNRAS.475..676S} {475, 676}

\bibitem[\protect\citeauthoryear{Stanley, Harrison, Alexander, Swinbank, Aird,
  Moro, Hickox  \& Mullaney}{Stanley et~al.}{2015}]{Stanley2015}
Stanley F.,  Harrison C.~M.,  Alexander D.~M.,  Swinbank A.~M.,  Aird J.~A.,
  Moro A.~D.,  Hickox R.~C.,   Mullaney J.~R.,  2015, \mn@doi [\mnras]
  {10.1093/mnras/stv1678}, 453, 591

\bibitem[\protect\citeauthoryear{{Stanley} et~al.,}{{Stanley}
  et~al.}{2017}]{Stanley2017}
{Stanley} F.,  et~al., 2017, \mn@doi [\mnras] {10.1093/mnras/stx2121}, \href
  {https://ui.adsabs.harvard.edu/abs/2017MNRAS.472.2221S} {472, 2221}

\bibitem[\protect\citeauthoryear{{Starkenburg}, {Tonnesen}  \&
  {Kopenhafer}}{{Starkenburg} et~al.}{2019}]{Starkenburg2019}
{Starkenburg} T.~K.,  {Tonnesen} S.,   {Kopenhafer} C.,  2019, \mn@doi [\apjl]
  {10.3847/2041-8213/ab0f34}, \href
  {https://ui.adsabs.harvard.edu/abs/2019ApJ...874L..17S} {874, L17}

\bibitem[\protect\citeauthoryear{{Sturm} et~al.,}{{Sturm}
  et~al.}{2011}]{Sturm2011}
{Sturm} E.,  et~al., 2011, \mn@doi [\apjl] {10.1088/2041-8205/733/1/L16}, \href
  {https://ui.adsabs.harvard.edu/abs/2011ApJ...733L..16S} {733, L16}

\bibitem[\protect\citeauthoryear{Tacconi et~al.,}{Tacconi
  et~al.}{2018}]{Tacconi2018}
Tacconi L.~J.,  et~al., 2018, \mn@doi [\apj] {10.3847/1538-4357/aaa4b4}, 853,
  179

\bibitem[\protect\citeauthoryear{{Tacconi}, {Genzel}  \& {Sternberg}}{{Tacconi}
  et~al.}{2020}]{Tacconi2020}
{Tacconi} L.~J.,  {Genzel} R.,   {Sternberg} A.,  2020, \mn@doi [\araa]
  {10.1146/annurev-astro-082812-141034}, \href
  {https://ui.adsabs.harvard.edu/abs/2020ARA&A..58..157T} {58, 157}

\bibitem[\protect\citeauthoryear{{Talbot}, {Bourne}  \& {Sijacki}}{{Talbot}
  et~al.}{2021}]{Talbot2021}
{Talbot} R.~Y.,  {Bourne} M.~A.,   {Sijacki} D.,  2021, \mn@doi [\mnras]
  {10.1093/mnras/stab804}, \href
  {https://ui.adsabs.harvard.edu/abs/2021MNRAS.504.3619T} {504, 3619}

\bibitem[\protect\citeauthoryear{{Tanner} \& {Weaver}}{{Tanner} \&
  {Weaver}}{2022}]{Tanner2022}
{Tanner} R.,  {Weaver} K.~A.,  2022, \mn@doi [\aj] {10.3847/1538-3881/ac4d23},
  \href {https://ui.adsabs.harvard.edu/abs/2022AJ....163..134T} {163, 134}

\bibitem[\protect\citeauthoryear{Terrazas, Bell, Woo  \& Henriques}{Terrazas
  et~al.}{2017}]{Terrazas2017}
Terrazas B.~A.,  Bell E.~F.,  Woo J.,   Henriques B. M.~B.,  2017, \mn@doi
  [\apj] {10.3847/1538-4357/aa7d07}, 844, 170

\bibitem[\protect\citeauthoryear{Terrazas et~al.,}{Terrazas
  et~al.}{2020}]{Terrazas2020}
Terrazas B.~A.,  et~al., 2020, \mn@doi [\mnras] {10.1093/mnras/staa374}, 493,
  1888

\bibitem[\protect\citeauthoryear{{Thacker}, {MacMackin}, {Wurster}  \&
  {Hobbs}}{{Thacker} et~al.}{2014}]{Thacker2014}
{Thacker} R.~J.,  {MacMackin} C.,  {Wurster} J.,   {Hobbs} A.,  2014, \mn@doi
  [\mnras] {10.1093/mnras/stu1180}, \href
  {https://ui.adsabs.harvard.edu/abs/2014MNRAS.443.1125T} {443, 1125}

\bibitem[\protect\citeauthoryear{Thomas, Davé, Anglés-Alcázar  \&
  Jarvis}{Thomas et~al.}{2019}]{Thomas2019}
Thomas N.,  Davé R.,  Anglés-Alcázar D.,   Jarvis M.,  2019, \mn@doi
  [\mnras] {10.1093/mnras/stz1703}, 487, 5764

\bibitem[\protect\citeauthoryear{Trayford, Theuns, Bower, Crain, del P.~Lagos,
  Schaller  \& Schaye}{Trayford et~al.}{2016}]{Trayford2016}
Trayford J.~W.,  Theuns T.,  Bower R.~G.,  Crain R.~A.,  del P.~Lagos C.,
  Schaller M.,   Schaye J.,  2016, \mn@doi [\mnras] {10.1093/mnras/stw1230},
  460, 3925

\bibitem[\protect\citeauthoryear{{Trump} et~al.,}{{Trump}
  et~al.}{2015}]{Trump2015}
{Trump} J.~R.,  et~al., 2015, \mn@doi [\apj] {10.1088/0004-637X/811/1/26},
  \href {https://ui.adsabs.harvard.edu/abs/2015ApJ...811...26T} {811, 26}

\bibitem[\protect\citeauthoryear{\VAN{Vlugt}{Van der}{van der}~Vlugt \&
  {Costa}}{\VAN{Vlugt}{Van der}{van der}~Vlugt \&
  {Costa}}{2019}]{vanderVlugt2019}
\VAN{Vlugt}{Van der}{van der}~Vlugt D.,  {Costa} T.,  2019, \mn@doi [\mnras]
  {10.1093/mnras/stz2944}, \href
  {https://ui.adsabs.harvard.edu/abs/2019MNRAS.490.4918V} {490, 4918}

\bibitem[\protect\citeauthoryear{{Valentino} et~al.,}{{Valentino}
  et~al.}{2021}]{Valentino2021}
{Valentino} F.,  et~al., 2021, \mn@doi [\aap] {10.1051/0004-6361/202141417},
  \href {https://ui.adsabs.harvard.edu/abs/2021A&A...654A.165V} {654, A165}

\bibitem[\protect\citeauthoryear{{Veilleux}, {Maiolino}, {Bolatto}  \&
  {Aalto}}{{Veilleux} et~al.}{2020}]{Veilleux2020}
{Veilleux} S.,  {Maiolino} R.,  {Bolatto} A.~D.,   {Aalto} S.,  2020, \mn@doi
  [\aapr] {10.1007/s00159-019-0121-9}, \href
  {https://ui.adsabs.harvard.edu/abs/2020A&ARv..28....2V} {28, 2}

\bibitem[\protect\citeauthoryear{{Vietri} et~al.,}{{Vietri}
  et~al.}{2021}]{Vietri2021}
{Vietri} G.,  et~al., 2021, arXiv e-prints, \href
  {https://ui.adsabs.harvard.edu/abs/2021arXiv211108730V} {p. arXiv:2111.08730}

\bibitem[\protect\citeauthoryear{Vogelsberger, Genel, Sijacki, Torrey, Springel
   \& Hernquist}{Vogelsberger et~al.}{2013}]{Vogelsberger2013}
Vogelsberger M.,  Genel S.,  Sijacki D.,  Torrey P.,  Springel V.,   Hernquist
  L.,  2013, \mn@doi [\mnras] {10.1093/mnras/stt1789}, 436, 3031

\bibitem[\protect\citeauthoryear{{Vogelsberger} et~al.,}{{Vogelsberger}
  et~al.}{2014a}]{Vogelsberger2014b}
{Vogelsberger} M.,  et~al., 2014a, \mn@doi [\mnras] {10.1093/mnras/stu1536},
  \href {https://ui.adsabs.harvard.edu/abs/2014MNRAS.444.1518V} {444, 1518}

\bibitem[\protect\citeauthoryear{{Vogelsberger} et~al.,}{{Vogelsberger}
  et~al.}{2014b}]{Vogelsberger2014a}
{Vogelsberger} M.,  et~al., 2014b, \mn@doi [Nature] {10.1038/nature13316}, 509,
  177–182

\bibitem[\protect\citeauthoryear{Weinberger et~al.,}{Weinberger
  et~al.}{2017}]{Weinberger2017}
Weinberger R.,  et~al., 2017, \mn@doi [\mnras] {10.1093/mnras/stw2944}, 465,
  3291

\bibitem[\protect\citeauthoryear{Weinberger et~al.,}{Weinberger
  et~al.}{2018}]{Weinberger2018}
Weinberger R.,  et~al., 2018, \mn@doi [\mnras] {10.1093/mnras/sty1733}, 479,
  4056

\bibitem[\protect\citeauthoryear{Whitaker et~al.,}{Whitaker
  et~al.}{2014}]{Whitaker2014}
Whitaker K.~E.,  et~al., 2014, \mn@doi [\apj] {10.1088/0004-637X/795/2/104},
  795, 104

\bibitem[\protect\citeauthoryear{Wild, Almaini, Dunlop, Simpson, Rowlands,
  Bowler, Maltby  \& Mclure}{Wild et~al.}{2016}]{Wild2016}
Wild V.,  Almaini O.,  Dunlop J.,  Simpson C.,  Rowlands K.,  Bowler R.,
  Maltby D.,   Mclure R.,  2016, \mn@doi [MNRAS] {10.1093/mnras/stw1996}, 463,
  832

\bibitem[\protect\citeauthoryear{{Wylezalek} \& {Zakamska}}{{Wylezalek} \&
  {Zakamska}}{2016}]{Wylezalek2016}
{Wylezalek} D.,  {Zakamska} N.~L.,  2016, \mn@doi [\mnras]
  {10.1093/mnras/stw1557}, \href
  {https://ui.adsabs.harvard.edu/abs/2016MNRAS.461.3724W} {461, 3724}

\bibitem[\protect\citeauthoryear{{Xie}, {Ho}, {Zhuang}  \& {Shangguan}}{{Xie}
  et~al.}{2021}]{Xie2021}
{Xie} Y.,  {Ho} L.~C.,  {Zhuang} M.-Y.,   {Shangguan} J.,  2021, \mn@doi [\apj]
  {10.3847/1538-4357/abe404}, \href
  {https://ui.adsabs.harvard.edu/abs/2021ApJ...910..124X} {910, 124}

\bibitem[\protect\citeauthoryear{Yesuf \& Ho}{Yesuf \& Ho}{2019}]{Yesuf2019}
Yesuf H.~M.,  Ho L.~C.,  2019, \mn@doi [\apj] {10.3847/1538-4357/ab4202}, 884,
  177

\bibitem[\protect\citeauthoryear{York et~al.,}{York et~al.}{2000}]{York2000}
York D.~G.,  et~al., 2000, \mn@doi [\apj] {10.1086/301513}, 120, 1579

\bibitem[\protect\citeauthoryear{Zhuang \& Ho}{Zhuang \&
  Ho}{2019}]{ZhuangHo2019}
Zhuang M.-Y.,  Ho L.~C.,  2019, \mn@doi [\apj] {10.3847/1538-4357/ab340d}, 882,
  89

\bibitem[\protect\citeauthoryear{Zhuang, Ho  \& Shangguan}{Zhuang
  et~al.}{2019}]{Zhuang2019}
Zhuang M.-Y.,  Ho L.~C.,   Shangguan J.,  2019, \mn@doi [\apj]
  {10.3847/1538-4357/ab0650}, 873, 103

\bibitem[\protect\citeauthoryear{Zhuang, Ho  \& Shangguan}{Zhuang
  et~al.}{2021}]{Zhuang2021}
Zhuang M.-Y.,  Ho L.~C.,   Shangguan J.,  2021, \mn@doi [\apj]
  {10.3847/1538-4357/abc94d}, 906, 38

\bibitem[\protect\citeauthoryear{{Zinger} et~al.,}{{Zinger}
  et~al.}{2020}]{Zinger2020}
{Zinger} E.,  et~al., 2020, \mn@doi [\mnras] {10.1093/mnras/staa2607}, \href
  {https://ui.adsabs.harvard.edu/abs/2020MNRAS.499..768Z} {499, 768}

\makeatother
\end{thebibliography}


\appendix

\section{sSFR Plots} \label{ap: ssfr}

In this work we have mostly focussed on the simulations' predictions for the molecular gas as this is a common quantity observers study when looking at the impact of AGN feedback on galaxy quenching. For completeness, here we present the equivalent plots for the specific star formation rate.

\begin{figure*}
    \centering
    \includegraphics[width=0.9\textwidth]{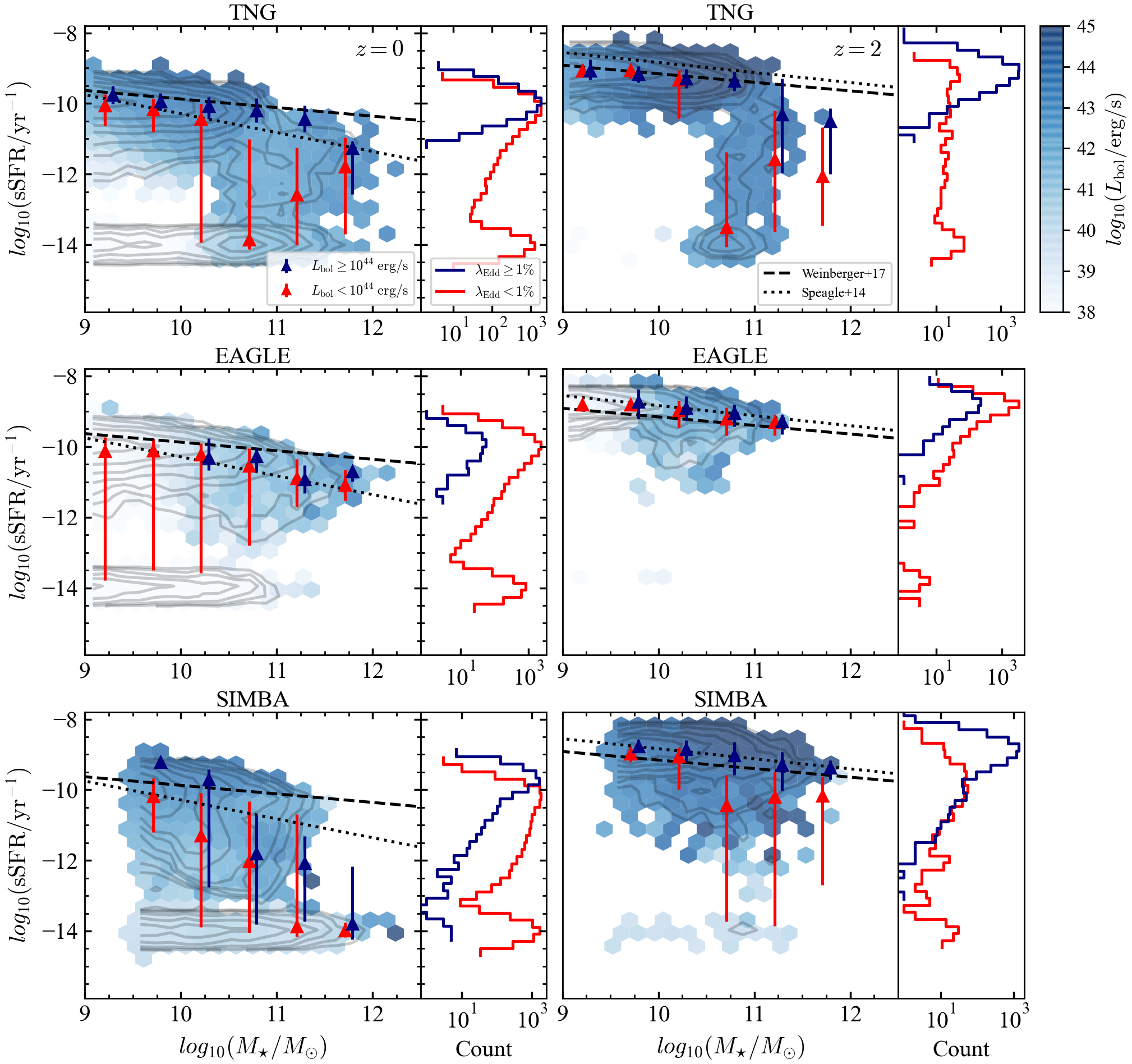}
    \caption{Galaxies in the sSFR$-M_\star$ plane coloured according to mean {\Lbol} within each pixel. The triangular points show the median and 16\ts{th}-84\ts{th} percentiles of the sSFR in mass bins of 0.5 dex, split by {\Lbol} in AGN (blue) and non-AGN (red). The panels on the right show the logarithmic number density for AGN selected by Eddington ratio - AGN are shown in blue and non-AGN in red.}
    \label{fig4sSFR}
\end{figure*}

Figure \ref{fig4sSFR} shows the number density of galaxies in the sSFR$-M_\star$ plane with pixels coloured by the mean {\Lbol}. We show the observational main sequence from \cite{Weinberger2017} and \cite{Speagle2014} as dashed and dotted lines respectively (note, we use \citealt{Weinberger2017} as our model throughout the paper, although we find no qualitative difference between the two). The sSFR bimodality is clear at $z=0$ and is beginning to be established at $z=2$ in TNG and SIMBA. Both high-{\Lbol} and high-{\fedd} AGN are preferentially located in high-sSFR galaxies.

\begin{figure}
    \centering
    \includegraphics[width=0.45\textwidth]{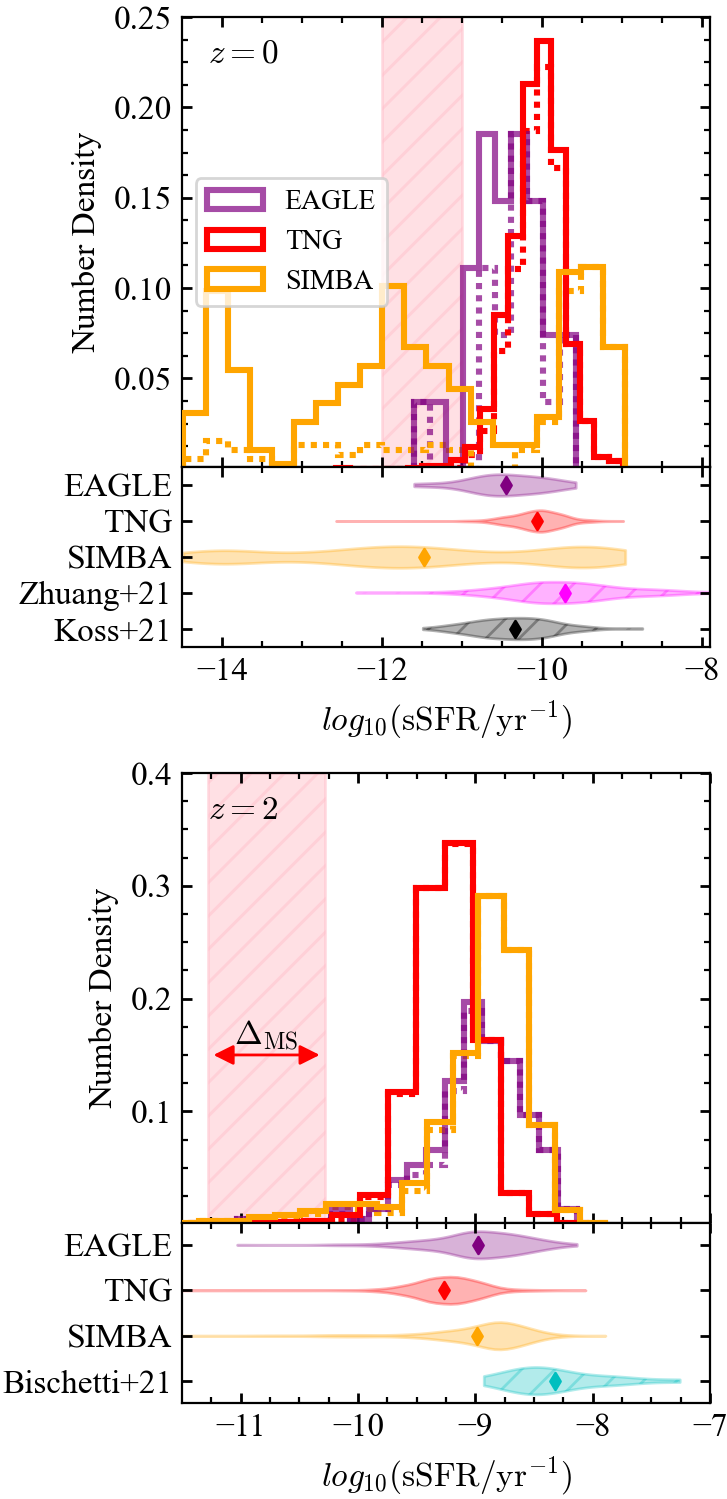}
    \caption{The distribution of sSFR for luminous AGN in the three simulations. Solid lines show all AGN with {\LbolEq{\geq}{44}} and dotted lines show the additional requirement of having {\feddEq{\geq}{1\%}}. The bottom panels show the smoothed density plot for each simulation compared to the comparison observational sample.}
    \label{fig5sSFR}
\end{figure}

In Figure \ref{fig5sSFR} we show the distribution in sSFR. At $z=0$ we see some clear differences between the simulations, with SIMBA showing a bimodal distribution. The lowest peak at sSFR$=10^{-14}\rm{\ yr^{-1}}$ is due to the galaxies which have unresolved star formation rates and should be considered part of the quiescent population at sSFR$<10^{-11}\rm{\ yr^{-1}}$. At $z=2$, the simulations all show high-sSFR with the peaks of all three simulations falling within $0.5 \rm{\ dex}$ of each other.

\begin{figure}
    \centering
    \includegraphics[width=0.45\textwidth]{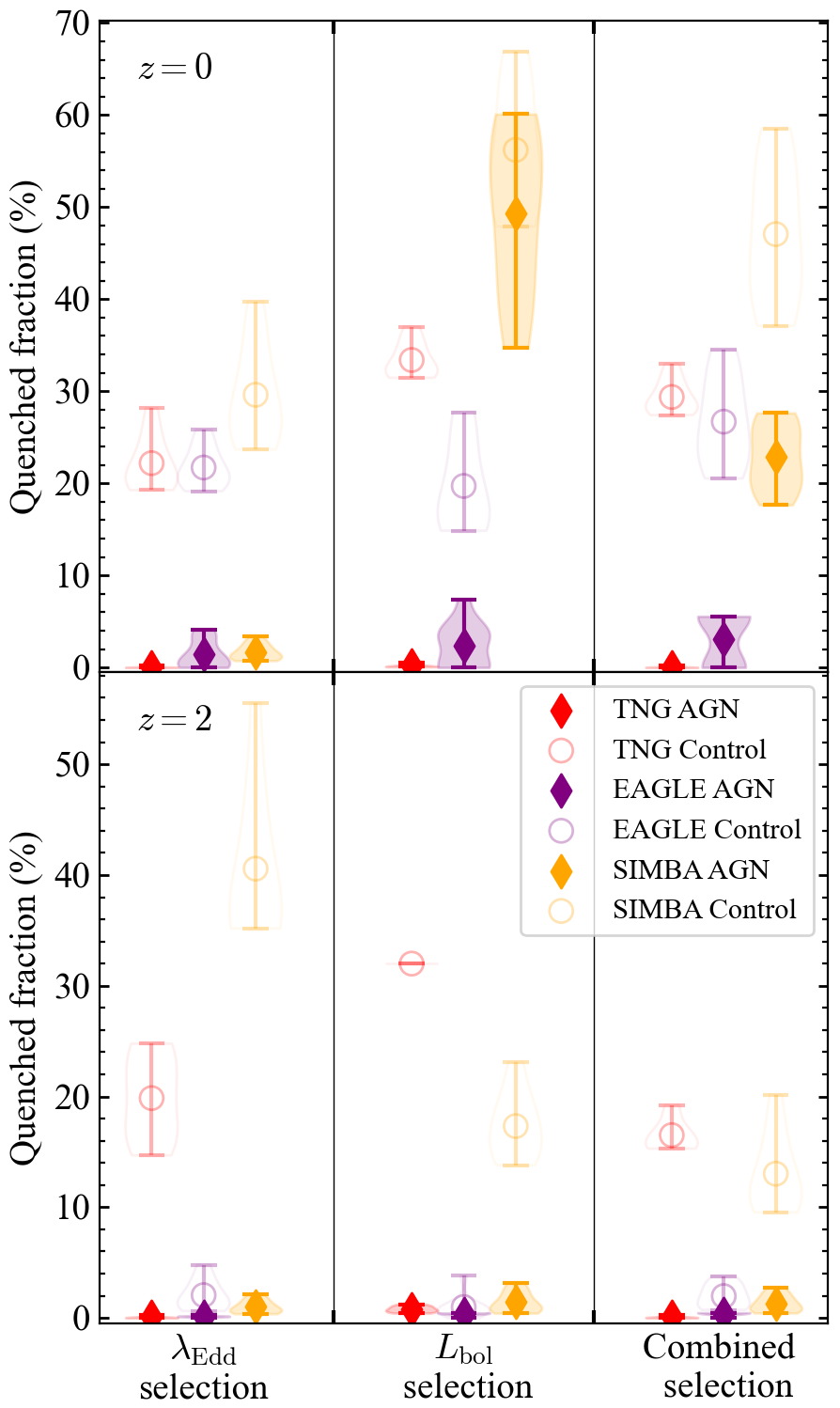}
    \caption{Quenched fraction for sSFR. We use three AGN selections: {\LbolEq{\geq}{44}}, {\fedd}$\geq 1\%$, and a selection combining the two. We also compare to a stellar-mass-matched control sample taken for each AGN selection criteria.}
    \label{fig6sSFR}
\end{figure}

Figure \ref{fig6sSFR} shows the quenched fractions for three different AGN selections and their corresponding mass-matched control sample. We can see that the quenched fraction for EAGLE and TNG is low for all three AGN selections at $z=0$, however, SIMBA shows a significantly higher value for the {\Lbol} selection. This is due to the fact that SIMBA's `jet' mode feedback can operate at high accretion rates and is efficient at quenching galaxies whilst still retaining such accretion rates (see Section~\ref{sim differences}).

\begin{figure*}
    \centering
    \includegraphics[width=\textwidth]{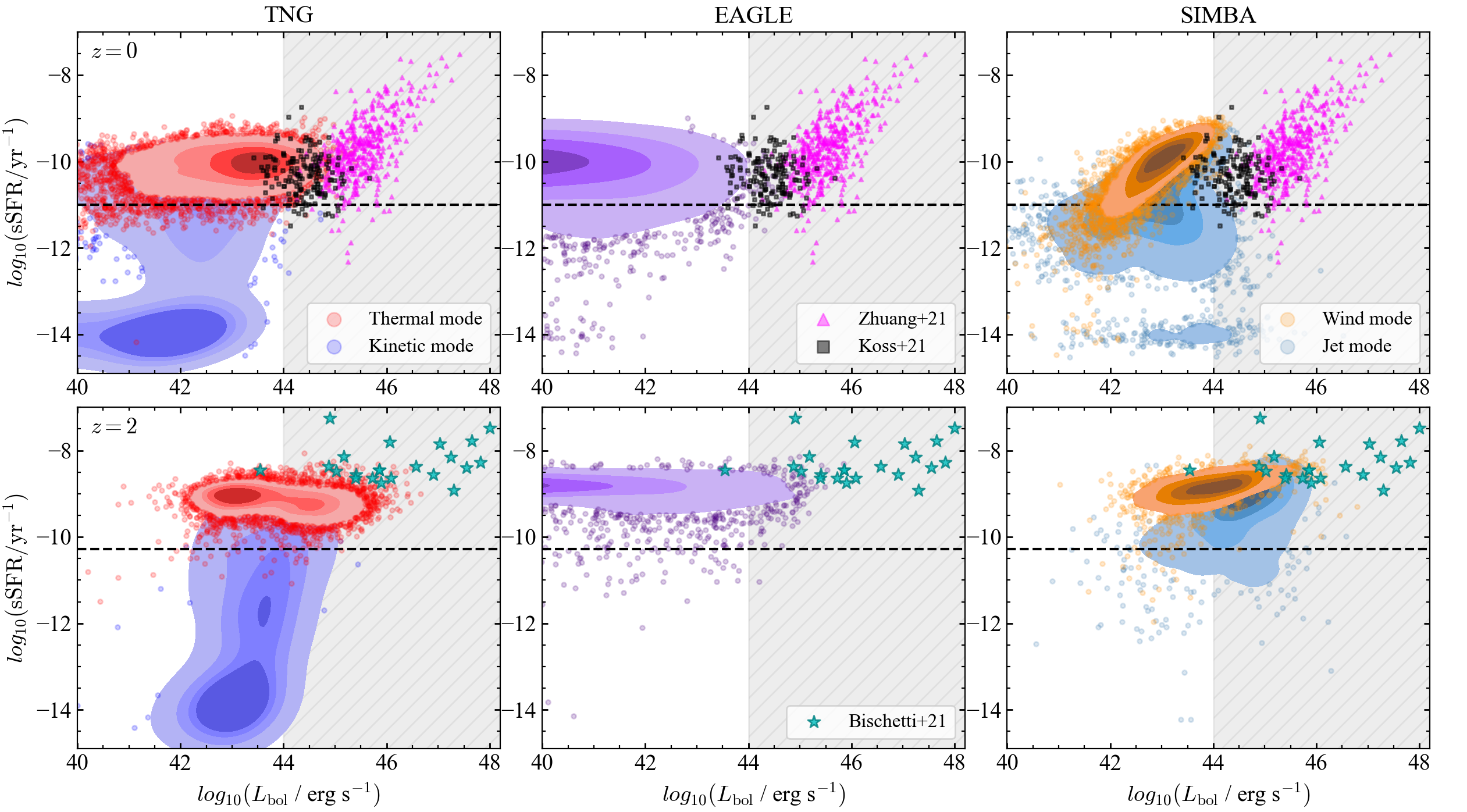}
    \caption{sSFR against {\Lbol} for the simulations at $z=0$ and $z=2$. We split TNG and SIMBA into their two main modes of feedback and also plot the observational sample. The blue shaded region represents our high-{\Lbol} selection and the black dotted line shows a rough quenching definition based on a fiducial galaxy of {\MstarEq{=}{10.5}} and a quenching distance from the main sequence of $\Delta_{\rm{MS}}=-1$.}
    \label{fig7sSFR}
\end{figure*}

In Figure \ref{fig7sSFR} we show the sSFR$-${\Lbol} plane for all three simulations. We also split TNG and SIMBA into their two primary feedback modes. The black dotted line shows a quenching definition for a {\MstarEq{=}{10.5}} lying $\Delta_{\rm{MS}}=-1$ below the main sequence. We can see that in TNG, almost all of the thermal mode (in red) sources lie above this line and that the quenched population is exclusively composed of kinetic mode sources (blue). Neither TNG or EAGLE have significant numbers of AGN in both the quenched region and the high-{\Lbol} region. In SIMBA, however, some quenched galaxies do have high luminosity and these are all in the low-{\fedd} `jet' mode.

\bsp	
\label{lastpage}
\end{document}